\documentclass[a4paper,10pt]{article}

\pdfoutput=1
\usepackage{jheppub}
\usepackage{array}
\usepackage{multirow}

\definecolor{green}{rgb}{0.1,0.8,0.2}

\def\JB#1{{\color{red}{ #1}}}

\makeatletter
\newcommand\footnoteref[1]{\protected@xdef\@thefnmark{\ref{#1}}\@footnotemark}
\makeatother

\title{An entropy current for dynamical black holes in four-derivative theories of gravity}
 
\author[a]{Jyotirmoy Bhattacharya}
\affiliation[a]{Department of Physics, Indian Institute of Technology
Kharagpur, Kharagpur 721302, India.}
 
\author[b]{, Sayantani Bhattacharyya}
\affiliation[b]{National Institute of Science Education and Research, HBNI, Bhubaneshwar 752050, Odisha, India.}
 
\author[b]{, Anirban Dinda}
 
\author[c]{, Nilay Kundu}
\affiliation[c]{Indian Institute of Technology, Kanpur 208016, India.}

\emailAdd{jyoti@phy.iitkgp.ac.in, sayanta@niser.ac.in, anirban.dinda@niser.ac.in, nilayhep@iitk.ac.in}

\abstract{We propose an entropy current for dynamical black holes in a theory with arbitrary four derivative corrections to Einstein's gravity, linearized around a stationary black hole.  The Einstein-Gauss-Bonnet theory is a special case of the class of theories that we consider.  Within our approximation, our construction allows us to write down a completely local version of the second law of black hole thermodynamics, in the presence of the higher derivative corrections considered here.  This ultra-local, stronger form of the second law is a generalization of a weaker form, applicable to the total entropy, integrated over a compact `time-slice' of the horizon, a proof of which has been recently presented in \cite{Wall:2015raa}.  We also provide a general algorithm to construct the entropy current for the four derivative theories, which may be straightforwardly generalized to arbitrary higher derivative corrections to Einstein's gravity. This algorithm highlights the possible ambiguities in defining the entropy current.}
 

\begin{document}

\begin{flushright} 
\end{flushright}

\def\Kt{\mathcal{K}}
\def\Kn{\overline{\mathcal{K}}}
\maketitle
\flushbottom
%

\section{Introduction and summary}\label{sec:intro}
It is widely believed that Einstein's theory of gravity must admit an adequate UV completion when we approach length scales comparable to Planck length. 
Such a putative UV complete theory of quantum gravity, at large length scales, must reduce to the weakly coupled, two derivative Einstein's theory, which has been exhaustively verified in the IR, by several experiments. However, at intermediate length scales, we may encounter a regime, 
where gravity is still weakly coupled and the quantum corrections are suppressed, 
but higher derivative corrections to Einstein equations cannot be ignored
\footnote{Such an intermediate regime exists, for instance, in string theory, which is a prominent candidate for the UV complete theory of quantum gravity. 
In string theory, within this regime, the string coupling $g_s \rightarrow 0$, implying the quantum corrections are suppressed. While, the ratio $ \ell_s/ \mathfrak R$ is non-negligible, 
$\mathfrak R$ being the length scale associated with space-time curvatures, while $\ell_s$ is the string length. 
Whether such an intermediate regime exists in the real world, or whether Einstein's description is a good description right up to the Planck scale, can only be answered with precision experiments of the future.}. 
In this regime, gravity would be described by an arbitrary diffeomorphism invariant classical theory  
\begin{equation}\label{arbdifth}
\mathcal L = \mathcal L_g \left( g_{\mu \nu} , R_{\mu \nu \lambda \sigma} , \nabla_\alpha R_{\mu \nu \lambda \sigma}, 
\nabla_\beta \nabla_\alpha R_{\mu \nu \lambda \sigma} + \dots \right) + \mathcal L_m, 
\end{equation} 
where, the gravity part of the Lagrangian would admit a derivative expansion, as follows 
\begin{equation}\label{arbdifth2}
\mathcal L_g = R + \left( \alpha_1 R^2 + \alpha_2 R_{\mu \nu} R^{\mu \nu} + \alpha_3 R_{\mu \nu \alpha \beta}R^{\mu \nu \alpha \beta} \right) + \text{higher derivatives}
\end{equation} 
Here, in \eqref{arbdifth}, $\mathcal L_m$ represents the matter part the theory 
\footnote{Note that, there would certainly be some phenomenological restrictions on $\mathcal L_m$. For example, it should reduce 
to the standard model Lagrangian at large length scales. Besides this, $\mathcal L_m$ here also incorporates all the matter fields, that may be required 
to make the higher derivative theory of gravity \eqref{arbdifth2} well defined. For instance, It was pointed out in \cite{Camanho:2014apa}, 
that causality constraints on tree level graviton three point functions imply that $\mathcal L_m$ should incorporate higher spin fields.}. Any non-minimal coupling to gravity, i.e., the terms involving curvature and the matter fields are also included in ${\mathcal L}_m$.
Note that, the specific values of the coefficients $\alpha_1$, $\alpha_2$ and $\alpha_3$ depend on the details of the UV complete theory. These are dimension full constants and therefore must be proportional to the (square of) some fundamental length scale of the UV complete theory. For example, in the case of string theory, all of these coefficients will have the form  
$\alpha_i \sim l_s^2 \tilde \alpha_i$ where $l_s$ is the string length and $\tilde \alpha_i$s are some numbers. The limit that we shall consider here, is the one, where the length scale associated with the curvatures of the space-time and those associated with the variations of the matter fields, are much larger compared to $l_s$. In other words, all the higher derivative corrections will be more and more suppressed, with the increase in the number of derivatives.
 
We know that the two derivative Einstein's theory admits black holes solutions with Killing event horizons. These solutions can be understood as macroscopic manifestation of an ensemble of many microscopic degrees of freedom of the more fundamental theory of gravity, in thermodynamic equilibrium, 
at finite temperature. If this statistical picture of the black hole is correct, then stationary black hole solutions should exist even after we add higher derivative corrections to the gravity action. Also, within these higher derivative theories of gravity, we should be able to construct macroscopic quantities, such as entropy, for the black hole solutions, which will satisfy the laws of thermodynamics.

In two derivative theory of Einstein's gravity, we have a candidate for entropy that satisfies both the first and second laws of thermodynamics. 
The entropy, in this case, is given by the area of a `time-slice' of the event horizon \cite{PhysRevLett.26.1344, PhysRevD.7.2333, Hawking1972}
(also see \cite{waldbook, Wald:1995yp, blau2011lecture}). 
We shall denote the even horizon with $\mathcal H$ and the time-slice of it with $\mathcal H_v$. 
The second law, for this entropy, followed from the famous area increase theorem for black holes \cite{PhysRevLett.26.1344, waldbook}, 
which assumes that the matter energy-momentum tensor obeys the 
null-energy condition. Throughout our discussion in this note, we shall assume this condition to be valid for the matter part of the Lagrangian $\mathcal L_m$.

This concept of entropy was generalized to stationary black hole solutions in higher derivative theories of gravity in \cite{PhysRevD.48.R3427, Iyer:1994ys}, in such a way that it satisfied the first law of thermodynamics \footnote{See \cite{Wall:2018ydq, Sarkar:2019xfd} for the latest reviews of black hole thermodynamics in higher derivative theories of gravity.}. 
We shall refer to this construction of entropy as Wald entropy $S_W$. Now, the first law of thermodynamics relates the infinitesimal shifts in the parameters of two different but nearby equilibrium configurations. Therefore, the Wald entropy, whose construction was solely based on consideration of first law alone, 
does not unambiguously extend to dynamical situations. Indeed, as it was pointed out in \cite{Jacobson:1993xs, Jacobson:1993vj, Jacobson:1995uq}, there were ambiguities associated with Wald entropy, for non-stationary black hole solutions with dynamical event horizons. All these ambiguities vanished for stationary solutions. 
We shall refer to these ambiguities as the JKM ambiguities. 

Unlike, two derivative Einstein's theory, it is not a priori clear whether Wald entropy satisfies the second law of thermodynamics. In this note, we would like 
to explore this question further.

In the weakest version, the second law could be stated as follows. 
Consider two equilibrium configurations (in our case two black hole solutions, not necessarily close by in any sense) 
$B_1$ and $B_2$ such that if one perturbs $B_1$ in certain ways it is possible to reach $B_2$ eventually. 
Then the entropy evaluated on the solution $B_2$ must be strictly greater than the entropy of $B_1$.

Though the above formulation of the second law does not really need a definition of entropy away from stationarity, clearly it refers to dynamics. One way towards a proof would be to show that there exists some extension of Wald entropy to dynamical situations so that the second law is satisfied. It is natural to expect that this extension (if at all possible) might fix those ambiguities related to the definition of entropy, which only arises in non-stationary situations (i.e. the JKM ambiguities).

Now it is very difficult to analyze dynamical black hole solutions even in two derivative Einstein's theory of gravity. People usually take recourse to several perturbation schemes, around the stationary solutions that are known exactly.
In this context, the simplest situation that comes to mind is the case where some stationary black hole is slightly perturbed by some external agent so that the resulting dynamical black hole metric could be decomposed as a sum of stationary part and the time-dependent part with small amplitude. Then one can analyze the equations in an expansion in terms of the amplitude. Note that under this approximation, it is not possible to study violent processes such as the formation of black holes or the merger of two black holes. Throughout this note, we shall work only up to linear order in the amplitude of fluctuations, and therefore, our results would not directly apply to these violent scenarios. See \cite{Liko:2007vi, Sarkar:2010xp, Chatterjee:2013daa}, where similar questions relating to entropy and the second law, for processes involving the merger of black holes have been addressed. 

In \cite{Wall:2015raa} the author has used this expansion to construct one `out of equilibrium' extension of Wald entropy, which, up to linear order in amplitude expansion, is monotonically increasing at every instant of `time' and therefore satisfies the second law in a stronger sense 
(also see \cite{Sarkar:2013swa, Bhattacharjee:2015yaa, Bhattacharjee:2015qaa, Wall:2011hj}).
 This locality in time is not entirely unexpected in this type of set-up where the space-time is `near' some equilibrium or stationary solution at every instant of time. 
Following the same intuition, we could also say that for such slow enough, `near-equilibrium' time evolution, where we could assume that different sub-regions of a large macroscopic system are in approximate equilibrium with its immediate neighbourhood, at every instant of time, we should also expect a spatial locality, in the formulation of second law. 
This expectation is completely consistent with the scenario in Einstein's theory of gravity, where the area increase theorem is valid locally, for every infinitesimal area element of a `time-slice' of the horizon. Our expectation for a local, stronger form of second law, is very much motivated by this example of Einstein's theory. 
Now, in a more general setting involving higher derivative corrections to Einstein's equation, 
during slow time evolutions, besides entropy production in every infinitesimal sub-region, we have to be also open to the possibility 
that entropy could be redistributed between the neighbouring regions, by flowing in or out via some spatial current.
The necessity of having such non-zero spatial current for entropy, and the existence of a strong ultra-local form of the second law of thermodynamics, in higher derivative theories of gravity as well, are the key points of our investigation here. 

In this note, we shall demonstrate that it is possible to formulate the second law in its strongest form,
 so that at least for `slow enough' dynamical situations, entropy is produced at every point of the evolving space-time, up to a possible inflow and outflow via some spatial current\footnote{An entropy current with non-negative divergence certainly exists in near-equilibrium states for theories that do not include dynamical gravity
 \cite{Bhattacharyya:2008xc, Bhattacharyya:2012nq, Bhattacharyya:2014bha, Bhattacharyya:2013lha}. 
 In the case of gravity, where the space-time itself becomes the fundamental dynamical object, the concept of locality might become a bit confusing. The locality in space-time in some sense becomes analogous to some form of locality in the space of fundamental fields of non-gravitational theories.
However, the kind of perturbation that we are considering here, there is always a stationary base metric which could play the role of the background and the above-mentioned issues could be avoided.}. We will explicitly construct this spatial current for entropy flow, in the most general four derivative theory of gravity.

Let us now outline the organization of this note, along with a brief summary of the key arguments and results in the various sections. 

At first, in \S\ref{sec:review}, we shall review the paper \cite{Wall:2015raa} in detail.
The author in \cite{Wall:2015raa} has shown that at the leading order in amplitude expansion, certain `time-time' component of the equation of motion of any higher derivative theories of gravity could always be written as two `time derivatives' acting on some quantity. Then he could further argue that if one identifies the integral of this quantity (the expression on which the two time derivatives are acting) over $\mathcal H_v$, with the entropy of the gravitational theory, 
then it will satisfy the second law at least at the leading order within this approximation.
Following \cite{Wall:2015raa} we have first set up an appropriate coordinate system, thus defining the `time' mentioned above. Next, by using a symmetry of the horizon geometry (referred to as `boost symmetry' in \cite{Wall:2015raa}), we have classified the terms that can appear in that particular `time-time' component of the equations of motion, according to their weight under this boost transformation. We shall see that the argument and construction in \cite{Wall:2015raa} smoothly 
goes through for all the higher weight terms except the one that appears at zero boost weight.

This point was noted in \cite{Wall:2015raa} and it has been argued that if these `zero boost terms' are not of the correct form (i.e., two time derivatives acting on some quantity) it would amount to the violation of the first law itself, once viewed in the `physical version' formulation of it \cite{Jacobson:1995uq,Gao:2001ut} 
(also see \cite{Amsel:2007mh,Bhattacharjee:2014eea,Chakraborty:2017kob,Chatterjee:2011wj,Kolekar:2012tq}). 
Therefore, though the central argument in \cite{Wall:2015raa} naively break down for these special `zero boost terms', it must work out in actual theories, where the physical process version of the first law is valid. 

Note that the formulation of the `physical process version ' of the first law uses exactly the same setup as the one used in \cite{Wall:2015raa}. Here also one perturbs the stationary black hole out of equilibrium and lets it settle to another nearby stationary solution with slightly shifted parameters. The first law is a relation between these shifts of parameters, which characterize the two equilibrium solutions\footnote{We would like to emphasize that though the proofs of both first law and second law use the same set-up, they are very different in terms of details. In particular, Wall's construction could fix many more terms in entropy (usually denoted as JKM ambiguities in literature) that do not contribute to the first law at all.}. 
In the arguments leading to this physical version of the first law, the external agent which drives the system out of equilibrium is a very specific one - some small matter (associated with a small shift in matter stress tensor) entering the system through asymptotic infinity. The similarity between the two set-up of the second law and the physical process version of the first law is very suggestive of the fact that the structural nature of the terms in entropy, which play a major role in the proof of the physical process version of first law, would also be extremely important in the proof of the second law. We shall refer to such terms as `zero boost terms', the justification of such terminology would be explained later in the main text.

After this extensive review of \cite{Wall:2015raa},
we shall closely study how the physical version of the first law constrains these `zero boost terms'. 
We shall find that locally the required `time-time' component of the equation of motion 
(let us denote this `time as $v$ and the relevant component of the equation of motion as $E_{vv}$) 
need not have the form specified form which is naively implied by the physical process version of the first law. This naive expectation would be that the zero boost terms in $E_{vv}$, has two `time' derivatives acting on some local quantity (let us denote it as $J^v$) defined on the horizon. 
This naive expectation is not accurate, since any term that could be expressed as a single `time' derivative acting on the spatial divergence of some 
space current (denoted here as $J^i$) may also be present in $E_{vv}$, without affecting the first law.
This is because the physical process version of the first law deals with a total change in entropy ( along with the charge and mass) as the black hole evolves from one equilibrium to another. The total entropy always comes with integration overall spatial section of the horizon $\mathcal H_v$.  
In that case, any such total divergence term would just integrate to zero.
 
It should be noted that such a term will not affect the argument of \cite{Wall:2015raa}, which proves a weaker version of the second law, in which the total entropy (integrated over $\mathcal H_v$) has been considered. This weaker form is a local statement in time (i.e, total entropy increases at every instant of time) but not in space. 
At every stage of the arguments in \cite{Wall:2015raa}, the integration over the spatial sections of the horizon played an important role.

Once we have realized that, it is possible to introduce the notion of a spatial entropy current, without affecting the proof of both the first law and the second law (even a strong ultra-local form of it), the next immediate question is whether such a spatial entropy current is necessary. In other words, we should investigate that, if we were to write down an ultra-local version of the second law, largely following the procedure of \cite{Wall:2015raa}, can we do it without introducing the entropy current, 
in any higher derivative theory of gravity. In more practical terms, we need to check whether the relevant 
 `zero boost terms' in the equation of motion $E_{vv}$ for a given higher derivative theory of gravity, does indeed have terms which give rise to the spatial entropy current 
 $J^i$. To answer this question, we specialize to four-derivative theories of gravity. 
 In \S\ref{ssec:4dercor}, we explicitly compute the relevant component of the equations of motion and we see that there has to be a spatial entropy current in some of these four derivative theories, if we want a completely local version of the second law to be true. This is the central result of our note. 
Besides achieving manifest locality, our procedure of constructing the entropy current
might play a crucial role in providing an alternative proof of the second law, 
without invoking the `physical process' version of the first law, the use of which has been a necessary input for the proof presented in \cite{Wall:2015raa}. We would like to investigate this possibility further in future work.

In this context, the four-dimensional Gauss-Bonnet theory requires a special mention. We have discussed this case in details in \S\ref{ssec:gb3p1}. 
It is well known that in four dimensions Gauss-Bonnet action is a total derivative and therefore does not contribute to the equation of motion. However, from Wald's analysis, we know that the entropy of the black holes in Gauss-Bonnet theory does receive correction which is proportional to the intrinsic Ricci-scalar evaluated on the two-dimensional spatial section of the horizon. Integration of this quantity over a compact two dimensional manifold results in a topological quantity, the Euler characteristics, that does not change under small continuous deformation of the horizon caused due to the perturbation. 
This is perfectly consistent with the fact that 4-d Gauss-Bonnet term does not introduce any correction to the equation of motion 
and so (following the argument of \cite{Wall:2015raa}) no correction to the change in total entropy during time evolution.
However, if we are thinking of in terms of the entropy density (i.e., the same intrinsic Ricci scalar without the integration over all spatial sections of the horizons), 
it does evolve with time. But, this entropy density, clearly, would not satisfy the ultra-local version of the second law. However, 
the validity of the local version of the second law is restored, if we also consider a spatial entropy current. It is satisfying to check that, for the $3+1$ dimension
the $v$-derivative of the Ricci scalar is identical to the time derivative of the divergence of a spatial current (given in terms of the extrinsic curvatures of $\mathcal H_v$). These
two contributions from the entropy density and the spatial entropy current, to the equation of motion, cancel out each other. This cancellation is off-shell 
and specific to $3+1$ dimensions only. Thus, in this simple example, it is easy to recognize the necessity of the entropy current, even before performing the detailed calculation. As we will see in more detail in \S \ref{ssec:gb3p1}, this example also helps us to identify an ambiguity present in the definition of the entropy current. 

In \S \ref{ssec:evvgenstr}, \S \ref{ssec:evvgenstr1} and \S \ref{ssec:entcurgenpro}, we go on to develop a general 
algorithm for constructing the spatial components of the entropy current for arbitrary four derivative theory. 
From this exercise, we learn that the most general form of the relevant equation of motion $E_{vv}$, which is consistent with the boost symmetry, 
has a structure that is more general compared to what would be essential for defining the entropy current. 
In other words, the fact we have an entropy current and consequently a local second law, puts very non-trivial 
constrains on the most general possible structure of $E_{vv}$ \footnote{As we have mentioned above, an explicit calculation for the four derivative theories demonstrates that, these constraints are automatically met for these theories.}. 
Although at the moment we do not have a precise explanation regarding the physical origin of these constraints, we believe that these may arise due to some residual gauge freedom. We think understanding the exact mathematical reason behind these constrains, would lead us to a proof of the local second law through the construction of the entropy current, 
without invoking the first law at all. 

This general algorithm has also helped us to understand the ambiguities related to the construction of the current more clearly. In \S\ref{ssec:entcurgenpro1}, we report on one of the primary sources of such ambiguities. Finally in \S\ref{sec:disco}, 
we conclude and discuss possible future directions.

Before concluding this section, we would like to mention that, the notion of an entropy current for black hole dynamics in higher derivative theories of gravity, 
is not completely new in this note. This idea has been previously introduced in \cite{Chapman:2012my, Eling:2012xa, Dandekar:2019hyc}.
In \cite{Chapman:2012my, Eling:2012xa} it was primarily motivated by the entropy current, constructed in the context of the fluid gravity 
correspondence \cite{Bhattacharyya:2008xc}. While in \cite{Dandekar:2019hyc}, the entropy current was constructed exploiting the membrane-gravity duality, using an expansion in inverse powers of space-time dimension. 
Although, the exact context of these constructions are different from our considerations here, but there are some similarities in the basic idea (see \S\ref{sec:disco} for further discussions on this). 
The exact relation between our construction and that reported in these papers 
is a topic of our current investigation and we hope to report on it in the near future. 
%

\section{A comprehensive review of \cite{Wall:2015raa}}\label{sec:review}
%
%
As we have discussed in \S\ref{sec:intro} in an arbitrary diffeomorphism invariant theory of gravity, a proof of the second law for dynamical black holes was provided in \cite{Wall:2015raa}. Let us  review the details of the proof here, which would serve as a useful prelude to the subsequent discussion of our entropy current. 

Let us first choose a coordinate system. 
Let $\partial_v$ be the null generators of the event horizon, where $v$ is the affine parameter. Let $\partial_i$s denote the rest of the 
spatial tangents of the horizon. Integral curves of $\partial_i$s are the spatial coordinates along the constant $v$ slices of the horizon. Then from every point on the horizon, we shoot off a set of null geodesics, making a precise angle with the coordinates on the horizon. We label each of these new sets of null geodesics (null everywhere) by the coordinates of the point at which it intersects the horizon. We denote the affine parameter along the null geodesics to be $r$ which is the coordinate,  away from the horizon. The most general metric with this choice of coordinates would have the following structure (see appendix of \cite{Bhattacharyya:2016xfs})
\begin{equation}\label{Wallmet}
\begin{split}
&~ds^2 = 2 dv~ dr - r^2X(r,v,x^i)~ dv^2 + 2 r~\omega_i(r,v,x^i) ~dv ~dx^i 
+ h_{ij} (r,v,x^i)~dx^i~ dx^j\\
\end{split}
\end{equation}
Here we have chosen the horizon to be at $r=0$ (a choice for the origin of the affine parameter along each null geodesic $\partial_r$). 
Note that this choice of gauge is slightly different from that of \cite{Wall:2015raa}. We have set $g_{rv} = 1$ throughout space-time, but in 
\cite{Wall:2015raa} this condition was set only on the future horizon $\mathcal H$. In Appendix A of \cite{Bhattacharyya:2016xfs}, it was demonstrated that this 
choice of metric \eqref{Wallmet}, is possible without any loss of any generality, even for dynamical black holes.  
We would also like to emphasize that this difference in gauge choice, do not affect the arguments in \cite{Wall:2015raa}, in any way. 
We shall work with this slight difference in this note, simply because we prefer to work with a metric where the gauge fixing is more complete. 

Given the form of the metric \eqref{Wallmet}, let us now outline the broad strategy of the proof of second law provided in \cite{Wall:2015raa}.

\subsection{Strategy of the proof of second law of black hole thermodynamics}\label{ssec:stgprf}

The general strategy of the proof follows that of area increase theorem for dynamical black holes in Einstein gravity \cite{PhysRevLett.26.1344,waldbook}. 
Following \cite{Wall:2015raa} we shall consider small time-dependent fluctuations about stationary black holes. Let us denote the amplitude of the fluctuation to be $\epsilon$. All the analysis would be linear in the amplitude (denoted by $\epsilon$) of this fluctuation.

Let us denote a $v$-slice of the horizon of the dynamical black hole to be $\mathcal H_v$. In \cite{Wall:2015raa} $\mathcal H_v$ 
has been considered to be compact, an assumption which played an extremely important 
role in the proof. This assumption ensured certain boundary terms to vanish. Therefore, 
even if the horizon was non-compact, but those boundary terms continued to vanish, the proof of \cite{Wall:2015raa} is completely valid. However, in the present paper, our local statement of the second law should not be sensitive to the compact nature of the horizon or depend on the vanishing of such boundary terms.  
Then, under the approximations considered here, let us schematically write down the entropy of the black hole, in an out of equilibrium scenario, to be
\begin{equation} \label{entsch1}
S = \int_{\mathcal H_v} \sqrt{h} (1 + s_n)
\end{equation} 
where 
$$ s_n = s_w^{\text{HD}} + s_c.$$
Here $s_w^{\text{HD}}$ are the corrections to the area law, coming from the Wald entropy formula due to the presence of the higher derivative corrections to the Einstein-Hilbert action.
Note that, in this notation the Wald entropy is given by 
\footnote{Here, we have treated the area term corresponding to Einstein theory separately, to facilitate comprehension for our readers who are familiar with the area increase theorem. }
\begin{equation} \label{entsch2}
S_W = \int_{\mathcal H_v} \sqrt{h} (1 + s_w^{\text{HD}}) = \int_{\mathcal H_v} 
{\partial \mathcal{L} \over \partial R_{\mu\nu\rho\sigma}} \, \epsilon_{\mu\nu} \epsilon_{\rho\sigma},
\end{equation} 
where $\epsilon_{\mu\nu}$ are the bi-normal to $\mathcal{H}_v$, the co-dimension$-2$ spatial slicing of the horizon.
Here, $s_c$ are further corrections to the Wald entropy, which are a part of the JKM ambiguity. One of the central idea of the proof, is to choose an appropriate $s_c$  so as 
to ensure $\partial_v S \geq 0$. This in turn, therefore, fixes the ambiguity. 

Now let us act \eqref{entsch1} with a $v$-derivative, which can be moved inside the integral in the RHS, since the integral is over a $v$-slice of the horizon. We have 
\begin{equation}
\partial_v S = \int_{\mathcal H_v} \partial_v \left( \sqrt{h} (1 + s_n) \right) \equiv \int_{\mathcal H_v} \sqrt{h} ~\vartheta,
\end{equation}
where $h_{ij}$ is the induced metric on $\mathcal H_v$, and 
$$\vartheta = \vartheta_E + \frac{1}{\sqrt{h}}\partial_v \left(  \sqrt{h}  s_n\right),$$
with $\vartheta_E$ being the contribution coming from the area form, which is present even in pure Einstein gravity without any higher derivative corrections. 
We can show that $\vartheta_E =  \frac{1}{2} h^{ij} \partial_v h_{ij}$ is the expansion of the congruence of the null generators of the horizon 
\cite{blau2011lecture}.

Following the proof of the black hole area increase theorem\cite{PhysRevLett.26.1344,waldbook} for Einstein gravity, the general strategy for proving $\partial_v S \geq 0$, is to demonstrate 
that $\partial_v \vartheta \leq 0$. This, together with the additional physical expectation  
\footnote{Here, the physical expectation is that the dynamical black hole will settle to a stationary metric with a Killing horizon at $v \rightarrow \infty$, leading to the vanishing of 
$\partial_v S$ at $v \rightarrow \infty$.}
$\vartheta|_{v\rightarrow +\infty} \rightarrow 0$, 
implies that $\vartheta \geq 0$, for all $v \geq 0$.

Now at linear order in amplitude both inequalities (i.e., $\partial_v S \geq 0$ and $\partial_v \vartheta \leq 0$),  must be some equality relation since terms linear in amplitude ($\epsilon$) could have any sign depending on the sign of  $\epsilon$. In other words, the only way the inequalities could be satisfied is to set them to zero at linear  order in $\epsilon$.
\begin{equation}\label{eq:arg1}
\partial_v \vartheta= {\cal O}\left(\epsilon^2\right)
\end{equation}

We shall try to choose $s_n$  such that $\partial_v \vartheta = {\cal O}\left(\epsilon^2\right)$ is ensured.
More precisely, by equation \eqref{eq:arg1}, what we mean is the following. 
\begin{itemize}
\item We shall consider only those dynamics where every metric component $G_{AB}$, that are not already fixed by our gauge choice, could be decomposed as
$$G_{AB} = G^{(0)}_{AB} + \epsilon~ \delta G_{AB}$$
where $G^{(0)}_{AB}$ is the non-dynamical part of the metric and has a time-like Killing vector  with a Killing horizon. $\delta G_{AB}$ is time-dependent. $\epsilon$ is the small parameter encoding the amplitude of dynamics, which could be of either sign but always small. All terms quadratic or higher-order in $\epsilon$ would be neglected.
\item We further demand that $G^{(0)}_{AB}$ is an exact solution of the Einstein equation with appropriate higher derivative corrections and also relevant matter stress tensor. Additionally, 
$G_{AB}$ also solves the same equations but up to corrections of order ${\cal O}\left(\epsilon^2\right)$.
\item Now our goal is to construct  an $s_n$ out of $G^{(0)}_{AB}$ and $\delta G_{AB}$ such that if we just blindly evaluate $\partial_v \vartheta$ and impose equations of motion it turns out to be order ${\cal O}\left(\epsilon^2\right)$ (or just vanishes within our approximation).
\end{itemize}
One of the key points of \cite{Wall:2015raa} is to provide an algorithm to construct such an $s_n$ in all possible higher derivative theories of gravity.

At this stage we would like to emphasize that equation \eqref{eq:arg1} is a necessary condition for second law, but certainly not sufficient, even within this perturbative treatment. Sufficiency would demand a particular sign for the coefficient of the ${\cal O}\left(\epsilon^2\right)$ term and in those special space-time points where this coefficient also vanishes, one has to keep track of even the higher-order terms. However, as it is the case in \cite{Wall:2015raa}, in this note we shall confine ourselves to computations only up to order ${\cal O}(\epsilon)$. They themselves turn out to be constraining enough to fix a large part of ambiguities that are there in the form of gravitational entropy for higher derivative theories.

Now let us process equation \eqref{eq:arg1} little further, which will finally tell us how,  manipulating a particular component of equations of motion, we could construct some $s_n$ that satisfies equation \eqref{eq:arg1}.
\begin{equation} \label{dvth}
\begin{split}
\partial_v \vartheta &= \partial_v \vartheta_E + \partial_v \left( \frac{1}{\sqrt{h}} \partial_v \left(  \sqrt{h}  s_n\right) \right)  \\ 
&= - R_{vv} + \partial_v \left( \frac{1}{\sqrt{h}} \partial_v \left(  \sqrt{h}  s_n\right) \right) + {\cal O}\left(\epsilon^2\right)\\ 
&= - T_{vv} + E_{vv}^\text{HD}+ \partial_v \left( \frac{1}{\sqrt{h}} \partial_v \left(  \sqrt{h}  s_n\right) \right) + {\cal O}\left(\epsilon^2\right)\\
\end{split}
\end{equation} 
Here $T_{vv}$ denotes the $vv$ component of the matter stress tensor and $E_{vv}^\text{HD}$ is the $vv$ component of the higher derivative corrections to the gravity part of the equations of motion.
In the second line we have used the fact  
$$\partial_v \vartheta_E= - R_{vv} + {\cal O}\left(\epsilon^2\right)$$
This is essentially the Raychaudhury equation for the congruence of null geodesics and this is an off-shell equation - it does not require the metric to satisfy any particular equation of motion.
We have used the equation of motion while going from second to the last line of equation \eqref{dvth}
\begin{equation}\label{eq:graveq}
R_{vv} + E_{vv}^\text{HD} = T_{vv}
\end{equation}
In all of our analysis this the only place where we shall use the on-shell condition on the metric components.

Now let us analyze the $\epsilon$ dependence of $T_{vv}$. We would like to argue that $T_{vv}$ is also of order ${\cal O}(\epsilon^2)$ and therefore does not contribute within our approximation.

In the case of higher derivative theory, the definition of matter stress tensor might become a bit confusing. Our convention is the following.
If we vary the action with respect to the metric fluctuation, the resultant two-indexed tensors could be categorized in two different classes; terms that  depend only on the metric components and terms that along with the metric components, also depend on the matter fields. All the higher derivative terms that are of the first category, are together called as $E_{AB}^\text{HD}$ and the matter stress tensor $T_{AB}$ consists of all the terms in the second category.

 Clearly, if we want to know the $\epsilon$ dependence of $T_{vv}$, we need to fix the $\epsilon$ dependence of the matter fields. Let $\Phi$ denotes all the matter fields (collectively) and let's assume that it also admits the following expansion.
 $$\Phi = \Phi^{(0)} + \epsilon~ \delta\Phi$$
 Here $\delta\Phi$ encodes the dynamics and $\Phi^{(0)}$ is the value of $\Phi$ on the stationary situation i.e., when all field configurations, including both metric and the matter fields, admit a Killing vector. As in the case of metric, we want  $\Phi^{(0)}$ to satisfy the equations of motion (on the background of stationary metric $G^{(0)}_{AB}$) for the matter field exactly and $\delta \Phi$ up to linear order in $\epsilon$.
 
 We shall consider only those matter stress tensors that satisfy the null energy condition. In our context, it implies that as long as the matter fields satisfy their equations of motion (in any smooth background geometry that need not be dynamical), the $vv$ component of the stress tensor is always non-negative.
 $$T_{vv}\geq 0$$
 We would like to stress again that the validity of the above condition requires only the matter fields to be on-shell, but the metric need not be.
 Now one can argue that in a stationary situation (i.e., in the limit of $\epsilon\rightarrow 0$) $T_{vv}$ simply vanishes (see appendix \S\ref{app:boostinv} for the details). Any quantity that satisfies some positivity condition and also vanishes in a stationary situation, must be quadratic in the amplitude of dynamics since the linear term could have either sign. It follows (exactly for the same reason as in equation-\eqref{eq:arg1})  that
$T_{vv}$ is also of order ${\cal O}\left(\epsilon^2\right)$ at every order in derivatives.

Now equation \eqref{dvth} could be  simply satisfied for some choice of $s_n$, provided $E_{vv}^\text{HD}$ has the following off-shell form\footnote{Note it is very important that the form predicted in \eqref{eomgbint1} is an off-shell requirement on $E_{vv}^\text{HD}$. Since we have already argued that $T_{vv}$ is of order ${\cal O}\left(\epsilon^2\right)$, equations of motion for the metric \eqref{eq:graveq} ensures that on-shell $E_{vv}^\text{HD}$ must be of order ${\cal O}\left(\epsilon^2\right)$. In other words, just like the solutions of any other differential equations, on-shell we do not have the freedom of determining the $\epsilon$ dependence of terms involving large number of derivatives, once the lower derivatives are fixed. 
However, our final goal is to construct an expression for $s_n$ and we can actually achieve this goal by treating $E_{vv}^\text{HD}$ off-shell, where  the naive $\epsilon$ counting works.}
\begin{equation} \label{eomgbint1}
E_{vv}^{\text{HD}}\Vert_\text{offshell} = \partial_v \left( \frac{1}{\sqrt{h}} \partial_v \left(  \sqrt{h} ~\varsigma\right) \right)  + \mathcal O(\epsilon^2). 
\end{equation}

If equation \eqref{eomgbint1} is true, then  one could just choose $\left[\int_{{\cal H}_v}\sqrt{h}~s_n\right]$ to be minus of $\left[-\int_{{\cal H}_v}\sqrt{h}~\varsigma\right]$ up to correction of order ${\cal O}(\epsilon^2)$.

Let us pause here for a moment, to make an important observation about a very special situation. Imagine a situation in which, a fluctuation in the  matter field with a very small amplitude sources the metric, through its energy-momentum tensor. The first correction to the zeroth-order stationary metric 
is entirely due to back-reaction from this source. In such a situation, the first change 
in the matter fields is of order $\epsilon$, and the energy-momentum tensor is of order $\epsilon^2$; but, the first correction to the metric would be 
of  order $\epsilon^2$. The $\mathcal O({\epsilon})$ piece for the metric would be trivially zero, in this special case, when the boundary conditions 
are chosen suitably.  So everything we have said so far, for the
$\mathcal O({\epsilon})$ coefficient in the metric remains true, but trivially true.

Now, in this special case, since the first correction to the metric occurs at $\mathcal O({\epsilon^2})$, all our conclusions in this note, regarding the linearized 
corrections to metric, would then be applicable to the $\mathcal O({\epsilon^2})$ terms. The only crucial difference would be that,
 instead of the equality \eqref{eq:arg1}, we would 
now get an inequality for the coefficient of $\epsilon^2$ in $\partial_v \vartheta$, i.e. $\partial_v \vartheta |_{\epsilon^2} \leq 0$. 
This happens because of \eqref{dvth}, where, due to the null-energy condition, $T_{vv}$ now contributes positively, at the same order at which the 
metric receives its first corrections. 
The cancellation between $E_{vv}^\text{HD}$ and $ \partial_v \left( \frac{1}{\sqrt{h}} \partial_v \left(  \sqrt{h}  s_n\right) \right) $ is realized in exactly  
the same way as discussed earlier. 
This, in turn,  implies that $\partial_v S \geq 0$, the inequality being 
important even for the linear (first non-trivial) corrections to the metric. Note that, this situation 
is perhaps physically important, since this is one of the simplest situations where we can realize a dynamical event horizon, by throwing a tiny amount of matter 
towards the black hole.

In \cite{Wall:2015raa}, author has explicitly shown that equation \eqref{eomgbint1} is true. As we have mentioned in the introduction, he has used the transformation property of $E_{vv}^\text{HD}$ under certain boost symmetry for his proof. His key argument works barring few `leading terms' in $E_{vv}^\text{HD}$, for which the author has used the `physical process formulation' of the first law as an extra input. As we shall see below, for particularly these terms the integration over the constant $v$ slices of the horizon  turns out to be very crucial.

\subsection{An entropy for non-stationary horizons obeying the second law} \label{ssec:jkmfix}

In this subsection, we shall review the arguments in \cite{Wall:2015raa}, which establishes that most of the terms in $E^{\text{HD}}_{vv}$ could be recast in the form \eqref{eomgbint1}.

\subsubsection{A residual coordinate redefinition freedom}\label{sssec:scaling}
Let us recall that in \eqref{Wallmet}, the coordinate $v$ constitutes 
an affine parameter along the null generators of the horizon, while the coordinate $x^i$ labels the individual 
generators. But this definition does not completely fix the coordinates on the horizon $\mathcal H$. We still have the freedom of the following two classes of coordinate redefinition.
\begin{enumerate}
\item We can perform an affine re-parametrization of the generators of horizon $\mathcal H$, through the transformation 
\begin{equation}\label{genrepara}
v \rightarrow \tilde v = v~ p_1 (x^j) + p_2(x^j). 
\end{equation}
Note that, an arbitrary re-parametrization of this form \eqref{genrepara}, may not be compatible with 
the gauge choice as used in \eqref{Wallmet}. Therefore, along with the transformation \eqref{genrepara}, 
it may be required to transform the $r$ coordinate as well, so that the gauge choice of \eqref{Wallmet} is retained, even after the coordinate transformation \eqref{genrepara}. 
This point is explicated further below, in a special case of \eqref{genrepara}. 
\item We can also relabel the generators as follows 
\begin{equation}\label{genrelabl}
x^i \rightarrow \tilde x^i  = f^i(x^j). 
\end{equation}
\end{enumerate}
The transformation \eqref{genrelabl} does not change the constant $v$ slices of the horizon; consequently $\mathcal H_v$ is 
invariant under it. Also, for the choice of the metric \eqref{Wallmet}, covariance under  \eqref{genrelabl} may be implemented  
by ensuring that all the spatial $i$-indices are covariant; especially, the covariant derivatives along $x^i$, should be compatible 
with the metric $h_{ij}$. For this reason, ensuring invariance (covariance) of entropy (or the second law) 
under \eqref{genrelabl} is relatively easy. However, covariance under \eqref{genrepara} is extremely non-trivial 
and leads to constraints, that were exploited in \cite{Wall:2015raa} to fix the form of $E^{\text{HD}}_{vv}$ and hence the correction to the entropy.

In \cite{Wall:2015raa} only a special case of \eqref{genrepara} was considered under which 
$p_1 = a, ~p_2 = 0$, so that $v$ is re-scaled as $v \rightarrow \tilde v = a v$, $a$ being a constant. 
Now, in order to ensure that our coordinate 
redefinition is compatible with the gauge choice of \eqref{Wallmet}, we must  rescale the $r$ coordinate suitably. For instance, 
in order to ensure that $g_{rv} =1$ everywhere, even after rescaling $v$, we must simultaneously rescale
\footnote{In \cite{Wall:2015raa}, this rescaling has been referred to as `boosts', while the quantities invariant under this rescaling has been referred to as 
`boost invariant'.}
\begin{equation}\label{rescale}
v \rightarrow \tilde v = a v, ~~r \rightarrow \tilde r = \frac{1}{a} r. 
\end{equation}
In the new coordinates \eqref{rescale} the metric takes the following form
\begin{equation}\label{eq:newcoord}
\begin{split}
ds^2 = &2~d\tilde v~d\tilde r -\tilde r^2~ X(\lambda \tilde r, \frac{\tilde v}{\lambda},x^i) d\tilde v^2  \\ \qquad \qquad \qquad
& + 2\tilde r~ \omega_i(\lambda\tilde r, \frac{\tilde v}{\lambda}, x^i) d\tilde v~dx^i 
+ h_{ij}(\lambda \tilde r, \frac{\tilde v}{\lambda}, x^i)  dx^i~ dx^j
\end{split}
\end{equation}
Note that, for the parametrization  \eqref{Wallmet}, the metric looks almost invariant under this coordinate transformation \eqref{rescale}, 
however, the arguments of the metric functions are appropriately scaled. 
In particular, on the horizon $\mathcal H$, the induced metric in the new coordinates takes the following form 
\begin{equation}\label{eq:metrictr}
\begin{split}
&~ds^2_\mathcal H = 2~ d\tilde v~ d\tilde r + h_{ij}dx^i~ dx^j~,\\
\end{split}
\end{equation}
which has an identical structure as compared to that in the old coordinates. 

\subsubsection{Structural form of $E^{\text{HD}}_{vv}$}\label{sssec:strfrmehd}

At first, let us enlist the various derivatives and functions that may occur in $E^{\text{HD}}_{vv}$, for any general diffeomorphism invariant theory of gravity 
\eqref{arbdifth}. These building blocks for constructing $E^{\text{HD}}_{vv}$ include 
\begin{enumerate}
\item The metric functions $X$, $\omega_i$ and $h_{ij}$.
\item The covariant derivative $\nabla_i$ with respect to $x^i$, compatible with the metric $h_{ij}$, which can act on the above metric functions. 
\item The partial derivatives $\partial_r$ and $\partial_v$, which may also act on the  metric functions. 
\end{enumerate} 
Let us immediately note that among these building blocks, it is only $\partial_v$ and $\partial_r$ that transform 
non-trivially under the coordinate rescaling \eqref{rescale}. These transform as 
\begin{equation}\label{drdvscl}
\partial_r \rightarrow \partial_{\tilde r} = \lambda \partial_r , ~~ \partial_v \rightarrow \partial_{\tilde v} = \frac{1}{\lambda} \partial_v
\end{equation}
As is apparent from \eqref{eq:newcoord}, the rest of the building blocks of $E^{\text{HD}}_{vv}$, which include the metric 
functions and the covariant derivative $\nabla_i$, remain invariant under \eqref{rescale}. 

Let us now note that under the coordinate rescaling \eqref{rescale}, $E^{\text{HD}}_{vv}$, being the $vv$-component of a covariant tensor, must transform as 
\begin{equation} 
E^{\text{HD}}_{vv} \rightarrow E^{\text{HD}}_{\tilde v \tilde v} = \frac{1}{\lambda^2} E^{\text{HD}}_{vv}. 
\end{equation}
Let us define the weight of a quantity to be the power of $\lambda$ by which the quantity rescales  under the transformation \eqref{rescale}. In this sense, 
the weight of $E^{\text{HD}}_{vv}$ is $-2$. 

Now, from the transformation property of the building blocks,  it is clear that only $\partial_v$ has a negative weight under \eqref{rescale}. Hence, it follows that, every term in $E^{\text{HD}}_{vv}$ 
must have at least two $\partial_v$. At linear order in $\epsilon$, the most general schematic structure of any term in $E^{\text{HD}}_{vv}$  would be
\begin{equation}\label{eq:schemeHvv1}
\begin{split}
E^{(m,n,k)}= \partial_r^k\left[(\partial_v\partial_r)^m P \right]~\partial_v^{k+2}~\left[(\partial_v\partial_r)^n Q \right] + {\cal O}(\epsilon^2), 
\end{split}
\end{equation}
where $m,n$ and $k$ are positive integers including zero. 
Here we have kept all the $\partial_r$ and $\partial_v$ derivatives explicit 
\footnote{\label{fn:ordzero} It may seem at first glance that the term in \eqref{eq:schemeHvv1} is already second order in $\epsilon$ for any non-zero $m$, since in that case, it becomes 
a product of two terms, on which some $v$-derivative has acted. However, it should be noted that when both the 
derivatives $(\partial_v\partial_r)$ act together on $P$, it can be non-zero on $\mathcal H$ even in equilibrium, i.e. it can be an $\mathcal O (\epsilon^0) $ term (see appendix \ref{app:vdep}).
}.
$P$ and $Q$ are appropriate structures built out of rest of the building blocks, which consist of the metric functions and $\nabla_i$ acting on them.  
They do not contain any further $\partial_v$ or $\partial_r$ derivatives. Thus the most general structure of $E^{\text{HD}}_{vv}$ would be 
\begin{equation}\label{ehdsum}
E^{\text{HD}}_{vv} = \sum_{m,n,k} E^{(m,n,k)}
\end{equation}
The upper limits of these sum would be fixed by the number of derivative on the metric in the gravity Lagrangian \eqref{arbdifth}. 

Now we shall manipulate $E^{(m,n,k)}$ to demonstrate that $E^{\text{HD}}_{vv}$ as given by \eqref{ehdsum} can be cast into the form \eqref{eomgbint1}. 
At first, we shall consider $E^{(m,n,k)}$ for $k\neq 0$, and derive a recursion relation for this quantity. This recursion relation would be used to 
derive a general structure for $E^{(m,n,k)}$ and hence  for $E^{\text{HD}}_{vv}$. 
Certain terms corresponding to the case $k=0$ would require special treatment, and after invoking 
the `physical process' version of the first law, we shall demonstrate that entire sum \eqref{ehdsum}, and hence the most general form of $E^{\text{HD}}_{vv}$, can be cast 
into the form \eqref{eomgbint1}. This would complete our objective as laid out in \S \ref{ssec:stgprf}, thus proving the second law of thermodynamics in the linearized case. 
This would also provide us with an explicit construction to compute the corrections to the Wald entropy, for an arbitrary theory \eqref{arbdifth}.

At first, let us note that, any term of the form $\partial^m_v\partial_r^n P$, where $n\geq m$ and  $P$ does not contain any further $\partial_r$ or $\partial_v$ derivative, would generically be non-zero once evaluated on the Killing horizon of the stationary solution. On the other hand, such terms would vanish on the Killing horizon if $m>n$ (see appendix \ref{app:vdep} for a demonstration of this fact). 
Therefore, it follows that, 
in a dynamical situation, when the amplitude of the time-dependent perturbation 
$\epsilon$ is small, $(\partial^m_v\partial_r^n P)$ must be of  order ${\mathcal O}\left(\epsilon\right)$, whenever 
$m>n$. Consequently, all terms of the product form $ ( \partial^m_v\partial_r^n P ) ( \partial^{m'}_v\partial_r^{n'} Q )$ with $m>n$ and  $m'>n'$, are of order ${\cal O}\left(\epsilon^2\right)$, and therefore, can be neglected in our linearized analysis.

Now let us turn our attention back to the expression \eqref{eq:schemeHvv1}, for $k\neq 0$. In  \eqref{eq:schemeHvv1}, we can take the $v$-derivatives acting on the term involving $Q$ and transfer them onto the term involving $P$ at the expense of a minus sign and a total derivative term. By repeating this operation, even on the terms that are generated due to previous such operations, it is possible to reduce \eqref{eq:schemeHvv1} to the following recurrence relation 

\begin{equation}\label{recrel1}
\begin{split}
&E^{(m,n,k)}
= \partial_v^2\bigg(\sum_{p=0}^{k-1} (-1)^p ~\partial_r^{k-p}\left[(\partial_v\partial_r)^{m+p} P \right]~\partial_v^{k-p}~\left[(\partial_v\partial_r)^n Q \right]\bigg)\\ & + (-1)^k \partial_v \bigg(\left[(\partial_v\partial_r)^{m+ k } P \right]~\partial_v~\left[(\partial_v\partial_r)^n Q \right]\bigg)-E^{(m+1,n,k-1)}
\end{split}
\end{equation}
\JB{}
Now, we can use this recursion relation \eqref{recrel1} itself, to evaluate $E^{(m+1,n,k-1)}$ 
\begin{equation}
\begin{split}
E^{(m+1,n,k-1)} =&~~ \sum_{p=0}^{k-2}(-1)^{p}~\partial_v^2\bigg( \partial_r^{k-p-1}\left[(\partial_v\partial_r)^{m+p+1} P \right]~\partial_v^{k-p-1}~\left[(\partial_v\partial_r)^n Q \right]\bigg)\\ & +(-1)^{k-1} \partial_v\bigg(\left[(\partial_v\partial_r)^{m+k} P \right]~\partial_v~\left[(\partial_v\partial_r)^n Q \right]\bigg)-E^{m+2,n,k-2}
\end{split}
\end{equation}
Using the recursion relation repeatedly, we can recast $E^{(m,n,k)}$ into the form 
\begin{equation}
\begin{split}
E^{(m,n,k)}  =&~~ \sum_{q=0}^{k-1}\sum_{p=0}^{k-q-1} (-1)^{q+p} ~\partial_v^2\bigg(\partial_r^{k-q-p}\left[(\partial_v\partial_r)^{m+q+p} P \right]~\partial_v^{k-q-p}~\left[(\partial_v\partial_r)^n Q \right]\bigg)  \\
& + \bigg( \underbrace{ (-1)^{k}-(-1)^{k-1}+(-1)^{k-2}
- \dots  }_{\text{$k$-terms}}\bigg)   \partial_v\bigg(\left[(\partial_v\partial_r)^{m+ k} P \right]~\partial_v~\left[(\partial_v\partial_r)^n Q \right]\bigg)
\end{split}
\end{equation}
Hence, after performing the sum in the second term, we have  
\begin{equation}\label{emnkexp}
\begin{split}
E^{(m,n,k)} 
=&~~ \sum_{q=0}^{k-1}\sum_{p=0}^{k-q-1}  (-1)^{q+p~} ~\partial_v^2\bigg(\partial_r^{k-q-p}\left[(\partial_v\partial_r)^{m+q+p} P \right]~\partial_v^{k-q-p}~\left[(\partial_v\partial_r)^n Q \right]\bigg)  \\
& + k(-1)^{k} \partial_v\bigg(\left[(\partial_v\partial_r)^{m+k} P \right]~\partial_v~\left[(\partial_v\partial_r)^n Q \right]\bigg)
\end{split}
\end{equation}
Note that, inside the sum, the first term has the structure  $\partial_v^2 \left(\partial_r^{\ell}X^{(\ell)}\right)~\left(\partial_v^{\ell}~Y^{(\ell)} \right)$, where the lowest value of $\ell = 1$. 
If we now, plug in \eqref{emnkexp} back into the sum  \eqref{ehdsum}, a similar structure of this first term would be maintained, again with $\ell$ starting from $1$. Note that, here, we are only considering 
values of $k$ starting from $1$, in the sum \eqref{ehdsum}. As we mentioned previously, the $k=0$ terms needs to be treated separately. 

But now, let us also note from \eqref{eq:schemeHvv1}, that the $k=0$ term can always be recast into the same form as the second term in \eqref{emnkexp}. 
This is true only in the linearized approximation.

Thus, from  \eqref{ehdsum} and \eqref{emnkexp} it is clear that, at linear order in amplitude,  $E^{\text{HD}}_{vv}$ could always be written in the following form
\begin{equation}\label{eq:schemeHvv}
\begin{split}
E^{\text{HD}}_{vv}  &=  \partial_v\left[A~\partial_v~B\right]+\partial_v^2\left[\sum_{k=1} \left(\partial_r^{k}A^{(k)}\right)~\left(\partial_v^{k}~B^{(k)}\right)\right] + {\cal O}(\epsilon^2)\\
&=\partial_v\left[A~\partial_v~B\right]+\partial_v\left[\left(1\over\sqrt{h}\right)\partial_v\sum_{k=1} \sqrt{h}\left(\partial_r^{k}A^{(k)}\right)~\left(\partial_v^{k}~B^{(k)}\right)\right] + {\cal O}(\epsilon^2)\\
\end{split}
\end{equation}
where $A$, $B$, $A^{(k)}$ and $B^{(k)}$ are appropriate structures as implied by \eqref{ehdsum} and \eqref{emnkexp}, which
 do not transform under rescaling \eqref{rescale}. Let us re-emphasize that, although \eqref{emnkexp} has been derived under the assumption $k \neq 0$, the form of $E^{\text{HD}}_{vv}$
 in \eqref{eq:schemeHvv}, also incorporates the $k=0$ term, in the sum \eqref{ehdsum}. 
 
 As argued above (also see appendix \ref{app:vdep}), we know that the action of a $v$-derivative, which does not appear with a compensating $r$-derivative, on a  quantity that is invariant under the rescaling \eqref{rescale}, must be of $\mathcal O (\epsilon)$ on $\mathcal H$, in the amplitude expansion. This is because, such a quantity should vanish on the 
 Killing horizon, and so must be at least $\mathcal O (\epsilon)$ for dynamical horizons. 
  Consequently, whenever  one or more $\partial_v$ act on any one of these $A$, $B$, $A^{(k)}$ or $B^{(k)}$, it must be $\mathcal O (\epsilon)$. 
  This also justifies appropriate incorporation of the  factors of $\sqrt{h}$ in \eqref{eq:schemeHvv}. We do not get any additional terms due to these factors of $\sqrt{h}$, 
  since we are working in the linearized approximation in $\epsilon$. 
 
 Thus to conclude, we have obtained a precise structural form of $E^{\text{HD}}_{vv}$ in \eqref{eq:schemeHvv}. At this stage, 
 we observe that the second term of $E^{\text{HD}}_{vv}$
 in \eqref{eq:schemeHvv} is already in the desired form \eqref{eomgbint1}. 
 So our objective would be accomplished, if we are able to argue that the first term in \eqref{eq:schemeHvv}, can also be written 
 in the form \eqref{eomgbint1}, i.e. as two $v$-derivatives acting on a quantity which is invariant under the rescaling \eqref{rescale}.  

\subsubsection{The physical process version of First law and its implications}\label{sssec:ppfrstlaw}
In \cite{Wall:2015raa} it was argued that the structure of the quantities $A$ and $B$ in \eqref{eq:schemeHvv}, must be such, that $E^{\text{HD}}_{vv}$ has the form \eqref{eomgbint1}. 
This conclusion followed from the \emph{physical process} version of the first law of black hole thermodynamics  \cite{Gao:2001ut, Jacobson:1995uq} 
(also see \cite{Chakraborty:2017kob, PhysRevD.86.021501}), 
which was assumed to be applicable to the theory of gravity under
consideration. It was demonstrated in \cite{Wall:2015raa}, that if $A$ and $B$ in \eqref{eq:schemeHvv} did not have the requisite structures, then the physical process 
version of the first law would be invalidated.  Let us now review this argument, as presented in \cite{Wall:2015raa}.
Let us consider a stationary black hole, which is perturbed by small fluctuations in the matter sector. The amplitude of such perturbations is assumed to be small.
 For instance, this could be some small amount of matter falling into the black hole.
The matter stress tensor would back-react onto the metric and produces fluctuations in it, which would also be small. 
These fluctuations would result in a non-stationary fluctuating black hole. 
However, in a physical situation, it may be expected that, at late times, these fluctuations would die down and the black hole would again become stationary. 
Such a dynamical process, where the black hole is stationary, both at early and late times, is referred to as a `physical process'. 

The new stationary black hole at late times would have slightly different parameters compared to the one at early times (such as mass or angular momentum). 
The overall shift in the mass (energy) of the black hole would be given by integrating specific component of the energy-momentum tensor over the horizon. This shift in mass, must be related to 
the shift in entropy of the black hole through the first law of black hole thermodynamics $T \Delta S = \Delta \mathcal E$. 
Therefore, we can express the change in entropy during this physical process $\Delta S$, to the integrated energy-momentum tensor, in the following way
 \begin{equation}\label{eq:prevwor}
 \begin{split}
 \Delta S = -{2\pi\over \kappa}\int_{\mathcal H}\Delta T_{ab}~\xi^a~d\Sigma^b 
 \end{split}
 \end{equation}
Here, $\Delta T_{ab}$ is the part of the energy-momentum tensor that has initiated the dynamics of the black hole horizon.  $\xi^a$ is the generator of the future horizon $\mathcal H$; it is a Killing generator at early and late times, when the black hole is stationary. Also, $d\Sigma^b$ is the area element along the horizon. The parameter $\kappa$ is the surface gravity of the black hole and is proportional to the temperature of the black hole\footnote{It turns out that the combination $\Delta T_{ab}\xi^a d\Sigma^b$ itself is of the order of amplitude of the perturbation.
Therefore, as long as we are working at linear order in the amplitude of perturbation, the difference in the value of $\kappa$ for the initial and  final stationary black holes is negligible. 
Thus, within this approximation, $\kappa$ can be taken to be constant throughout the duration of the physical process.}. 

The equation \eqref{eq:prevwor} is referred to as the physical process version of the first law of black hole thermodynamics. For a more complete and 
detailed discussion of this, see section (2) of \cite{Jacobson:1995uq}. We should note that because the initial and final states are stationary,
the $\Delta S$ in \eqref{eq:prevwor} is expected to be given by the change in Wald entropy, which, by construction, satisfies the \emph{usual} form of the first law for stationary black holes. 
However, whether Wald entropy does satisfy the physical process version of the first law \eqref{eq:prevwor} does not immediately follow from its construction, and 
we require additional arguments to establish this, see \cite{Gao:2001ut, Chakraborty:2017kob} for recent developments. 

This version of the first law \eqref{eq:prevwor} now enables us to make further deductions regarding the structural 
form of $A$ and $B$ appearing in \eqref{eq:schemeHvv}. With our choice of 
coordinates \eqref{Wallmet}, $\xi^a$ is related to the affinely parametrized null generators of the horizon $\partial_v$, in the following way
\begin{equation} \label{xidef}
\xi^a\partial_a  = \kappa ~v~\partial_v
\end{equation}
While, in our coordinates \eqref{Wallmet}, the area element on the horizon $\mathcal H$ is given by 
\begin{equation}\label{areadef}
d\Sigma^b\partial_b = -\sqrt{h}~d^{d-2}x ~dv~\partial_v
\end{equation}
Using \eqref{xidef} and \eqref{areadef} back in \eqref{eq:prevwor}, we have 
\begin{equation}\label{eq:prevworm1}
 \begin{split}
\Delta S &= {2\pi}\int_{\mathcal H}\sqrt{h}~d^{d-2}x ~dv~v~\Delta T_{vv} = {2\pi}\int_{\mathcal H}\sqrt{h}~d^{d-2}x ~dv~v~( R_{vv} + E_{vv}^{\text{HD}})\\
 \end{split}
 \end{equation}
 Here, we have used the equation of motion  to rewrite the stress tensor in terms of geometric quantities. Now, if entropy $S$ has the form \eqref{entsch1}, 
 $\Delta S$ in \eqref{eq:prevworm1} can be
 split into two parts $\Delta S = \Delta S_{E}+ \Delta S_{\text{HD}}$. $\Delta S_{E}$ being the change in the integrated area of $\mathcal H_v$,
  responsible for the change in entropy in two derivative Einstein theory, while 
$\Delta S_{HD}$ is the change in entropy due to higher derivative terms. 
 Clearly, the terms proportional to $R_{vv}$ on the RHS of \eqref{eq:prevworm1}, must be equal to $\Delta S_{E}$ on the LHS of
 \eqref{eq:prevworm1}. This is manifest in the limit when the higher derivative corrections to Einstein's gravity vanish. 
 Using \eqref{eq:schemeHvv}, we can therefore write 
 \begin{equation}\label{eq:prevworm2}
 \begin{split}
\Delta S_{\text{HD}} &= {2\pi}\int_{\mathcal H}\sqrt{h}~d^{d-2}x ~dv~v~( E_{vv}^{\text{HD}})\\
  &= {2\pi}\int_{H}\sqrt{h}~d^{d-2}x ~dv~v~\partial_v\left[ A~\partial_v ~B+\left(1\over\sqrt{h}\right)\sum_{k=1} \partial_v\left(\sqrt{h}~\partial_r^kA^{(k)}~\partial_v^{k}B^{(k)}\right)\right]\\
  &+ {\cal O}\left(\epsilon^2\right) 
 \end{split}
 \end{equation}
It is extremely important for the subsequent arguments to realize that $\Delta S_{\text{HD}}$ must be non-zero in general, and should be expressible in terms of some geometrical 
quantity integrated over $\mathcal H_v$. This is clear from the fact that for arbitrary higher derivative corrections to Einstein's gravity $s_n$ in \eqref{entsch1} is non-zero even for stationary black holes. In the stationary case, it is expected to be given by Wald entropy \eqref{entsch2}, which in general is different than area of the horizon. Thus, if $s_n$ is non-trivial, it is expected that, in general, in a dynamical scenario, $\Delta S_{\text{HD}}$ must be non-trivial. In fact, as pointed out earlier, $\Delta S_{\text{HD}}$ should be given by  change in the corresponding Wald entropy, since the state both at early and late times are stationary states. 
 
Now the contribution from the second term in \eqref{eq:prevworm2} vanishes. This can be seen  by manipulating the term as follows. 
\begin{equation}\label{eq:2ndterm}
\begin{split}
&\int_{\mathcal H}\sqrt{h}~d^{d-2}x ~dv~v~\partial_v\left[{1\over \sqrt{h}}~\partial_v (\sqrt{h}~ X )\right]\\
&= \int_{\mathcal H}~d^{d-2}x ~dv ~\partial_v \left[ v~ \partial_v (\sqrt{h}~ X ) - (\sqrt{h}~ X )\right]  + {\cal O}\left(\epsilon^2\right) \\
&= \int_{v=-\infty}^{v=\infty}~dv ~\partial_v\bigg(\int_{\mathcal H_v}~d^{d-2}x  \left[ v~ \partial_v (\sqrt{h}~ X ) - (\sqrt{h}~ X )\right]\bigg)  + {\cal O}\left(\epsilon^2\right) \\
&=\bigg[\int_{\mathcal H_v}~d^{d-2}x ~ v~ \partial_v (\sqrt{h}~ X ) -\int_{\mathcal H_v}~d^{d-2}x(\sqrt{h}~ X )\bigg]_{v=-\infty}^{v=\infty}
\end{split}
\end{equation}
where 
\begin{equation}
\begin{split}
X = \sum_{k=1}^N \left[\left(\partial_r^kA^{(k)}\right)~\left(\partial_v^{k}~B^{(k)}\right)\right] + {\cal O}\left(\epsilon^2\right) 
 \end{split}
 \end{equation}
Note that both the terms in the last line of equation \eqref{eq:2ndterm} contain more than one $\partial_v$ derivatives on expressions that are invariant under the rescaling \eqref{rescale}.  Therefore, they must vanish in the two limits of far past and far future, where we have a stationary black hole with a Killing horizon. It follows that, these terms do not contribute  to $\Delta S_{\text{HD}}$ in \eqref{eq:prevworm2}. Hence, \eqref{eq:prevworm2} can be always reduced to an integral of the form 
  \begin{equation}\label{eq:1stterm1}
 \begin{split}
\Delta S _\text{HD} = {2\pi}\int_{\mathcal H}\sqrt{h}~d^{d-2}x ~dv~v~\partial_v\left[ A~\partial_v ~B\right]\\
 \end{split}
 \end{equation}
Again by performing an integration by parts, we obtain 
 \begin{equation}\label{eq:1stterm2}
 \begin{split}
\Delta S _\text{HD} = {2\pi}\left[\int_{\mathcal H_v}\sqrt{h}~d^{d-2}x ~v~\left( A~\partial_v ~B\right)\right]_{v=-\infty}^{v=\infty}-{2\pi}\int_{\mathcal H_v}\sqrt{h}~d^{d-2}x ~dv~\left[ A~\partial_v ~B\right]
 \end{split}
 \end{equation}
 However, to have a consistent first law, the second term also should be write-able as a total $\partial_v$ derivative so that the $v$ integration  of this term from infinite past to infinite future finally would give the net change of some geometric quantity - entropy,  defined on the constant $v$ slices of the horizon\footnote{At this stage, by `geometric quantity' we simply mean some expression in terms of the metric components and their derivatives  that is invariant under any diffeomorphism, mixing only the spatial coordinates of the constant $v$ slices of the horizon.}. Naively this is possible  if $A$ is of the form $\bigg[{\text{some constant}/\sqrt{h}}\bigg]$ and $B$ has the form $\bigg[{\sqrt h} ~\tilde B\bigg]$ where $\tilde B$ is a scalar under $\{x^i\} \rightarrow \{y^i\} = y^i \left(\{\vec x\}\right)$.

If it is indeed the case that $A$ is always a constant times $\left(1/\sqrt{h}\right)$ (let's choose the constant to be one without any loss of generality), then the schematic form of $E^\text{HD}_{vv}$
\begin{equation}\label{eq:schemeHvv2}
\begin{split}
E^\text{HD}_{vv}  
=&~\partial_v\left({\partial_v \left(\sqrt{h}~ \tilde B\right)\over\sqrt{h}}\right) +\partial_v\left[\left(1\over\sqrt{h}\right) \partial_v \bigg(\sqrt{h}\sum_{k=1}^N \partial_r^k A^{(k)}~\partial_v^{k}B^{(k)}\bigg)\right] + {\cal O}(\epsilon^2)\\
\end{split}
\end{equation}
If $E^\text{HD}_{vv}$ does have the form of equation \eqref{eq:schemeHvv2}, not only the `physical process version'  of the first law but also the second law as argued in \cite{Wall:2015raa} will be true with the following identification for correction to the total entropy (see equation \eqref{eomgbint1}).
\begin{equation}\label{eq:scor}
\begin{split}
\delta S_\text{HD}=\int_{{\cal H}_v}\bigg[\tilde B+\sum_{k=1}^N \partial_r^{k}A^{(k)}~\partial_v^{k}B^{(k)}\bigg]
\end{split}
\end{equation}
Note that $\tilde B$ is the only term that is non-zero even in equilibrium. This term must match with Wald entropy\footnote{Though we have  explained the argument here,  specializing to higher derivatives theories, it is trivially true for two derivative theories of gravity where $\tilde B$ is simply 0, and entropy is simply given by 
$S = \int_{\mathcal H_v} \sqrt{h}$\, . }. Rest of the terms (for $k\geq 1$ ) vanish on stationary metric  and therefore are part of JKM ambiguities.

So in summary, \cite{Wall:2015raa} has given a constructive proof for the second law for dynamical black hole solutions in higher derivative theories of gravity provided the physical process formulation of the first law is true for these solutions. The validity of the `physical process formulation of the first law' 
requires a very specific structure for a certain term   in the equation of motion 
(the first term in \eqref{eq:schemeHvv}  must take the form of the first term in \eqref{eq:schemeHvv2}), 
which does not follow from the boost-symmetry \eqref{rescale} alone (the only symmetry that is considered in \cite{Wall:2015raa})
\footnote{\label{note1}This special structure of the first term in \eqref{eq:schemeHvv2}  has only been verified in specific theories of gravity 
where the physical process version of the first law has been proven (for instance, see \cite{Jacobson:1995uq}). 
To our knowledge, a complete proof demonstrating this special structure of the zero boost term in a general higher derivative theory of 
gravity does not exist. It would be interesting to explore, if it is possible to arrive at such a proof using 
the residual gauge transformations \eqref{genrepara}, which is more general than the boost symmetry \eqref{rescale} (see \S\ref{sec:disco} for further discussion). 
%
%
}. 
For the convenience of reporting, we shall refer to the first term in \eqref{eq:schemeHvv2} (or the first term in \eqref{eq:schemeHvv}) as the `zero boost term'. This nomenclature is inspired by the fact that $\tilde B$ in \eqref{eq:schemeHvv2} (or $A$ and $B$ in \eqref{eq:schemeHvv}) does not have any $v$ or $r$ derivative. Apart from this, there is absolutely no other physical motivation behind this nomenclature. 
The reader must not confuse the phrase `zero boost term' to be a synonym for 
boost invariant term. By boost invariant terms we shall continue to mean such terms which 
are invariant under the rescaling symmetry \eqref{rescale}.

\section{An entropy current for four derivative theories of gravity
} \label{sec:gbexp}

In the previous section \S \ref{sec:review}, we have reviewed the proof of the second law for linearized fluctuations, following \cite{Wall:2015raa}. 
As emphasized earlier, this proof is designed to prove a second law for the `total entropy' of the system. 
The proof crucially involves an integration over the full spatial slice of the horizon, which defines the `total entropy'. Therefore, 
it is insensitive to any total (spatial) derivative term, that may be present in the integrand, which is derived from the equation of motion. 
This drawback exists even in the proof for the  physical version of the first law.

In this section, we shall carefully re-examine this particular subtlety. 
In explicit examples of four-derivative theories of gravity, we shall demonstrate that such total derivative terms do exist if we follow the algorithm of \cite{Wall:2015raa}, and 
their inclusion would naturally lead to a construction of an entropy current. With the help of this entropy current, we can immediately 
prove an ultra-local version of second law, associated with any dynamical horizon $\mathcal H$.

Let us now elaborate this point further. In \eqref{eq:schemeHvv2}, 
we have shown that  a special structure for $E_{vv}^\text{HD}$ is necessary for the validity of both the second law, as well as the physical process version of the first law. As we have explained in \S\ref{sec:review}, the  structure of the second term in \eqref{eq:schemeHvv2} is fixed by the `boost symmetry' \eqref{rescale}, up to higher-order corrections in the amplitude of fluctuations $\epsilon$. But the same is not true for the first term in \eqref{eq:schemeHvv2}, which we have named  as `zero-boost terms' of $E_{vv}^\text{HD}$. 

In \S \ref{sec:review}, we have argued that the physical version of the first law, and consequently, the second law, 
would be true if the `zero boost term' in $E_{vv}^{\text{HD}}$  has the following schematic structure (see \eqref{eq:schemeHvv2})
\begin{equation}\label{eq:hvvzero1}
E_{vv}^{\text{HD}} \big\vert_\text{zero boost} \sim \partial_v\left({1\over\sqrt{h}}\partial_v \left(\sqrt{h}~ \widetilde B\right)\right) \, ,
\end{equation} where $\widetilde B$ is some scalar, which is invariant  under spatial diffeomorphism.

However, the above form of the zero-boost term, 
though sufficient for the validity of the physical process version of the first law \footnoteref{note1} and the second law, 
it is neither necessary,  nor does it  follow in any way, from the boost-symmetry \eqref{rescale}.
In this section, to begin with, our goal is just  to verify \eqref{eq:hvvzero1}. We shall explicitly compute the equation of motion, and in particular the zero-boost terms, in all possible  four-derivative theories of gravity.
From this explicit computation in four derivative theories of gravity we will 
show that \eqref{eq:hvvzero1} is not true in general.  
There exist cases where the zero boost terms in $E_{vv}^{\text{HD}}$ could not be recast in the above form. In fact, the zero boost terms in $E_{vv}^{\text{HD}}$
consists of additional terms, which can never be cast into the form \eqref{eq:hvvzero1}. 

Being motivated by this observation, we investigate the 
structural nature of the  zero boost terms of $E_{vv}^{\text{HD}}$, to understand why such additional terms do not affect validity of the first and the second law. 
The possibility of non-zero spatial components of the entropy current arises here very naturally. Finally, 
through a general algorithm, we shall establish that, the zero boost terms of $E_{vv}^\text{HD}$ for every four derivative theories of gravity, could be rendered into a form, which 
guarantees an ultra-local version of the second law, in terms of an entropy current with non-zero spatial components.

\subsection{Explicit calculation of $E_{vv}^\text{HD}$ and the entropy current for theories with four derivative corrections to Einstein gravity} \label{ssec:4dercor}
%
In this subsection, we shall compute the `$vv$'-component of the equation of motion, $E_{vv}$, for all possible  four derivative theories of gravity. We  shall immediately find  that it is possible to  rearrange the terms so that up to corrections of order ${\cal O}(\epsilon^2)$ it takes the form
\begin{equation}\label{eq:entrecast}
E_{vv}=-\,\partial_v\left[{1\over\sqrt{h}}\partial_v \left(\sqrt{h} \, J^v\right) + \nabla_i J^i\right] + {\cal O}\left(\epsilon^2\right) \, .
\end{equation}
Once we could rewrite $E_{vv}$ in this form
\footnote{Note that for Einstein gravity $E_{vv}$ takes the simple form
\begin{equation*} 
E_{vv}^{\text{\tiny{Einstein}}} = -\,\partial_v\left[{1\over\sqrt{h}}\partial_v \left(\sqrt{h} \right) \right]. 
\end{equation*} 
Hence, when we consider higher derivative corrections to Einstein's equations, the terms in this equation arising out of these corrections also has a similar form
\begin{equation*} 
E_{vv}^{\text{HD}} = -\,\partial_v\left[{1\over\sqrt{h}}\partial_v \left(\sqrt{h} \, \tilde J^v\right) + \nabla_i J^i\right] + {\cal O}\left(\epsilon^2\right),
\end{equation*} 
where $J^v - \tilde J^v = 1$. For most of our analysis, especially in the abstract manipulations, we have used $E_{vv}^{\text{HD}}$, instead of $E_{vv}$.}
, it is very natural to identify $J^v$ with the entropy density and $J^i$ as the spatial entropy current, capturing the in-flow and out-flow of entropy. Vanishing of $E_{vv}$ at order ${\cal O}\left(\epsilon\right)$ would then correspond to a locally conserved entropy current and therefore an ultra-local version of the second law 
(see \S\ref{sec:review} for details of this argument)\footnote{The calculations here clearly suggest about the existence of an entropy current for some of  these higher derivative theories. However, it requires a bit of clever manipulation. See the following subsections for a more algorithmic method which clearly exhibits that we need the spatial entropy current, which in turn provides us with an ultra-local version of the second law.}. More explicitly, once $E_{vv}$ has the form \eqref{eq:entrecast}, the standard arguments 
outlined in \S\ref{sec:review} would imply that 
\footnote{ If we consider special processes where the metric is entirely sourced by a small matter energy-momentum tensor, so that both the first correction to the metric 
as well as the matter energy-momentum tensor are of $\mathcal O(\epsilon^2)$ (the $\mathcal O(\epsilon)$ correction to the metric being zero), then 
for the $\epsilon^2$ coefficient, \eqref{locsec} would be modified to the inequality 
\begin{equation} \label{locsecine}
\left( \frac{1}{\sqrt{h}} \partial_v \left(\sqrt{h} \,  J^v\right) + \nabla_i J^i  \right) \Big|_{\epsilon^2} \geq 0. 
\end{equation} 
Note that, while deducing this inequality, we have assumed that there exist other matter fields satisfying the null energy condition. 
See  \S \ref{ssec:stgprf} for a more detailed discussion of this point. 
}
\begin{equation} \label{locsec}
\frac{1}{\sqrt{h}} \partial_v \left(\sqrt{h} \,  J^v\right) + \nabla_i J^i =  \mathcal O(\epsilon^2).
\end{equation} 

There are only three possible covariant terms which can appear in the gravity Lagrangian with 4-derivatives 
on the metric. These are given by:  
$ R^2, ~R_{\mu \nu} R^{\mu \nu}, ~R_{\mu \nu \sigma \lambda} R^{\mu \nu \sigma \lambda}.$ 
In the following subsections, we shall  separately consider three different four derivative theories of gravity 
\begin{align*}
1. &~ \text{Ricci scalar squared theory:} ~~\mathcal I^{(1)} =\int d^{d}x \sqrt{-g} \, (R+a_{1} \, R^{2}), \\
2. &~ \text{Ricci tensor squared theory:} ~~ \mathcal I^{(2)}=\int d^{d}x \sqrt{-g}(R+a_{2}\, R_{\mu \nu }R^{\mu \nu}), \\
3. &~ \text{Riemann tensor squared theory:} ~~ \mathcal I^{(3)}=\int d^{d}x \sqrt{-g}(R+a_{3}\, R_{\mu \nu \rho \sigma }R^{\mu \nu \rho \sigma}),
\end{align*}
and explicitly compute the `$vv$'-component of the respective equations of motion, $E_{vv}$, for each of them. 
After some algebraic manipulations on $E_{vv}$, in each of these cases,  we shall write down the  entropy current, 
It is then trivial to combine these results, to give us the entropy current 
for any arbitrary four derivative theory of gravity. The final result is tabulated in  Table-(\ref{table:4derres}).

For each of the three four-derivative theories mentioned above, if we just evaluate the equation of motion on our gauge fixed metric \eqref{Wallmet}, it turns out to be an extremely complicated expression, even after we restrict it to the horizon. In general, just by inspection, 
it is quite difficult to rearrange the terms to arrive at the form \eqref{eq:entrecast}. However, we know that in stationary situations, at least $J^v$ should reduce to the well-known form of Wald entropy and the rest of the terms must be such that they vanish in a stationary situation.
We shall use this fact to guide our intuition about the form of the entropy density and then finally deduce the form of the entropy current. More precisely, we shall obtain the following constituents for $J^v$ 
\begin{equation}\label{eq:splitentropy}
J^v =\sqrt{h}\, \left( s_w + s_{c}\right)
\end{equation}
where $s_w$ is the Wald entropy density for the stationary black holes defined as 
\begin{equation}\label{eq:waldformula}
s_w = {\partial \mathcal{L} \over \partial R_{\mu\nu\rho\sigma}} \, \epsilon_{\mu\nu} \epsilon_{\rho\sigma}, 
\end{equation}
where $\epsilon_{\mu\nu}$ are the bi-normal to $\mathcal{H}_v$, the co-dimension$-2$ spatial slicing of the horizon. Also, $s_{c}$ is the non-stationary correction to $s_w$. As we have  argued before, $s_{c}$ will vanish once we take stationary limit. Let us also define the contribution to $s_w$ from the higher derivative part of the action as $s_w^{\text{HD}}$
\begin{equation} \label{eq:waldentdensityHD}
s^{\text{HD}}_{w} = {\partial \mathcal{L}^{\text{HD}} \over \partial R_{\mu\nu\rho\sigma}} \, \epsilon_{\mu\nu} \epsilon_{\rho\sigma}.
\end{equation}
At this point, let us clarify one subtlety regarding the split mentioned in \eqref{eq:splitentropy}. It turns out that if we evaluate $s_w$ on any dynamical metric, along with the terms that contribute in stationary situation, it will also have terms that vanish in the stationary limit. For convenience, let us name such terms as `off-equilibrium' structures. Such off-equilibrium terms in the entropy suffer from the well-known class of JKM ambiguities, which arises as soon as we try to extrapolate Wald's formalism to non-stationary solutions. 
In our identification of the entropy density, we have used the fact that, in the stationary limit, it should reduce to Wald entropy. As we will see in the later sections, this requirement also fixes one class of ambiguities, in defining the entropy current. 

For convenience, let us separate out the contribution of Wald entropy density $s_w^{\text{HD}}$ to $E_{vv}^\text{HD}$ and define the quantity  ${\mathbb{E}^{\text{HD}}_{vv}}^*$
as follows 
\begin{equation}\label{eq:esdiff}
{\mathbb{E}^{\text{HD}}_{vv}}^* \equiv E_{vv}^\text{HD} + \partial_v\left[{1\over\sqrt{h}}~\partial_v \left(\sqrt{h} ~s^\text{HD}_{w}\right)\right]
\end{equation}
Then, from the definition \eqref{eq:splitentropy} it follows that 
\begin{equation}\label{eq:Estar}
{\mathbb{E}^{\text{HD}}_{vv}}^*=-\,\partial_v\left[{1\over\sqrt{h}}\left(\partial_v \sqrt{h} \, s_{c}\right) + \nabla_i J^i\right] + {\cal O}\left(\epsilon^2\right).
\end{equation}
It turns out that, in the examples that we consider, algebraically it is comparatively easier to recast ${\mathbb{E}^{\text{HD}}_{vv}}^* $ in the form \eqref{eq:Estar}, 
instead of dealing with the full $E_{vv}^\text{HD}$. 

We would like to emphasize here that, the procedure adopted in this subsection, is a set of intuitive manipulations and educated guess-work. It gives us an explicit demonstration
that for the theories that we consider here, it is possible to lift both the first and the second law to an ultra-local form, 
by entertaining the possibility of non-zero spatial components of the entropy current, which captures the effect of the inflow and outflow of the entropy from any arbitrary 
local sub-region. However, at this stage, we would not be able to say, whether the spatial components of the current  is an absolute necessity, 
or there exist other possible rearrangements of terms, such that we can avoid the spatial components of the current altogether. 
In the later subsections, we shall repeat the same analysis more systematically, and for the four derivative theories, we shall be able to quantify these ambiguities 
involved in defining the entropy current more precisely. We shall conclude that, 
although there are some ambiguities in defining the entropy current, its non-zero spatial components are an unavoidable feature of the ultra-local form of the second law.

\subsubsection{Ricci scalar square theory} \label{sssec:ricciscal}

The action for Ricci scalar square theory is 
\begin{equation}\label{riccisqth}
\mathcal I^{(1)} =\int d^{d}x \sqrt{-g} \, (R+a_{1} \, R^{2}) 
\end{equation}
where $a_1$ is an arbitrary constant. 
The equations of motion which follows from the action \eqref{riccisqth}, is given by
\begin{equation}\label{riccisqeom}
E_{\mu \nu}=R_{\mu \nu} -\frac{1}{2} g_{\mu \nu} R + E_{\mu \nu}^{\text{HD}}=0,
\end{equation}
where 
\begin{equation}
E_{\mu \nu}^{\text{HD}}= a_1 \left( 2 RR_{\mu\nu }- 2D_{\mu}D_{\nu}R +2g_{\mu\nu}D^{\rho}D_{\rho}R-\frac{1}{2}g_{\mu\nu}R^{2} \right)
\end{equation}
are the higher derivative corrections to the Einstein equation. 
The explicit form of the vv-component of the equations of motion, on the horizon, is 
\begin{equation}
E_{vv}=R_{vv} + E_{vv}^{\text{HD}},
\end{equation}
where 
\begin{equation}\label{evvhdth1}
E_{vv}^{\text{HD}}= a_1 \left( 2 RR_{vv}-2 D_{v}D_{v}R \right).
\end{equation}

The Wald entropy for this theory happens to be
\begin{equation}\label{eq:WaldRicci}
\begin{split}
S_w=\int_{\mathcal{H}_v} d^{d-2}x \, \sqrt{h} \, (1 +2 \,a_1 \, R).
\end{split}
\end{equation}
Once we have the Wald entropy, we could compute ${\mathbb{E}^{\text{HD}}_{vv}}^*$. In this case it simply vanishes implying that we do not need to add any current, neither do we get any correction to entropy density, beyond what is given by the Wald entropy.

\subsubsection{Ricci tensor square theory} \label{sssec:ricciten}
In this theory, the Ricci tensor square is added to the Einstein-Hilbert action, as a higher derivative correction. We have 
\begin{equation}
I=\int d^{d}x \sqrt{-g} \, (R+a_{2} \, R_{\mu \nu}R^{\mu \nu})  
\end{equation}
The equations of motion, for this theory are given by 
\begin{equation}
\begin{split}
 & E_{\mu \nu} =  R_{\mu \nu} -\frac{1}{2} g_{\mu \nu} R + E_{\mu \nu}^{\text{HD}} =0 , ~~ \text{where} \\
 & E_{\mu \nu}^{\text{HD}}  = a_2 \bigg(  2 R^{\alpha \beta}R_{\mu \alpha \nu \beta}- D_{\mu}D_{\nu}R + D^{\alpha}D_{\alpha} R_{\mu \nu} 
 + \frac{1}{2}\, g_{\mu\nu}\, D^{\alpha}D_{\alpha}R -\frac{1}{2}\, g_{\mu \nu}\, R_{\alpha \beta}R^{\alpha \beta}  \bigg)
\end{split}
\end{equation}
The explicit form of the vv-component of the equations of motion on the horizon, is as given below
\begin{equation}
\begin{split}
E_{vv} =& R_{vv}+E_{vv}^{\text{HD}}=0, \\
E_{vv}^{\text{HD}} =&\, a_2 \big(\, 2 R^{\alpha \beta}R_{v \alpha v \beta}- D_{v}D_{v}R + D^{\alpha}D_{\alpha} R_{vv} \big).
\end{split}
\end{equation}

The Wald entropy for this theory is given by
\begin{equation}\label{eq:WaldRicciT}
\begin{split}
S_w=\int_{\mathcal{H}_v} d^{d-2}x \, \sqrt{h} \, (1 +2 \,a_2 \, R_{rv}), 
\end{split}
\end{equation}
so that  $s^{\text{HD}}_w = 2 \,a_2 \, R_{rv}$ . Once we have obtained the Wald entropy, we can compute ${\mathbb{E}^{\text{HD}}_{vv}}^*$. Using the form of the metric \eqref{Wallmet} and the formulae provided in appendix \ref{app:notcon}, we evaluate ${\mathbb{E}^{\text{HD}}_{vv}}^*$ explicitly in terms of metric functions and their derivatives.
\begin{equation}\label{eq:sdiffrtensor}
\begin{split}
{\mathbb{E}^{\text{HD}}_{vv}}^*= \, a_2 \, \partial_v \left[ {1 \over \sqrt{h}} \, \partial_v \big( \sqrt{h} \, K\bar{K} \, \big) \right] +a_2 \, \partial_v \left[  \nabla_i \left( \nabla^i K +\,h^{ij} \partial_v \omega_j -2 \, \nabla_j K^{ij}\right)\right] . 
\end{split}
\end{equation}
Now, we could easily re-express ${\mathbb{E}^{\text{HD}}_{vv}}^*$ in the form of \eqref{eq:Estar}. Subsequently, it is straightforward to identify the current as 
\begin{equation}
\begin{split}
J^v =& -\, s^{\text{HD}}_w -a_2\, K\bar{K}, \\
J^i =& \, a_2 \, \left(2 \, \nabla_j K^{ij}-\nabla^i K -\,h^{ij} \partial_v \omega_j \right).
\end{split}
\end{equation}

\subsubsection{Riemann tensor square theory} \label{sssec:riesq}
The action for Riemann tensor square theory is 
\begin{equation}
I=\int d^{d}x \sqrt{-g}(R+a_{3}R_{\mu \nu \rho \sigma }R^{\mu \nu \rho \sigma})  \nonumber
\end{equation}
 The corresponding equations of motion are
\begin{equation}
E_{\mu \nu}=R_{\mu \nu} -\frac{1}{2} g_{\mu \nu} R + E_{\mu \nu}^{\text{HD}} = 0,
\end{equation}
where 
\begin{equation}
\begin{split}
E_{\mu \nu}^{\text{HD}}= a_3 \bigg( & 4 R^{\alpha \beta}R_{\mu \alpha \nu \beta}- 2D_{\mu}D_{\nu}R + 4 D^{\alpha}D_{\alpha} R_{\mu \nu}-4R_{\mu}^{\alpha}R_{\nu \alpha} 
\\ & 
 -\frac{1}{2}g_{\mu \nu}R_{\alpha \beta\gamma\sigma}R^{\alpha \beta\gamma\sigma} +2 R^{\alpha \beta \sigma}_{\mu}R_{\nu \alpha \beta \sigma} \bigg).
\end{split}
\end{equation}
The vv-component of the equations of motion is
\begin{equation} \label{rimehd}
\begin{split}
E_{vv} = & R_{vv}+ E_{vv}^{\text{HD}} =0, \\
E_{vv}^{\text{HD}} = & a_3 \bigg( 4R^{\alpha \beta}R_{v \alpha v \beta}- 2D_{v}D_{v}R + 4D^{\alpha}D_{\alpha}  R_{vv}  
-4R_{v}^{\alpha}R_{v \alpha}\\& \quad +2 R_{v}^{\alpha \beta \sigma}R_{v\alpha \beta \sigma}  \bigg).
\end{split}
\end{equation} 
The Wald entropy for this theory will be
\begin{equation}
s_w = \int_{\mathcal H_{v}} d^{d-2}x \, \sqrt{h} \, (1-4 \, a_{3} \, R_{rvrv}), \nonumber
\end{equation}
such that $s^{\text{HD}}_w = -4 \, a_{3} \, R_{rvrv}$.

Once we have Wald entropy, it is easy to compute ${\mathbb{E}^{\text{HD}}_{vv}}^*$. Using the  form of the metric \eqref{Wallmet}, and the formulae provided in appendix \ref{app:notcon}, we can evaluate ${\mathbb{E}^{\text{HD}}_{vv}}^*$, explicitly in terms of metric functions and their derivatives. We find that 
\begin{equation}\label{eq:sdiffrimtensor}
\begin{split}
{\mathbb{E}^{\text{HD}}_{vv}}^*= \, 4 \,a_3 \, \partial_v \left[ {1 \over \sqrt{h}} \, \partial_v \left( \sqrt{h} \, K_{ij}\bar{K}^{ij}  \right) \right] +4\, a_3 \, \partial_v \left[  \nabla_i \left( h^{ij} \partial_v \omega_j - \, \nabla_j K^{ij}\right)\right] ,
\end{split}
\end{equation}
which has been expressed in the structural form \eqref{eq:Estar}. From this, it is again straightforward to read off the entropy current to be 
\begin{equation}
\begin{split}
J^v =& - \, s^{\text{HD}}_w -4\,a_3 \,K_{ij}\bar{K}^{ij}, \\
J^i =& -4\, a_3 \, \left(h^{ij} \partial_v \omega_j - \, \nabla_j K^{ij}\right).
\end{split}
\end{equation}

\begin{table}
\begin{center}
\begin{tabular}{|c|l||l|}\hline
 & &  \\
 & &  \\
 & &  \\
 1& Ricci scalar square theory: &   $E_{vv}^{\text{HD}} = - \partial_v \Theta + \mathcal{O}[\epsilon]^{2}$ \\
  & &  \\
  & $I = \int d^dx \sqrt{-g} \left( R + a_1 R^2 \right)$ & $\Theta = \frac{2\, a_1}{\sqrt{h}} \partial_{v}\left(  \sqrt{h} ~R \right)$  \\
 & &  \\
 & &  \\
  & &  \\
  \hline
  & &  \\
 & & \\
 & &   $E_{vv}^{\text{HD}} = - \partial_v \Theta -  \partial_v \left(\nabla_i J^i \right) + \mathcal{O}[\epsilon]^{2}$ \\
 2& Ricci tensor square theory: &   \\
  & & $\Theta = \frac{a_2 }{\sqrt{h}} \partial_{v}\left[ \sqrt{h}  \left( 2  R_{rv} -   \bar{K}K\right)\right] $ \\
  & $I = \int d^dx \sqrt{-g} \left( R + a_2 R_{\mu \nu}R^{\mu \nu} \right)$ &  \\
 & &  $J_i  = a_{2}\left[2 \nabla^{j}K_{ij}-\nabla_{i}K-\partial_{v}\omega_{i} \right]$ \\
 & &  \\
  & &  \\ 
  \hline
  & &  \\
 & &  \\
 & & $E_{vv}^{\text{HD}} = - \partial_v \Theta - \partial_v  \left(\nabla_i J^i\right)  + \mathcal{O}[\epsilon]^{2}$ \\
 3& Riemann tensor square theory: &  \\
  & &  $\Theta =  \frac{4 \, a_3}{\sqrt{h}} \partial_{v}\left(  \sqrt{h}\left(- R_{rvrv} + \bar{K}_{ij}K^{ij} \right) \right) $ \\
  & $I = \int d^dx \sqrt{-g} \left( R + a_3 R_{\mu \nu \sigma \lambda} R^{\mu \nu \sigma \lambda} \right)$ &  \\
 & &  $J_{i} = 4a_{3}\left[\nabla^{j}K_{ij}- \partial_{v}\omega_{i} \right]$ \\
 & &  \\
  & &  \\
  \hline
\end{tabular}
\end{center}
\caption{Table showing the higher derivative corrections to Einstein's equations, for all possible 4-derivative theories of gravity.} \label{table:4derres}
\end{table}

\subsection{The most general structure of the `zero boost term' in $E_{vv}^{\text{HD}}$ } \label{ssec:evvgenstr}
%
Determining the equation of motion and in particular, its `vv'-component, given the coordinate choice in \eqref{Wallmet}, are, in principle, a straightforward task.
But, it becomes increasingly tedious with the number of derivatives present in the action. 
Also, as we will see in \S \ref{ssec:entcurgenpro}, the unambiguous definition of the spatial components of the entropy current arises out of the zero 
boost term in $E_{vv}^{\text{HD}}$. This implies that, for the construction of the entropy current,
we do not need the equation of motion in its every detail. 
What we need is a very specific set of terms in $E_{vv}^{\text{HD}}$, namely the terms that could be written in the form of the first term in 
\eqref{eq:schemeHvv}. For convenience, we are re-writing \eqref{eq:schemeHvv} here again
\begin{equation}\label{eq:schemeHvvrep}
\begin{split}
E_{vv}^{\text{HD}}
&=\underbrace{\partial_v\left[A~\partial_v~B\right]}_{\left(=E_{vv}^{\text{HD}}\big \vert_\text{zero boost}\right)}+\underbrace{{1\over\sqrt{h}}\, \partial_v^2\left[\,\sum_{k \ge 1} \sqrt{h}\left(\partial_r^{k}A^{(k)}\right)\left(\partial_v^{k}~B^{(k)}\right)\right]}_{\left(=E_{vv}^{\text{HD}}\big \vert_\text{higher boost}\right)} + {\cal O}(\epsilon^2) \, . 
\end{split}
\end{equation}
In this subsection, we would like to develop an algorithm that would isolate out these zero-boost terms in $E_{vv}^\text{HD}$. The most important feature of these terms is that at linear 
order in amplitude expansion of the perturbations, it is always possible to rewrite them as
\begin{equation}\label{zerstu}
E_{vv}^{\text{HD}} \bigg\vert_\text{zero boost} = \, \partial_v\left[A\, \partial_vB\right]\sim A \, \partial_v^2 B +\mathcal{O}(\epsilon^2) \, ,
\end{equation} 
where both $A$ and $B$ are boost invariant quantities and they are non-vanishing  on the stationary solutions. Hence here our main focus would be to search for terms of the  form $A\,\partial_v^2 B$ in the `vv'-component of the linearized equation of motion, $E_{vv}^\text{HD}$. However, before proceeding to extract the 
zero boost terms from $E_{vv}^\text{HD}$, let first point out an important ambiguity in defining the zero boost terms, through the structure \eqref{zerstu}.

\subsubsection*{Generating terms like $A \, \partial_v^2 B$ from the $k=1$ terms in \eqref{eq:schemeHvvrep}} 
Before proceeding further with the zero boost terms, we would like to discuss one subtle point that will be important in our attempt to separate out the $k=0$ terms (i.e. the zero boost sector) from the $k \neq 0 $ ones in \eqref{eq:schemeHvvrep}. Recall that our final goal is to determine the form of the boost invariant terms $A$ and $B$ in \eqref{eq:schemeHvvrep} and we plan to do that by keeping track of the terms of form $A \, \partial_v^2 B$ in the linearized $E_{vv}^\text{HD}$. However, the strategy mentioned above to uniquely extract out the zero boost terms from linearized $E_{vv}^\text{HD}$ would be unsuccessful if there is a possibility of generating terms of the form  $A \,\partial_v^2 B$ (with $A$ and $B$ being boost invariant) from the second term in  equation \eqref{eq:schemeHvvrep}. As we will see now, there is indeed such a possibility of contamination arising from the term $k=1$ in the summation on the RHS of \eqref{eq:schemeHvvrep}. Let us analyze this term more carefully 
\begin{equation}\label{eq:someprocess}
\begin{split}
E_{vv}^{\text{HD}}\bigg\vert_{k=1} &= {1\over\sqrt{h}}\, \partial_v^2\left[ \sqrt{h}\left(\partial_rA^{(1)}\right)\left(\partial_v\,B^{(1)}\right)\right] + {\cal O}(\epsilon^2)\\
 &=2 \left(\partial_v\partial_r A^{(1)}\right) \left(\partial_v^2\,B^{(1)}\right) +\left(\partial_rA^{(1)}\right)\left(\partial_v^3\,B^{(1)}\right)+ {\cal O}(\epsilon^2)  \, .
 \end{split}
 \end{equation}
In \eqref{eq:someprocess} above the first term is precisely of the form $\sim X \partial_v^2 Y$, 
where $\left(\partial_v\partial_rA^{(1)}\right)$ and $ B^{(1)}$ respectively can be added to  $A$ and $B$  of the zero boost terms.
Thus, the terms of our interest $A \partial_v^2 B$, which we are looking for in $E_{vv}^\text{HD}$ can be contaminated by 
terms generated from $\partial_v\left[B^{(1)}~\partial_v\left(\partial_v\partial_rA^{(1)}\right)\right]$.
This clearly demonstrates that it is impossible to uniquely determine $A$ and $B$, appearing in the first term of \eqref{eq:schemeHvvrep}, 
just by looking at the terms of the form  $A \, \partial_v^2B$ alone in $E_{vv}^\text{HD}$; it would be difficult to know if they arise from 
$k=0$ or $k=1$ terms in our classification \eqref{eq:schemeHvvrep} for the terms in $E_{vv}^\text{HD}$. 

With this subtlety in mind, let us also comment on the way to tackle this issue. We can subtract off the contributions coming from the $k=1$ terms, that are of the same form as the $k=0$ terms in \eqref{eq:schemeHvvrep}. This could be done easily by noting that whenever such a term is generated from $k=1$ piece, it will also generate the second term in equation \eqref{eq:someprocess}. Hence to determine $A$ and $B$ unambiguously and construct $E_{vv}^{\text{HD}} \big\vert_\text{zero boost}$, 
we will have to isolate out few special terms of the form $(\partial_r X)(\partial_v^3 Y)$ in $E_{vv}^\text{HD}$, with $X$ and $Y$ being boost invariant.

Note that, due to the structural nature of the terms, a similar issue may also arise from the $k=2$ term in \eqref{eq:schemeHvvrep}. 
However, we are not discussing the $k=2$ case in greater detail here, because such terms would not arise in four derivative theories of gravity. 
This is because, there are a total of six derivatives in the $k=2$ terms. 

\subsubsection*{Algorithm to uniquely extract the terms like $A \, \partial_v^2 B$ from linearized $E_{vv}^\text{HD}$ } 

Our job will now be to develop an algorithm, to determine the most general structure of this $k=0$ `zero boost term' appearing in \eqref{eq:someprocess}, 
keeping in mind the above-mentioned subtlety. 

It is clear from our previous discussions that for constructing the entropy current, which satisfies the strongest form of the second law, 
we need the knowledge of the zero boost term in $E_{vv}^{\text{HD}}$, only on the horizon $\mathcal H$. 
This in turn means that, the $g_{vi} = r\, \omega_i$  component of the metric \eqref{Wallmet} can appear in 
the zero boost term, only after differentiation with one $\partial_r$, and the $g_{vv} = r^2 X$ component can appear only after the action of two $\partial_r$. The spatial components of the metric $h_{ij}$ can appear without any derivative acting on it. So the basic building blocks, for constructing the zero boost term on the horizon, are given in Table-(\ref{table:basicblock}).

\begin{table} [h!] 
\centering
 \begin{tabular}{|c|c|c|c|} 
  \hline
  & Candidate terms & Derivative counting & Boost weight \\ \hline
  1.& $h_{ij}$ & zero & zero \\ \hline
  2.& $\omega_i$ & one & zero \\ \hline
  3.& $X$ & two & zero \\ \hline
\end{tabular}
\caption{the basic building blocks}
\label{table:basicblock}
\end{table}

 Let us first concentrate on terms of the form $\left(X\partial_v^2 Y\right)$ and isolate such terms in $E_{vv}^{\text{HD}}$
 when we have a four derivative theory of gravity. These terms in $E_{vv}^{\text{HD}}$ can be constructed by applying $\partial_r$, $\partial_v$ and $\nabla_i$ 
 on these building blocks
 \footnote{To begin with the structures that appear in $E_{vv}^{\text{HD}}$ will have only $\partial_i$. However, we know that $E_{vv}^{\text{HD}}$ is a scalar with respect to the coordinate transformation that only mixes the $\{x^i\}$ coordinates among themselves. If we want to construct scalars out of the horizon data with spatial derivatives on the three building blocks, it must be combined with appropriate spatial derivatives of $h_{ij}$ so that it finally becomes a covariant derivative with respect to $h_{ij}$.
This covariance with respect to the mixing of $\{x^i\}$ tells us that just spatial derivatives of $h_{ij}$ need not be taken as any independent data.

Also note that in our set-up $r$ and $v$ are genuinely distinguished coordinates and we do not demand any covariance with respect to the transformation that mixes these two coordinates among themselves and others. Therefore the derivatives with respect to $r$ and $v$ would remain as simple $\partial_r$ and $\partial_v$.}, so that the total number of derivatives are always equal to four\footnote{In this derivative counting  $\omega_i$ and $X$ must be taken as one derivative and two derivative data, respectively},
when we restrict to the four derivative theories of gravity. 

For convenience, we shall now classify the data in two categories:
$$ 1. ~ \text{ \emph{equilibrium} data} ~~\text{ and} \quad 2.~ \text{\emph{off-equilibrium} data}. 
$$
As it is clear from the names, `equilibrium data' are those structures that are non-vanishing even in a stationary situation, whereas  `off-equilibrium data' vanishes when stationary limits are taken.
Now, from the discussion in appendix-(\ref{app:vdep}) it follows that `equilibrium data' must be `boost-invariant' and therefore could have definite structures and their appropriate products as listed in Table-(\ref{table:eqdata}).

\begin{table} [h!] 
\centering
 \begin{tabular}{|c|c|c|c|} 
  \hline
  & Candidate structures & Derivative counting & Boost weight  \\ \hline
  1.& $ \left(\nabla_{j_1}\cdots\nabla_{j_p}\right)(\partial_r\partial_v)^{m_1}h_{ij}$ & $p+2\,m_1$ & zero \\ \hline
  2.& $(\partial_r\partial_v)^{m_2}\left(\nabla_{j_1}\cdots\nabla_{j_q}\right)\omega_i$ & $q+2\,m_2$ & zero \\ \hline
  3.& $(\partial_r\partial_v)^{m_3}\left(\nabla_{j_1}\cdots\nabla_{j_r}\right)X$ & $r+2\,m_3$ & zero\\ \hline
\end{tabular}
\caption{`equilibrium' and `boost-invariant' structures built out of the basic building blocks.}
\label{table:eqdata}
\end{table}

On the other hand, the `off-equilibrium data' are not boost-invariant, i.e., the total number of $\partial_v$ should be more than the total number of $\partial_r$, when we consider these two derivatives as operators acting on the three basic building blocks listed above. In general there are many possibilities for such `off-equilibrium data'. However, here we are interested in a very specific term in $E_{vv}^{\text{HD}}$, where the total number of $\partial_v$'s is exactly two more than the number of $\partial_r$'s ( again considering them as operators on the basic building blocks and not directly on the metric components). Also both of these two extra $\partial_v$'s must be acting on the same structure, otherwise it would generate a term which is second order in terms of the amplitude expansion we are considering here. `Off-equilibrium data' with this property could have the following structures in general as given in Table-(\ref{table:offeqdata}): 

\begin{table} [h!] 
\centering
 \begin{tabular}{|c|c|c|c|} 
  \hline
  & Candidate structures & Derivative counting & Boost weight  \\ \hline
  1.& $ \left(\nabla_{j_1}\cdots\nabla_{j_p}\right)\partial_v^2(\partial_r\partial_v)^{m_1}h_{ij}$ & $p+2\,m_1+2$ & two \\ \hline
  2.& $\partial_v^2(\partial_r\partial_v)^{m_2}\left(\nabla_{j_1}\cdots\nabla_{j_q}\right)\omega_i$ & $q+2\,m_2+2$ & two \\ \hline
  3.& $\partial_v^2(\partial_r\partial_v)^{m_3}\left(\nabla_{j_1}\cdots\nabla_{j_r}\right)X$ & $r+2\,m_3+2$ & two\\ \hline
\end{tabular}
\caption{The list of `off-equilibrium' and `boost-weight$=2$' data built out of the basic building blocks.}
\label{table:offeqdata}
\end{table}

Finally, we have to contract the `equilibrium data' and `off-equilibrium data' appropriately to get the scalar term in $E_{vv}^{\text{HD}}$. Since in this note we are focusing only on the four-derivative theories of gravity, every term in $E_{vv}^{\text{HD}}$ contains four-derivatives on the metric components. So the relevant equilibrium data can have a maximum of two derivatives acting on the metric components, the possible structures are listed in Table-(\ref{table:list1}). Following the same argument to maintain the derivative counting, the relevant `off-equilibrium data' are listed  below in Table-(\ref{table:list2}). 

\begin{table} [h!] 
\centering
\begin{tabular}{|c|c|l|c|}
    \hline
    \multicolumn{3}{|c|}{Equilibrium and boost-invariant data} & Number of derivatives \\ \hline
   \multirow{3}{*}{1. } &\multirow{3}{*}{Tensor structures: }&$T^{(1)}_{ij} \equiv \partial_r\partial_v\, h_{ij}$ & 2\\ \cline{3-4}
    & &$T^{(2)}_{ij} \equiv \nabla_i\omega_j$ & 2 \\ \cline{3-4} 
    & &$T^{(3)}_{ij} \equiv \mathcal{R}_{ij}$ & 2\\
    \hline
    2. &Vector structure: & $V^{(1)}_i \equiv\omega_i$ & 1 \\
    \hline
    3. &Scalar Structure: &  $S^{(1)} \equiv X~$ & 2 \\
    \hline
\end{tabular}
\caption{Relevant equilibrium and boost invariant data with maximum number of derivatives$=2$, in four-derivative theories of gravity}
\label{table:list1}
\end{table}

\begin{table} [h!] 
\centering
\begin{tabular}{|c|c|l|c|}
    \hline
    \multicolumn{3}{|c|}{Off-equilibrium and boost-weight$=2$ data} & Number of derivatives  \\ \hline
   \multirow{5}{*}{1. } &\multirow{5}{*}{Tensor structures: } &$T^{(4)}_{ij} \equiv \nabla_{i}\nabla_{j}\left(\partial_v^2\, h_{kl}\right)$ & 4\\ \cline{3-4}
    & &$T^{(5)}_{ij} \equiv \nabla_{i}\left(\partial_v^2\, h_{jk}\right)$ & 3\\ \cline{3-4} 
    & &$T^{(6)}_{ij} \equiv \partial_v^2\left(\partial_r\partial_v h_{ij}\right)$ & 4 \\ \cline{3-4} 
    & &$T^{(7)}_{ij} \equiv \partial_v^2\left(\nabla_i\omega_j\right)$ & 4\\ \cline{3-4} 
    & &$T^{(8)}_{ij} \equiv \partial^2_v h_{ij}$ & 2\\
    \hline
    2. & Vector structure: & $V^{(2)}_i \equiv \partial_v^2\omega_i$ & 3\\
    \hline
    3. & Scalar Structure: &  $S^{(2)} \equiv \partial_v^2 X$ & 4\\
    \hline
\end{tabular}
\caption{Relevant off-equilibrium data with maximum number of derivatives$=4$ and boost-weight$=2$, in four-derivative theories of gravity. Let us emphasize that within the tensor structures there are three types of terms: $(i)~4$-index structure: $T^{(4)}_{ij}$, $(ii)~3$-index structure: $T^{(5)}_{ij}$, $(iii)~2$-index structure: $T^{(6)}_{ij}, \,T^{(7)}_{ij} $ and $T^{(8)}_{ij}$.
}
\label{table:list2}
\end{table}

Now our job is to contract these two sets of data as given in Table-(\ref{table:list1}) and Table-(\ref{table:list2}), to get the candidate scalar terms in $E_{vv}^{\text{HD}}$, maintaining the count of total number of derivatives equal to four. This could be done systematically as outlined below:
\begin{itemize}
\item The four-indexed tensor structure $T^{(4)}_{ij}$ itself has four derivatives. Therefore the free indices have to be contracted with zero derivative `equilibrium-data' or just among themselves. Now, there is no `equilibrium-data' that has zero derivatives, see Table-(\ref{table:list1}). Therefore, self contraction of the indices in $T^{(4)}_{ij}$ is the only possibility here and it could be done in two ways leading to two different scalar structures:
$$T_1 = h^{ij}h^{kl}\nabla_{i}\nabla_{j}\left(\partial_v^2~ h_{kl}\right),~~T_2 = h^{ik}h^{jl}\nabla_{i}\nabla_{j}\left(\partial_v^2~ h_{kl}\right)$$
\item The three-indexed `off-equilibrium data' $T^{(5)}_{ij}$ has three derivatives and therefore it has to be contracted with one derivative `equilibrium data' ($V^{(1)}_i=\omega_i$). Here also, two different types of contractions are possible leading to two different scalars:
$$T_3 = h^{ij}h^{kl}\omega_{i}\nabla_{j}\left(\partial_v^2\,h_{kl}\right),~~T_4 = h^{ik}h^{jl}\omega_{i}\nabla_{j}\left(\partial_v^2\,h_{kl}\right)$$
\item The two `off-equilibrium' tensor structures with $2$ indices, $T^{(6)}_{ij}$ and $T^{(7)}_{ij}$, themselves have four derivatives and therefore the free indices have to be contracted among themselves. In each case there is only one way the contraction could be done. The resultant scalars are
$$T_5=h^{ij}\partial_v^2\left(\partial_r\partial_v h_{ij}\right), \quad   T_6 = h^{ij} \partial_v^2\left(\nabla_i\omega_j\right)$$
\item The last `off equilibrium' tensor structure $T^{(8)}_{ij}$ has two derivatives. It has to be contracted with two derivative `equilibrium-data' and also the equilibrium data must have even number (in this case it could be either zero or two) of free indices so that contraction is possible. Here we get the following structures:
\begin{equation*}
\begin{split}
&T_7 = X ~h^{ij}\left(\partial_v^2 h_{ij}\right)\, ,\\
&T_8 = h^{ij}h^{kl} \left(\nabla_i\omega_j\right)\left( \partial_v^2 h_{kl}\right), \quad  T_9= h^{ik}h^{jl} \left(\nabla_i\omega_j\right)\left( \partial_v^2 h_{kl}\right)\, ,\\
&T_{10}= h^{ij}h^{kl} \left(\omega_i\omega_j\right)\left( \partial_v^2 h_{kl}\right), \quad  T_{11}= h^{ik}h^{jl} \left(\omega_i\omega_j\right)\left( \partial_v^2 h_{kl}\right)\, ,\\
&T_{12}= h^{ij}h^{kl} \left(\partial_r\partial_vh_{ij}\right)\left( \partial_v^2 h_{kl}\right), \quad  T_{13}= h^{ik}h^{jl} \left(\partial_r\partial_vh_{ij}\right)\left( \partial_v^2 h_{kl}\right)\\
\end{split}
\end{equation*}
\item The `off-equilibrium' vector data, $V^{(2)}_{i}$, is a three-derivative structure therefore it has to be contracted with one derivative `equilibrium data' $V^{(1)}_i=\omega_i$, leading to the following scalar structure
$$T_{14} = h^{ij}\omega_i\partial_v^2\omega_j$$
\item The `off-equilibrium' scalar data, $S^{(2)}$, itself is a four-derivative  and no contraction is needed.
$$T_{15} = \partial_v^2 X$$
\item Considering possible contractions between the `equilibrium data' $T^{(3)}_{ij}$ (given in terms of the intrinsic curvature of $\mathcal H_v$), and the `off equilibrium' tensor structure $T^{(8)}_{ij}$, we can also get two more terms as given below 
\begin{align*}
T_{16} = h^{ik}\, h^{jl}\, \mathcal{R}_{kl} \, \partial_v^2 h_{ij}; \quad 
T_{17} = h^{ij}\, h^{kl}\, \mathcal{R}_{kl} \, \partial_v^2 h_{ij} \, .
\end{align*}
\item Finally, as we have already mentioned in the beginning of this subsection, to determine the boost-invariant $A$ and $B$ in \eqref{eq:schemeHvvrep} unambiguously, we also need to keep track of terms of the form $(\partial_r X )(\partial_v^3 Y)$ where $X$ and $Y$ are boost invariant. These are the terms which will contribute to the $k=1$ sector of linearized $E_{vv}^{\text{HD}}$. Although, we are interested in finding out the $k=0$ zero-boost sector of the same, we need to track these specific $k=1$ terms (see \eqref{eq:someprocess})  as they will be needed to separate out the boost-invariant $A$ and $B$ in \eqref{eq:schemeHvvrep}. In case of four-derivative theories we have only two possibilities for these terms as listed below
\begin{equation}\label{eq:listmore}
\widetilde T_1 = h^{ij}h^{kl}(\partial_r h_{ij})(\partial_v^3 h_{kl}),  \quad\widetilde T_2 = h^{ik}h^{jl}(\partial_r h_{ij})(\partial_v^3 h_{kl}) \, .
\end{equation}
\end{itemize}
It is important to note that in this list of structures we have not counted $h_{ij}$ and the determinant of $h_{ij}$ as independent structures. All possible occurrences of these two pieces of data are automatically taken care of in the way we have listed our data. For example $h_{ij}$ could only occur in contraction of other indices and all possible contractions of indices are already counted in our listing. Finally, all the nineteen possible candidate terms (seventeen of the $T_i$'s and two of the $\widetilde{T}_i$'s) to appear in $E_{vv}^{\text{HD}} \big\vert_{\text{zero boost}}$, are listed in Table-(\ref{table:T17list}).

\begin{table} [h!] 
\centering
 \begin{tabular}{|l | l|} 
  \hline
  $T_1 = h^{ij}h^{kl}\nabla_{i}\nabla_{j}\left(\partial_v^2~ h_{kl}\right)$ & $T_2 = h^{ik}h^{jl}\nabla_{i}\nabla_{j}\left(\partial_v^2~ h_{kl}\right)$ \\ \hline 
  $T_3 = h^{ij}h^{kl}\omega_{i}\nabla_{j}\left(\partial_v^2~ h_{kl}\right)$ & $T_4 = h^{ik}h^{jl}\omega_{i}\nabla_{j}\left(\partial_v^2~ h_{kl}\right)$ \\  \hline 
  $T_5 = h^{ij}\partial_v^2\partial_r\partial_v h_{ij}$ & $T_6 = h^{ij} \partial_v^2\left(\nabla_i\omega_j\right)$ \\ \hline 
  $T_{7} = X ~h^{ij}\left(\partial_v^2 h_{ij}\right)$ & $T_{8} = h^{ij}h^{kl} \left(\nabla_i\omega_j\right)\left( \partial_v^2 h_{kl}\right)$ \\ \hline 
  $T_{9} = h^{ik}h^{jl} \left(\nabla_i\omega_j\right)\left( \partial_v^2 h_{kl}\right)$ & $T_{10} = h^{ij}h^{kl} \left(\omega_i\omega_j\right)\left( \partial_v^2 h_{kl}\right)$ \\ \hline 
  $T_{11} = h^{ik}h^{jl} \left(\omega_i\omega_j\right)\left( \partial_v^2 h_{kl}\right)$ & $T_{12} = h^{ij}h^{kl} \left(\partial_r\partial_vh_{ij}\right)\left( \partial_v^2 h_{kl}\right)$ \\ \hline 
  $T_{13} = h^{ik}h^{jl} \left(\partial_r\partial_vh_{ij}\right)\left( \partial_v^2 h_{kl}\right)$ & $T_{14} = h^{ij}\omega_i\partial_v^2\omega_j$ \\ \hline
  $T_{15} = \partial_v^2 X$ & $T_{16} = h^{ik}\, h^{jl}\, \mathcal{R}_{kl} \, \partial_v^2 h_{ij}$ \\ \hline
  $T_{17} = h^{ij}\, h^{kl}\, \mathcal{R}_{kl} \, \partial_v^2 h_{ij}$ &    \\
 \hline \hline
  $\widetilde{T}_{1} =h^{ij}h^{kl}(\partial_r h_{ij})(\partial_v^3 h_{kl}) $ &  $\widetilde{T}_{2} = h^{ik}h^{jl}(\partial_r h_{ij})(\partial_v^3 h_{kl}) $ \\ \hline
\end{tabular}
\caption{Listing the seventeen $T_i$'s and two $\widetilde{T}_{i}$'s, the possible $4$-derivative scalar data with boost weight $=2$. They are candidate terms that appear in $E_{vv}^{\text{HD}}$ for $4$-derivative theories of gravity. The seventeen $T_i$ terms  will contribute to $k=0$ sector, and the two $\widetilde{T}_{i}$ terms will contribute to $k=1$ sector of $E_{vv}^{\text{HD}}$.}
\label{table:T17list}
\end{table}

At this stage our claim is that the first term in \eqref{eq:schemeHvvrep}, i.e., the term of the form $\partial_v\left(A \,\partial_v B\right)\sim A\,\partial_v^2 B + \mathcal{O}(\epsilon^2)$, for any four-derivative theory could always be expressed as a sum of these seventeen terms listed in Table-(\ref{table:T17list}) with constant coefficients. Further, we claim that the contribution of the $k=1$ piece from the second term of \eqref{eq:schemeHvvrep} (written in the form of a sum over several $k$ values) could also be expressed in terms of these seventeen structures plus two more, listed in equation \eqref{eq:listmore}
\begin{equation}\label{eq:evvhdgen}
E_{vv}^{\text{HD}} =-\sum_{i=1}^{17}a_i \, T_i- \sum_{i=1}^2 \tilde a_i \, \widetilde T_i+\cdots \, ,
\end{equation}
where $\cdots$ denote the terms that do not matter for the proof of the physical process version of the first law. The negative sign on the RHS of \eqref{eq:evvhdgen} is chosen for convenience. The specific values of these seventeen $a_i$ and two $\tilde a_i$ coefficients appearing in \eqref{eq:evvhdgen}, will of course vary from theory to theory.  
As we have mentioned before, the above classification of terms have been done keeping in mind the four derivative theories of gravity. The most general four derivative theory of pure gravity could have three more terms apart from the standard two derivative term in Einstein gravity. In Table-(\ref{table:const2}) we are listing the values of $a_i$'s and $\tilde a_i$'s for each of these three cases. These set of values of the $a_i$ coefficients are obtained by comparing \eqref{eq:evvhdgen} with the explicit calculation of $E_{vv}^{\text{HD}}$ for each of the three four derivative theories of gravity, which was performed in \S \ref{sssec:ricciscal}, \S \ref{sssec:ricciten} and \S \ref{sssec:riesq}.

\begin{table} [h!]
\centering
 \begin{tabular}{||c| m {4.5cm} | m{6.5cm}||} 
 \hline \hline
 & Different theories \newline$\big(\mathcal{I}^{(i)} = \int d^{d}x \sqrt{-g}~ \mathcal{L}^{(i)}\big)$ & The calculated values of the coefficients $a_i$ and $\tilde a_i$'s\\ [0.5ex] 
 \hline
 & & \\
1. & $\mathcal L^{(1)} = R^{2}$ \newline (Ricci scalar squared)   & $a_1 = 2,~  a_2 = -2, ~ a_3 = 0,~  a_4 = 0,~ \newline a_5 = 4,~  a_6 = -4,~  a_7 = 1,~ \newline a_8 =-2 ,~  a_9 = 4,~  a_{10} =3/2 ,~  \newline a_{11} = -3,~  a_{12} = 4,~\newline  a_{13} =-10 ,~  a_{14} =6 ,~  a_{15} =2 ,~ \newline a_{16} = 2,~  a_{17} = -1 , \newline \tilde{a}_1 = 1, ~ \tilde{a}_2= -3. $ \\ 
 & & \\
 \hline \hline
 & & \\
2. & $\mathcal L^{(2)} = R_{\mu \nu}R^{\mu \nu} $ \newline (Ricci tensor squared) & $a_1 = 1/2,~ a_2 = -1,~ a_3 = -(1/2), ~\newline a_4 = 1,~ a_5 = 1, ~a_6 = 0, ~ \newline a_7 = 1/2, ~ a_8 = -(1/2), ~  a_9 = 1, ~ \newline a_{10} = 1/2, ~a_{11} = -1, ~ a_{12} = 1, ~\newline a_{13} = -2 , ~a_{14} = 2 , ~ a_{15} = 1, ~ \newline a_{16} = 0, ~a_{17} = 0  , \newline \tilde{a}_1 = 1/4 \,, ~ \tilde{a}_2= -(1/2). $  \\
 & & \\
 \hline \hline
 & & \\
3. &$\mathcal L^{(3)} =R_{\mu \nu \rho \sigma }R^{\mu \nu \rho \sigma}$ \newline(Riemann tensor squared) & $a_1 = 0,~  a_2 = -2,~  a_3 = -2, ~ \newline a_4 = 4, ~  a_5 = 0,~  a_6 = 4, ~ a_7 = 1, ~ \newline a_8 = 0, ~ a_9 = 0, ~ a_{10} = 1/2, ~ \newline a_{11} = -1, ~ a_{12} = 0,  a_{13} = 2,~  ~\newline a_{14} = 2, ~ a_{15} = 2, ~ a_{16} = 0, ~ a_{17} = 0,  \newline \tilde{a}_1 = 0, ~ \tilde{a}_2= 1. $.  \\
 & & \\
 \hline \hline
\end{tabular}
\caption{Explicit calculation for each of the three theories produces these values of the coefficients $a_i$, appearing in $E_{vv}^{\text{HD}} =-\sum_{i=1}^{17}a_i \, T_i- \sum_{i=1}^2 \tilde a_i \, \widetilde T_i$\, , for $4$-derivative theories of gravity.}
\label{table:const2}
\end{table}

\subsection{Constraints on the `zero boost terms' in $E_{vv}^{\text{HD}}$ 
} \label{ssec:evvgenstr1}
%
A very specific structure for the zero boost terms in $E_{vv}^\text{HD}$ is predicted in \eqref{eq:hvvzero1}. This structure  does not follow automatically just from the boost transformation property, which we have used to classify terms in the previous subsection. Clearly, imposing \eqref{eq:hvvzero1} would impose further constraints on the seventeen coefficients mentioned above. In this subsection, we shall first find those constraints. We shall list the most general possible structure for $\widetilde B$, defined in \eqref{eq:hvvzero1}, which is a two-derivative  scalar with vanishing boost weight. According to the terminology of the previous subsection it must be an `equilibrium data'. It turns out that $\widetilde B$ could have only five independent structures which are non-vanishing at equilibrium (see table -(\ref{table:Btildelist})). 

\begin{table} [h!]
\centering
 \begin{tabular}{| m {4cm} | m{4cm}|} 
 \hline \multicolumn{2}{|c|}{Candidate terms for $\widetilde B$} \\ \hline
Equilibrium data: & $1. ~h^{ij} \, (\partial_r\partial_v h_{ij}),$ \\
& $2.~h^{ij} \, \nabla_i \,\omega_j,$ \\
& $3.~h^{ij} \, \omega_i \, \omega_j$,  \\
& $4.~\mathcal{R}$,\\ 
& $5.~X,$\\
 \hline
Off-equilibrium data: & $6.~h^{ij} \, h^{kl}\left(\partial_v{h_{ij}}\right)\left(\partial_r h_{kl}\right)$,\\
& $7.~h^{ik} \, h^{jl}\left(\partial_v{h_{ij}}\right)\left(\partial_r h_{kl}\right)$ \\
 \hline
\end{tabular}
\caption{Possible structures that can appear in $\widetilde B$: each of them has two derivatives and boost weight$=0$.}
\label{table:Btildelist}
\end{table}

Using linear combinations of the independent structures presented in Table-(\ref{table:Btildelist}) we can now write down the most general structure of $\widetilde B$, if it exists, as follows
\begin{equation}\label{eq:genstructure0}
\begin{split}
\widetilde B = &~A_1\, h^{ij}\partial_r\partial_v h_{ij} +A_2 \, h^{ij}\nabla_i\omega_j +A_3\, h^{ij}\omega_i\omega_j +A_4~ X+ A_5 \, \mathcal{R}  \,.
\end{split}
\end{equation}
The first term in $\widetilde B$ needs a special attention. This is the term whose contribution to $E_{vv}^\text{HD}$ could get mixed with some the $k=1$ term (see equation \eqref{eq:schemeHvvrep} and the discussion after that). To see this more explicitly, let us write down the contribution to $E_{vv}^\text{HD}$ coming from the term $\widetilde B = A_1\, h^{ij}\partial_r\partial_v h_{ij}$ 
\begin{equation}\label{eq:spterm}
\begin{split}
 \partial_v\left({1\over\sqrt{h}}\partial_v \left(\sqrt{h}~ \widetilde B \right)\right) & \sim A_1\, \partial_v\left({1\over\sqrt{h}}\partial_v \left(\sqrt{h}~ h^{ij}\partial_r\partial_v h_{ij}\right)\right) \\ & = A_1 \,\left( T_5 + \frac{T_{12}}{2} -T_{13} \right).
\end{split}
\end{equation}
It can be easily checked that the terms $T_{12}$ and $T_{13}$ could also be generated as $k=1$ terms in $E_{vv}^\text{HD}$ from the following two off-equilibrium candidates for $\widetilde B$ (see the list of off-equilibrium data in table-\ref{table:Btildelist})
$$ (i)~ h^{ij}h^{kl}\left(\partial_v{h_{ij}}\right)\left(\partial_r h_{kl}\right), \quad (ii)~h^{ik}h^{jl}\left(\partial_v{h_{ij}}\right)\left(\partial_r h_{kl}\right) \, .$$
We assume that in $E_{vv}^\text{HD}$, these two terms mentioned above contribute with coefficients $A_6$ and $A_7$ respectively as written below 
\footnote{To obtain the expressions in \eqref{eq:eqsp0} we have used the following relations 
\begin{align*}
\begin{split}
&\partial_v\left({1\over\sqrt{h}}\partial_v \left(\sqrt{h}~ h^{ij}h^{kl}\left(\partial_v{h_{ij}}\right)\left(\partial_r h_{kl}\right)\right)\right) = \,2\, \,T_{12}+ \widetilde T_1 \, ,\\
&\partial_v\left({1\over\sqrt{h}}\partial_v \left(\sqrt{h}~ h^{ik}h^{jl}\left(\partial_v{h_{ij}}\right)\left(\partial_r h_{kl}\right)\right)\right) = \,2\,T_{13}+ \widetilde T_2 \, .
\end{split}
\end{align*}
}
\begin{equation}\label{eq:eqsp}
\begin{split}
E_{vv}^\text{HD} \big\vert_{k=1}= -A_6\left(2 \,T_{12} + \tilde T_1\right)-A_7\left(2\, T_{13} +\tilde T_2\right).
\end{split}
\end{equation}
From the above equation it is clear that $A_6$ and $A_7$ could be simply  fixed by comparing the coefficients of $\tilde T_1$ and $\tilde T_2$ respectively in the $k=1$ sector of \eqref{eq:evvhdgen} and \eqref{eq:eqsp},
\begin{equation}
A_6=\tilde a_1 \,, \quad A_7 = \tilde a_2 \, .
\end{equation}
We have now extracted out the $k=1$ part of $E_{vv}^{\text{HD}}$ in \eqref{eq:eqsp} which has the form of $A\,\partial_v^2 B$, as desired from \eqref{eq:schemeHvvrep}. Next, we subtract off \eqref{eq:eqsp} from \eqref{eq:evvhdgen} and obtain the part of $E_{vv}^\text{HD}$ that is entirely generated from zero boost sector. This could be written as 
\begin{equation}\label{eq:eqsp0}
\begin{split}
E_{vv}^\text{HD} \big\vert_{\text{zero boost}}= 
&-\sum_{i=1}^{11}a_i \,  T_i - (a_{12} - 2\, \tilde a_1)\, T_{12} \\ 
& - (a_{13} - 2\, \tilde a_2)\, T_{13} -\sum_{i=14}^{17} a_i \,  T_{i} \, .
\end{split}
\end{equation}
At this point, it is important to note that although in \eqref{eq:evvhdgen} there were nineteen terms to begin with, the zero boost sector $E_{vv}^\text{HD} \big\vert_{k=0}$ is constructed out of seventeen terms $T_i$'s appearing on the RHS of \eqref{eq:eqsp0}. We, therefore, have to deal with seventeen coefficients as well. The easiest way to understand this is by realizing that the coefficients $\tilde{a}_1$ and $\tilde{a}_2$ do not count as additional ones since they will always appear in the combination $(a_{12} - 2\, \tilde a_1)$ and $(a_{13} - 2\, \tilde a_2)$ respectively \footnote{In what follows, whenever we refer to the seventeen coefficients $a_i$'s, it will be implied that there are actually nineteen coefficients (the $a_i$'s and the $\tilde{a}_i$'s) but the two coefficients $\tilde{a}_1$ and $\tilde{a}_2$ will always appear being paired with $a_{12}$ and $a_{13}$ respectively, see \eqref{eq:eqsp0}, and hence the independent coefficients will be counted as seventeen. }.

On the other hand, from \eqref{eq:genstructure0} we know that the number of free coefficients in $\widetilde B$ thus turns out to be five, namely the $A_i$'s (for $i=1, \cdots, 5)$. As of now the $A_i$'s are free coefficients and we want to solve them in terms of the $a_i$'s appearing in \eqref{eq:eqsp0}. To do that, we first substitute $\widetilde B$ from (\ref{eq:genstructure0}) in \eqref{eq:hvvzero1} and write it in terms of the basis of $T_i$ structures, as listed in Table-(\ref{table:T17list}) and obtain the following  
\begin{equation}\label{eq:hvvzero3}
\begin{split}
E_{vv}^{\text{HD}} &\big\vert_\text{zero boost} \sim -\, \partial_v\left({1\over\sqrt{h}}\partial_v \left(\sqrt{h}~ \widetilde B\right)\right)\\
& =  - \bigg[ A_1\left(T_5 + \frac{T_{12}}{2} -T_{13}\right)+A_2\left(T_6 + \frac{T_{8}}{2}-T_9 \right) 
\\
& \quad  +A_3\left(2T_{14} + \frac{T_{10}}{2}-T_{11} \right)  +A_4 \left(T_{15} + \frac{T_{7}}{2} \right)
 \\
& \quad - A_5 \left( T_{16} - {T_{17} \over 2} + T_1 - T_2 \right)  \bigg].
\end{split}
\end{equation}
In deriving \eqref{eq:hvvzero3}, we have used the following relations
\begin{equation}\label{eq:tildeBrel}
\begin{split}
&\partial_v\left({1\over\sqrt{h}}\partial_v \left(\sqrt{h} \, h^{ij} \, \partial_r\partial_v h_{ij}\right)\right) = \,T_5 + \frac{T_{12}}{2} -T_{13} \, , \\
&\partial_v\left({1\over\sqrt{h}}\partial_v \left(\sqrt{h} \, h^{ij} \, \nabla_i\omega_j\right)\right) = \,T_6 + \frac{T_{8}}{2}-T_9 \, , \\
&\partial_v\left({1\over\sqrt{h}}\partial_v \left(\sqrt{h} \, h^{ij} \, \omega_i\omega_j\right)\right) = \, 2\,T_{14} + \frac{T_{10}}{2}-T_{11} \, , \\
&\partial_v\left({1\over\sqrt{h}}\partial_v \left(\sqrt{h} \,  X\right)\right) = \,T_{15} + \frac{T_{7}}{2} \, , \\
&\partial_v\left({1\over\sqrt{h}}\partial_v \left(\sqrt{h} \, \mathcal{R}\right)\right) = - \left( T_{16} - {T_{17} \over 2} + T_1 - T_2 \right)\, . 
\end{split}
\end{equation}
We now compare \eqref{eq:hvvzero3} with \eqref{eq:eqsp0} and equate the coefficients of $T_i$'s on both sides, which gives us seventeen relations between the $A_i \,(i =1, \cdots, 5)$ and $a_j \, (j=1, \cdots, 17), \, \tilde{a}_i \, (i=1,\,2)$ \footnote{As we have mentioned before, we have seventeen coefficients on the RHS of \eqref{eq:eqsp0}, and not nineteen, because the coefficients $(a_{12}-2\, \tilde{a}_1)$ and $(a_{13}-2\, \tilde{a}_2)$ always comes in this particular combination.}.
We can solve the five $A_i$'s in terms of the $a_i$'s and then we will be left with twelve constraints on the coefficients $a_i$'s which ensure the consistency of \eqref{eq:hvvzero1}. These twelve constraints on $a_i$'s are listed below,
\begin{equation} \label{eq:aiconstraints1}
\begin{split}
&a_1 = a_{16},~ 2\, a_{10}=-a_{11},~ a_{14}=4\, a_{10},~ a_{15}=2\, a_7,~ 2\, a_{17}=-a_1,  \\   
&a_2 = -a_1,~ a_3=0,~ a_4=0,~ a_6=2\, a_8,~ 2\, a_8=-a_9 \, ,\\
&2 \, (a_{12}-2 \, \tilde{a}_1) = -(a_{13} - 2\, \tilde{a}_2),~ a_5 = 2 \, (a_{12} - 2 \, \tilde{a}_1) \, .
\end{split}
\end{equation}

Finally, we would like to check whether the $a_i$'s as given in Table-(\ref{table:const2}) satisfy the constraints given in \eqref{eq:aiconstraints1}. Remember that in the previous subsection, we have already calculated the allowed values of the $a_i$'s for each of the three different $4$-derivative theory of gravity, see Table-(\ref{table:const2}). Upon inspection, we can convince ourselves that for Ricci scalar squared theory the  constraints in \eqref{eq:aiconstraints1} are satisfied, where as for both of the other two four derivative theories of gravity, namely the Ricci tensor squared and the Riemann tensor squared theories, the constraints in \eqref{eq:aiconstraints1} are simply not satisfied. Therefore, we convince ourselves that the constraints obtained in \eqref{eq:aiconstraints1} are not correct, as they are not satisfied by the results obtained by explicit calculation of $E_{vv}^{\text{HD}}$ which is the content of Table-(\ref{table:const2}) for the most general four derivative theory of gravity. We should keep this in mind that these constraints were derived from demanding the consistency of \eqref{eq:hvvzero1}. As a result, we are led to the conclusion that the general structure of $E_{vv}^{\text{HD}}$ in the zero boost sector, as predicted in \eqref{eq:hvvzero1}, is not generically true for the most general four derivative theory of gravity.  
\subsection{The general strategy for constructing the entropy current maintaining the boost symmetry}\label{ssec:entcurgenpro}
%
From the analysis of the previous subsection, we have established the fact that the zero boost terms in $E_{vv}^\text{HD}$ does not always follow the structure predicted in \eqref{eq:hvvzero1}. Motivated by this observation, in this subsection our goal will therefore be to explore what are the further structures, if any, that we could allow for the zero boost terms in $E_{vv}^\text{HD}$ without affecting the proofs for the physical version of the first law and second law.

As we have explained before,  the first law is a statement about the total change in the thermodynamic parameters like entropy, energy etc., characterizing two nearby equilibrium solutions connected by dynamics. Hence its formulation always involves an integration over all space and therefore is usually  insensitive to any boundary terms. The same is true for black hole mechanics. The total change in entropy as described in \eqref{eq:1stterm2} has an integration over the spatial slices of the horizon. 
If the horizon is compact, this integration would be insensitive to any boundary term that appears in $E_{vv}^{\text{HD}}$. It follows that the zero boost term in $E_{vv}^{\text{HD}}$, in addition to the term already mentioned in \eqref{eq:hvvzero1}, could also have a structure of the form
\begin{equation}\label{eq:hvvzero2}
E_{vv}^{\text{HD}} \big\vert_\text{zero boost} \sim - \,\partial_v \left(\nabla_i J^i\right)=-\,\partial_v \left({1\over\sqrt{h}}\partial_i \left(\sqrt{h} \, J^i\right)\right) \, 
\end{equation}
where $J^i$ is some spatial current with boost weight 1 (i.e., it must contain an explicit $\partial_v$ that could not be paired up with any $\partial_r$). 
On compact horizons such a term would clearly integrate to zero and therefore will not contribute to the total change in entropy (see the derivation of
\eqref{eq:1stterm2}).

It is worth noting that the compatibility with the first law also allows a term, generically of the form $\nabla_i Y^i$ in $E_{vv}^{\text{HD}}$ where $Y^i$ is some arbitrary vector quantity, i.e. a spatial current with boost weight equal to $2$. However, the manipulation that follows from \eqref{eq:schemeHvv1} shows that working up to linear order of amplitude perturbations we could always re-arrange the terms in $E_{vv}^{\text{HD}}$ (including the possible $\nabla_i Y^i$ term) in a form where there is an overall $\partial_v$ outside\footnote{This can be schematically presented as 
\begin{equation} \label{eq:rearrangeK}
E_{vv}^{\text{HD}}\big\vert_\text{zero boost} \sim \nabla_i Y^i \sim \partial_v \left(\nabla_i \widetilde{Y}^i\right) + \mathcal{O}(\epsilon^2),
\end{equation}
where $\widetilde{Y}^i$ is some spatial vector with boost weight equal to one, since one $\partial_v$ is extracted from $Y^i$ which has boost weight equal to one.}.
It is important to stress that although the first law itself does not require this rearrangement as in \eqref{eq:rearrangeK}, it is a must to proceed towards an argument for the second law. Therefore, in our classification, we shall not consider such terms for which this rearrangement is not true. This, in particular, allows us not to consider the term that we have just mentioned above in \eqref{eq:hvvzero2} as a possible term in $E_{vv}^{\text{HD}}$.

Combining equations \eqref{eq:hvvzero1} and \eqref{eq:hvvzero2}, it follows that, both the first and the second law would be satisfied, at least at the linear order in amplitude of time-dependent perturbations, provided the zero boost terms in $E_{vv}^{\text{HD}}$ has the following form
 \begin{equation}\label{eq:hvvzero}
E_{vv}^{\text{HD}} \big\vert_\text{zero boost} \sim -\, \partial_v \left({1 \over\sqrt{h}}\partial_v \left(\sqrt{h}\, \widetilde B\right)+\nabla_i J^i\right).
\end{equation}
Interestingly, we should note that on the RHS of \eqref{eq:hvvzero}, the term inside the parenthesis (i.e. ignoring the overall $\partial_v$), looks exactly like the divergence of a `four-current', let us call it $S^A$, such that it's $v$ and $i$ components are respectively given by 
\begin{equation}\label{entcurkzero}
\begin{split}
E_{vv}^{\text{HD}} \big\vert_\text{zero boost}& \sim -\, \partial_v \left(\nabla_{A}S^A\right), \\ \text{such that} ~~ S_{(k=0)}^v &= \widetilde{B} \, , ~~ S_{(k=0)}^i =  J^i \, ,
\end{split}
\end{equation}
where, the index $A = v,\, x^i$ and we have also used $k=0$ as a  subscript in $S^A_{(k=0)}$ to denote the fact that we are only looking at the zero boost terms in $E_{vv}^{\text{HD}}$.

Next, we would like to see how the seventeen structures, listed in Table-(\ref{table:T17list}) in \S \ref{ssec:evvgenstr}, should combine so that the zero boost terms in $E_{vv}^\text{HD}$ could be recast in the form of \eqref{eq:hvvzero}. In other words, if the form of $E_{vv}^\text{HD}$ as proposed in \eqref{eq:hvvzero} is correct, we will be using that to derive the constraints that the seventeen coefficients $a_i$ should satisfy. As it appears in \eqref{eq:hvvzero}, $\widetilde B$ is a boost invariant scalar data and $J^i$ is a vector data with boost weight one. In the previous subsection, we have already argued the most general structure of $\widetilde B$ in \eqref{eq:genstructure0}. Now $J^i$ is an off-equilibrium data and from the counting of boost-weight we could see it must have exactly one $\partial_v$ derivative, which is not paired with an $\partial_r$ provided we are considering them as operators acting on the three basic building blocks, namely $h_{ij}$, $\omega_i$ and $X$. Taking all these facts into account, we could construct the five possible structures for a candidate term in $J_i$, as listed in Table-(\ref{table:Jilist}) below. 
\begin{table} [h!]
\centering
 \begin{tabular}{| m {4cm} | m{3cm}|} 
 \hline  \multicolumn{2}{|c|}{Candidate terms for $J_i$} \\ \hline
& $1.~ \partial_v\omega_i $  \\
& $2.~h^{jk}\nabla_j\left(\partial_v h_{ki}\right)$ \\
Off-equilibrium data: & $3.~h^{jk}\nabla_i\left(\partial_v h_{jk}\right)$ \\
& $4.~h^{jk}\,\omega_j\left(\partial_v h_{ki}\right)$\\ 
& $5.~h^{jk}\,\omega_i\left(\partial_v h_{jk}\right)$ \\
 \hline
\end{tabular}
\caption{Possible structures that can appear in $J_i$: each one of them has two derivatives and boost weight$=1$.}
\label{table:Jilist}
\end{table}

It is now straightforward to write down the most general form of $J^i$ using linear combinations of the structures written in Table-(\ref{table:Jilist})
 \begin{equation}\label{eq:current}
\begin{split}
J^i =& ~B_1 \, h^{ij}\partial_v\omega_j  +B_2 \,h^{il}h^{jk} \, \nabla_j\left(\partial_v h_{kl}\right) +B_3 \,  h^{il}h^{jk}\,\nabla_l\left(\partial_v h_{jk}\right)  \\
& + B_4 \, h^{il}h^{jk}\,\omega_j\left(\partial_v h_{kl}\right)+ B_5 \, h^{il}h^{jk}\,\omega_l\left(\partial_v h_{jk}\right), 
\end{split}
\end{equation}
where the coefficients $B_i$ for $i=1, \cdots, 5$, are, as of now, arbitrary constant coefficients. Our aim will now be to fix them in terms of the coefficients $a_i$'s ($i=1,\cdots, 17$), just like the coefficients $A_i$'s, appearing in \eqref{eq:genstructure0}, were fixed in the previous subsection. To achieve this we will calculate the second term on the RHS of \eqref{eq:hvvzero}, with $J^i$ being substituted from \eqref{eq:current}. We express the resulting expression in terms of the $T_i$'s, listed in Table-\ref{table:T17list}, and obtain the following relation 
\begin{equation}\label{eq:hvvzeroJi}
\begin{split}
E_{vv}^{\text{HD}} \big\vert^{(J^i~ \text{part})}_\text{zero boost} &\sim - \, \partial_v \left(\nabla_i J^i\right)  ~= - \bigg[ 
B_1\left(T_6+T_4-{T_3\over 2}\right)+ B_2 ~T_2 \\ & + B_3 ~T_1
 +B_4~ (T_4 +T_9) + B_5~(T_3 +T_8) \bigg].
\end{split}
\end{equation}
In deriving \eqref{eq:hvvzeroJi} we have used the following relations 
\begin{equation}\label{eq:DiJilist}
\begin{split}
&\partial_v\left[\nabla_i (h^{ij}\partial_v\omega_j)\right]= \, T_6 +T_4 - (T_3 / 2) \, , ~~
\partial_v\left[\nabla_i (h^{il}h^{jk}\nabla_j\left(\partial_v h_{kl}\right))\right] = \, T_2,\\
& \partial_v\left[\nabla_i (h^{il}h^{jk}\nabla_l\left(\partial_v h_{jk}\right))\right] = \, T_1 \, , ~~
\partial_v\left[\nabla_i (h^{il}h^{jk}\omega_j\left(\partial_v h_{kl}\right))\right] =\,  T_4 + T_9,\\
& \partial_v\left[\nabla_i (h^{il}h^{jk}\omega_l\left(\partial_v h_{jk}\right))\right] = \, T_3 + T_8 \, .
\end{split}
\end{equation}
Once we have obtained \eqref{eq:hvvzero3} and \eqref{eq:hvvzeroJi}, we shall combine them to obtain a complete expression for the zero boost part (i.e. $k=0$) of $E_{vv}^{\text{HD}}$ in terms of the $T_i$'s as follows

 \begin{equation}\label{eq:finfirst}
 \begin{split}
 &E_{vv}^{\text{HD}} \big\vert_\text{zero boost} 
 =~ - \bigg[ A_1\left(T_5 + \frac{T_{12}}{2} -T_{13}\right) +A_2 \left(T_6 + \frac{T_{8}}{2} -T_{9}\right)
\\
 &+A_3\left(2T_{14} + \frac{T_{10}}{2}-T_{11} \right)+A_4 \left(T_{15} + \frac{T_{7}}{2} \right) \\
 &
  - A_5 \left( T_{16} - {T_{17} \over 2} + T_1 - T_2 \right) + B_1\left(T_6+T_4-{T_3\over 2}\right)+ B_2 ~T_2 + B_3 ~T_1 \\
 &
 +B_4~ (T_4 +T_9) + B_5~(T_3 +T_8) \bigg].
 \end{split}
 \end{equation}
 
It is obvious from the RHS of \eqref{eq:finfirst} above that we still have ten undetermined coefficients, five of the $A_i$'s and five of the $B_i$'s. We therefore conclude that, if we want the first term in \eqref{eq:schemeHvvrep}, to have a form such that it is compatible with the physical process version of the first law, then it can have twelve independent coefficients ($A_i, \,B_i$) for any four derivative theories of gravity. On the other hand, just from the consideration of boost symmetry, a total of seventeen terms are allowed in $E_{vv}^{\text{HD}}$, see \eqref{eq:evvhdgen}. Clearly, even after the inclusion of the spatial current in \eqref{eq:hvvzero}, the compatibility with the physical version of the first law would imply some constraints between $a_i$'s (though it would certainly be less in number than what we have derived in the previous subsection). A naive counting suggests that there must be $(17-10)\, = \, 7$ relations among the seventeen possible coefficients $a_i$. However, as it turns out, there is a redundancy in our counting of independent structures that could appear in the expression of entropy density ($\widetilde{B}$) and spatial entropy current ($J^i$). In other words, not all of the ten $A_i, \, B_i$'s are independently and one of them, the term with $A_2$ as coefficient, can be absorbed into others by redefining some of the $B_i$ coefficients. It is easy to check that if we redefine the coefficients $B_1, \, B_4$ and $B_5$ in the following way,
\begin{equation} \label{eq:defBhat}
\widehat{B}_1 = B_1 + A_2 \, , \quad \widehat{B}_4 = B_4 - A_2 \, , \quad \widehat{B}_5 = B_5 + {A_2 \over 2} \, ,
\end{equation}
the term with coefficient $A_2$ in \eqref{eq:finfirst} disappears and we are left with
 \begin{equation}\label{eq:finfirstnew}
 \begin{split}
 &E_{vv}^{\text{HD}} \big\vert_\text{zero boost} 
 =~ - \bigg[ A_1\left(T_5 + \frac{T_{12}}{2} -T_{13}\right)
+A_3\left(2T_{14} + \frac{T_{10}}{2}-T_{11} \right)\\
 & +A_4 \left(T_{15} + \frac{T_{7}}{2} \right) - A_5\left( T_{16} - {T_{17} \over 2} + T_1 - T_2 \right)  \\
 &  + \widehat{B}_1\left(T_6+T_4-{T_3\over 2}\right)+ B_2 ~T_2 + B_3 ~T_1 +\widehat{B}_4~ (T_4 +T_9) \\
 &   + \widehat{B}_5~(T_3 +T_8) \bigg].
 \end{split}
 \end{equation}
 
As the independent terms on the RHS of \eqref{eq:finfirstnew} has now been reduced to nine, we should obtain eight relations among the coefficients $a_i$, which are given by
\begin{equation}
 \begin{split}\label{eq:relation}
 a_4 =&\, a_6 + a_9, \quad a_3 = -{a_6 \over 2} + a_8, \quad a_{16} = -2\, a_{17}, \quad a_{15} = \,  2 \, a_7,  \\  a_{11} = &-{a_{14} \over 2}, \quad a_{10} = {a_{14} \over 4} \, , \quad 2\, (a_{12} - 2\, \tilde{a}_1)=a_5\, ,  \quad a_{13}- 2\,\tilde{a}_2 =\, a_5 \,.
 \end{split}
 \end{equation}
Furthermore, once the $a_i$'s satisfy the identities given in \eqref{eq:relation}, we can solve the $A_i$'s and $B_i$'s in terms of the $a_i$'s, as given below \footnote{In \eqref{eq:solution} we are still writing in terms of the coefficients $B_1, \, B_2, \, B_3$, instead of writing them in terms of the redefined $\widehat{B}_1,\,\widehat{B}_2,\,\widehat{B}_3$. This makes the appearance of the undetermined coefficient $A_2$ explicit, and is just a matter of convenient choice for us.}
\begin{equation}\label{eq:solution}
\begin{split}
& A_1 = a_5 \, , ~~ A_3 = {a_{14} \over 2}, ~~ A_4 = 2\, a_7 \, , ~~ A_5 = 2\,  a_{17} \,, \\ &   \,  A_2 =  \text{free/undetermined}, \\ & B_1 =  a_6-A_2,~~ B_2 = a_2 -2 \, a_{17} \, ,  ~~ B_3 = a_1 + 2 \, a_{17}\, ,~~ B_4 =  a_9 +A_2 \, ,\\ & B_5 =  a_8-{A_2 \over 2} \,. 
\end{split}
\end{equation}
It is worth mentioning that in deriving the identities in \eqref{eq:relation} and the solutions in \eqref{eq:solution} we have not assumed any particular form of the four derivative gravity Lagrangian. In other words these relations are true for any four derivative theory of gravity. 

Once we have obtained the coefficients $A_i$ and $B_i$, one can readily derive the entropy density $\widetilde{B}$ and the entropy currents $J^i$ in terms of the coefficients $a_i$. Since specific values for the set of coefficients $a_i$ corresponds to specific four derivative theories of gravity, (see Table-(\ref{table:const2})), we can substitute them for $a_i$'s in \eqref{eq:solution} to obtain the specific values of $A_i$ and $B_i$ for each of the three individual four derivative theories of gravity. We present them in  Table-(\ref{table:const3}).

\begin{table} [h!]
\centering
 \begin{tabular}{||c| m {4.5cm} | m{6.5cm}||} 
 \hline \hline
 & Different theories \newline$\big(\mathcal{I}^{(i)} = \int d^{d}x \sqrt{-g}~ \mathcal{L}^{(i)}\big)$ & Values of the coefficients $A_i,\, B_i$ \\ [0.5ex] 
 \hline
 & & \\
1& $\mathcal L^{(1)} = R^{2}$ \newline (\footnotesize{Ricci scalar squared})   & $A_1 = 4,~  A_2 = \text{undetermined}, ~ \newline A_3 = 3,~  A_4 = 4 ,~ A_5 = -2,~
\newline B_1 = -4-A_2,~  B_2 = 0,~ B_{3} = 0,~  \newline  B_{4} = 4+A_2,~  B_{5} = -2-(A_2/2).$ \\ 
 & & \\
 \hline \hline
 & & \\
2& $\mathcal L^{(2)} = R_{\mu \nu}R^{\mu \nu} $ \newline (\footnotesize{Ricci tensor squared}) & $A_1 = 1,~  A_2 = \text{undetermined}, ~ \newline   A_3 =1,~  A_4 = 2,~ A_5 = 0,~ 
\newline B_1 = -A_2,~  B_2 = -1,~ B_{3} = 1/2,~  \newline  B_{4} = 1+A_2,~  B_{5} =-(1/2)-(A_2/2).$ \\
 & & \\
 \hline \hline
 & & \\
3&$\mathcal L^{(3)} =R_{\mu \nu \rho \sigma }R^{\mu \nu \rho \sigma}$ \newline(\footnotesize{Riemann tensor squared}) & $A_1 = 0,~  A_2 = \text{undetermined}, ~ \newline  A_3 =1,~ A_4 =4 ,~   A_5 = 0,~
\newline B_1 = 4-A_2,~  B_2 = -2,~B_{3} = 0,~  \newline  B_{4} = A_2,~  B_{5} = -(A_2/2).$  \\
 & & \\
 \hline \hline
\end{tabular}
\caption{$A_i, \, B_i$'s for different $4$-derivative theories of gravity.}
\label{table:const3}
\end{table}

Finally, we conclude this sub-section with the following remarks: 
\begin{itemize}
\item{\bf More details on the redundancy in the parameter $A_2$:} We have already mentioned before that our analysis in this subsection to classify possible candidate terms in the zero boost sector of $E_{vv}^{\text{HD}}$ solely based on boost symmetry, cannot fix the coefficient $A_2$ in \eqref{eq:finfirst}. As a result it remained undetermined in \eqref{eq:solution}. We have also seen that this redundancy in fixing $A_2$ is actually related to a proper count of the independent data in $\widetilde B$ and $J^i$. 

In order to make it explicitly manifest, let us now consider the specific terms written below and their combinations as candidates for $\widetilde B$ and $J^i$ 
\footnote{Note that this $\widetilde B_{(*)}$ appears in the expression of $\widetilde B$ in \eqref{eq:genstructure0} with the coefficient $A_2$.}
$$\widetilde B_{(*)} =  h^{ij}\nabla_i\omega_j,~~J_{(*)}^i =-h^{ij}\partial_v\omega_j + h^{il}h^{jk}\omega_j\partial_v h_{kl} -{1\over 2} h^{il}h^{jk}\omega_l\partial_v h_{jk} \, ,$$
and with this choices it can be shown that 
\begin{equation}\label{eq:ppes}
{1\over \sqrt{h}}\partial_v \left(\sqrt{h}~\widetilde B_{(*)}\right) + \nabla_i J_{(*)}^i =0 \, .
\end{equation}
The interesting thing to note about the combination written in \eqref{eq:ppes} is that it identically vanishes without any use of the gravity equations of motion and therefore we could add the $v$-derivative of this combination (so that it has the appropriate boost weight$=2$) to any expression for $E_{vv}^{\text{HD}}$, without affecting the equation of motion and dynamics. Because of this, among the twelve terms that appeared on the RHS of \eqref{eq:finfirst} above we could hope to fix only eleven of them by comparing with the $E_{vv}^{\text{HD}}$ of a given four derivative theory, \eqref{eq:evvhdgen} 
\footnote{
Actually, we can make use of this redundancy to reorganize \eqref{eq:genstructure0} and \eqref{eq:current} with the following re-definition of $\widetilde B$ and $J^i$
\begin{equation}\label{eq:genstructure2}
\widetilde B \rightarrow \widetilde B \,; \quad 
J^i \rightarrow J^i + \alpha_{*} \, A_2 \, J^i_{(*)},
\end{equation}
where $\alpha_{*}$ being a tunable free parameter and thus enabling us to fix the value of the coefficient $A_2$ to any specific number. In particular by making the choice of $\alpha_{*}=1$, we can even make the coefficient $A_2$ not contributing to \eqref{eq:finfirst}, as in that case, as $A_2$ disappears from $E_{vv}^{\text{HD}}$.
}. 
Also, for the same reason, we have seen that in each of the three cases tabulated in Table-(\ref{table:const3}), the coefficient $A_2$ could not be fixed as it could combine with few spatial currents to give vanishing contribution to $E_{vv}^\text{HD}$.

\item{\bf The redundancy in $A_2$ is fixed by matching the equilibrium limit of $\widetilde{B}$ with the equilibrium Wald entropy density:} Having realized the fact that only boost symmetry alone can not fix the coefficient $A_2$, let us now focus on the implications of this redundancy in the coefficient $A_2$ beyond boost symmetry and try to explore if there is any other principle that can fix it. Looking at the Table-(\ref{table:Btildelist}) and \eqref{eq:genstructure0}, we remind ourselves that, by construction, the scalar structures appearing in $\tilde B$ does not vanish when evaluated on a stationary solution. Most importantly, the term that appears in $\tilde{B}$, \eqref{eq:genstructure0}, with coefficient $A_2$ is generically non-zero in the equilibrium limit. Therefore the redundancy in the coefficient $A_2$ discussed in detail above, implies that possible different choices of $A_2$ would amount to having different expressions for the equilibrium entropy density, $s_w$, of the same configuration. Though the difference does not persist in the expression of total entropy $S_W$, since this density turns out to be a total derivative term: $\left(\nabla\cdot\omega\right)$ in this case. 

Motivated by the arguments given above and based on general grounds, we should, therefore, also require that once the equilibrium limit is considered, the entropy density $\widetilde{B}$ in \eqref{eq:genstructure0}, should reproduce the appropriate Wald entropy density. This should be satisfied by the $\widetilde{B}$ apart from being constructed following the boost symmetry. As we will see now, at least for the cases that we are studying in this note, this additional requirement uniquely fixes the ambiguity related to the coefficient $A_2$. Thus, the important point to note here is that the Wald's formula \eqref{eq:waldformula} picks up a very specific value for $A_2$ for every case that we have discussed here, and in some sense fixes this ambiguity which clearly could not be fixed just by imposing first or second law of thermodynamics even in its ultra-local version. For example, in $R^2$ theory, once we demand matching with \eqref{eq:waldformula}, $A_2$ gets fixed to a specific numerical value $A_2 = -4$, implying that there is no spatial current, which is actually consistent with what we have found in subsection \S\ref{ssec:4dercor} \footnote{A first glance at the non-zero values of the coefficients $B_1, \, B_4$ and $ B_5$ for the $R^2$ theory in Table-(\ref{table:const3}) might naively suggest that there is a non-zero current for the $R^2$ theory. However once we make the choice of $A_2 = -4$ in order to match with the equilibrium Wald entropy density $s_w^{\text{HD}}$, it can be verified that there is no spatial current in this case, but a finite non-equilibrium correction $s_{cor}$ to $s_w^{\text{HD}}$, see \eqref{eq:splitentropy}.
}.

The consistency with Wald's formula in the equilibrium limit, forces the entropy density $\widetilde{B}$, that we have obtained in this subsection, to reduce to the stationary limit of $s^\text{HD}_{Wald}$ (see \eqref{eq:waldentdensityHD}), which we derived in subsection \S\ref{ssec:4dercor}, up to the ambiguity of $A_2$. More precisely, if we take the expressions of $s^\text{HD}_{Wald}$ as computed in  subsections \S\ref{sssec:ricciscal}, \S\ref{sssec:ricciten} and \S\ref{sssec:riesq} and simply remove the terms that would vanish in stationary situations (for example, a term like ${\cal K}\bar{\cal K}$ would be ignored), the resultant expressions should exactly match with the corresponding $\tilde B$'s derived in this subsection with a specific choice of the coefficient $A_2$ for every case 
\footnote{Though a mismatch at this stage would have been a serious contradiction with the existing literature and Wald's formalism, we still do not have any abstract proof for it, applicable to any higher derivative theories of gravity. According to our understanding, this would essentially amount to showing a step by step equivalence between the proof of physical version of the first law and the Wald formalism. We could not find it in literature and leave it for future work.
}.
It turns out that they indeed match provided we choose the coefficient $A_2$ to be as follows:
\begin{enumerate}
\item for $R^2$ theory: $A_2 = -4$,
\item for $R_{\mu\nu}R^{\mu\nu}$ theory: $A_2 = -1$,
\item for $R_{\mu\nu\alpha\beta}R^{\mu\nu\alpha\beta}$ theory: $A_2 = 0$.
\end{enumerate}
This matching serves as a consistency check for our results. Therefore, once we use the values of the coefficient $A_2$ for different cases, as written above, in Table-(\ref{table:const3}) and further using \eqref{eq:genstructure0} and \eqref{eq:current} the specific expressions for $\widetilde{B}$ and $J^{i}$ can be derived as listed in Table-(\ref{table:const4}). 
\begin{table} [h!]
\centering
 \begin{tabular}{||c| m {4.2cm} | m{7cm}||} 
 \hline \hline
 & Different theories \newline$\big(\mathcal{I}^{(i)} = \int d^{d}x \sqrt{-g}~ \mathcal{L}^{(i)}\big)$ &  Expressions for $\widetilde{B}\, ,~ J^{i}$ \\ [0.5ex] 
 \hline
 \multirow{2}{*}{1. } & \multirow{2}{*}{$\mathcal L^{(1)} = R^{2}$}  & $\widetilde{B} =4 \, h^{ij} \, \partial_r\partial_v h_{ij} -4 \, h^{ij} \nabla_i \omega_j \, \newline + \, 3 \, h^{ij} \, \omega_i\omega_j +\, 4\, X
-2 \, \mathcal{R}$ \\ 
\cline{3-3} 
  & & $J^{i} =0$\\
\cline{3-3} 
 \hline \hline
 \multirow{5}{*}{2.} & \multirow{5}{*}{$\mathcal L^{(2)} = R_{\mu \nu}R^{\mu \nu} $} &  $\widetilde{B} = h^{ij}\,\partial_r\partial_v h_{ij}+\,h^{ij}\, \omega_i\omega_j + \,2\,X$ \\ 
\cline{3-3} 
 & & $J^{i} =-\,h^{il}h^{jk}\,\nabla_j\left(\partial_v h_{kl}\right)  \newline +{1 \over 2} \,h^{il}h^{jk}\nabla_l\left(\partial_v h_{jk}\right) 
+\,h^{il}h^{jk}\,\omega_j\,\left(\partial_v h_{kl}\right)\newline  - {1 \over 2}\, h^{il}h^{jk}\,\omega_l\,\left(\partial_v h_{jk}\right)$\\
\cline{3-3} 
 \hline \hline
\multirow{2}{*}{3.} & \multirow{2}{*}{$\mathcal L^{(3)} =R_{\mu \nu \rho \sigma }R^{\mu \nu \rho \sigma}$ } & $\widetilde{B} =\,h^{ij}\,\omega_i\omega_j+\,4\,X $ \\ 
\cline{3-3} 
& & $J^{i} =4\, h^{ij}\,\partial_v\omega_j  -2 \,h^{il}h^{jk} \, \nabla_j\left(\partial_v h_{kl}\right) $\\
\cline{3-3}
 \hline \hline
\end{tabular}
\caption{$\widetilde{B}$ and $ J^{i}$'s for different four derivative theory of gravity. While writing the expressions we have used the values for the coefficient $A_2$ in each of the three cases as following : (i) for $R^2$ theory: $A_2 = -4$, (ii) for $R_{\mu\nu}R^{\mu\nu}$ theory: $A_2 = 0$, and (iii) for $R_{\mu\nu\alpha\beta}R^{\mu\nu\alpha\beta}$ theory: $A_2 = 0$.
}
\label{table:const4}
\end{table}

\item{\bf Constraints on $a_i$'s satisfied: } In \S \ref{ssec:evvgenstr} we computed the specific values of the coefficients $a_i$'s and tabulated them in the Table-(\ref{table:const2}) for three different four derivative theories of gravity. It is now straightforward to check that the constraints derived in \eqref{eq:relation} are indeed satisfied by all of the four derivative theories of gravity. In other words, the physical process version of the first law holds for all of these theories once we allow for the spatial current term in $E_{vv}^{\text{HD}}$, \eqref{eq:hvvzero}.

\end{itemize}
\subsection{Einstein-Gauss-Bonnet gravity in $d \geq (4+1)$}
%
The Einstein-Gauss-Bonnet theory has been extensively studied as a prototype of higher derivative corrections to Einstein's gravity 
and has been accorded significant importance in the relevant literature. 
It is also a theory with 4-derivative correction to Einstein's gravity, where 
the 4-derivative term is a specific combination of the three terms, that has been discussed 
in \S \ref{sssec:ricciscal}, \S \ref{sssec:ricciten} and \S \ref{sssec:riesq}. 
This linear combination is such that, although the Einstein-Hilbert action has 4-derivatives corrections, the equations 
of motion that follow from it, only have two derivatives on the metric, just like Einstein equations. 
Since the Gauss-Bonnet term is simply a specific linear combination of the four derivative terms discussed in the previous sections, the analysis for the Einstein-Gauss-Bonnet theory can be done quickly by 
considering the same linear combination of the results we obtained before.
In this section, we state our results explicitly for this theory. 

The Einstein-Gauss-Bonnet theory is non-trivial in any dimensions greater than $3+1$. In $3+1$ dimension the 4-derivative term is a total derivative (and is, therefore, a topological surface term). In lower dimensions, it vanishes as an identity. Let us first consider this theory in space-time dimensions $d \geq 4+1$; we shall discuss the 
the special case of $d=3+1$ in the next subsection. 

The action for  Einstein-Gauss-Bonnet theory is given by 
\begin{equation}\label{eq:gauss1}
I=\int d^{d}x \sqrt{-g}\left(R+a_{gb}\left(R^{2}-4R_{\mu \nu}R^{\mu \nu}+R_{\mu \nu \rho \sigma }R^{\mu \nu \rho \sigma} \right) \right), 
\end{equation}
where $a_{gb}$ is a constant Gauss-Bonnet parameter. 
The corresponding equations of motion are
\begin{equation}
E_{\mu \nu}=R_{\mu \nu} -\frac{1}{2} g_{\mu \nu} R + E_{\mu \nu}^{HD}=0,
\end{equation}
where
\begin{equation}
\begin{split}
E_{\mu\nu}^{HD} = a_{gb} \bigg( & 2RR_{\mu\nu} - 4 R^{\alpha \beta}R_{\mu \alpha \nu \beta} -4{R_{\mu}}^{\alpha}R_{\nu \alpha} +2 R^{\alpha \beta \sigma}_{\mu}R_{\nu \alpha \beta \sigma} \\ 
&-\frac{1}{2}g_{\mu\nu} (R^{2}- 4 R_{\alpha\beta}R^{\alpha\beta} + R_{\alpha \beta\gamma\rho}R^{\alpha \beta\gamma\rho}) \bigg) \, .
\end{split}
\end{equation}
The explicit vv-component of the equations of motion is
\begin{equation}
\begin{split}
E_{vv}&=R_{vv} + E_{vv}^{\text{HD}} = 0, \\
 E_{vv}^{\text{HD}} & = a_{gb}\left( 2RR_{vv}-4R^{\alpha\beta}R_{v\alpha v \beta}-4R^{\alpha}_{v}R_{v\alpha}+2{R_{v}}^{\alpha\beta\sigma}R_{v\alpha\beta\sigma} \right)
\end{split}
\end{equation}

By explicitly computing  $E_{vv}^{\text{HD}}$ in terms of the metric components \eqref{Wallmet} and their derivatives, it is possible to rewrite 
$E_{vv}$ for the Einstein-Gauss-Bonnet theory into the form \eqref{eq:entrecast}. Subsequently, we can read off the entropy current from it and we have 
\begin{equation} 
\begin{split}
J^v &= \left(1+2a_{gb}(\mathcal{R} - 2\bar{K}_{AB}K^{AB} + 2K\bar{K})\right) \\
J^i &=  - 4a_{gb} \nabla^{j}\left(Kh_{ij}-K_{ij}\right)
\end{split}
\end{equation} 

%
%

Note that, this entropy density and spatial entropy current for the Einstein-Gauss-Bonnet theory 
has been constructed following the philosophy of \S\ref{ssec:4dercor}. 
In the next subsection we shall do a systematic study of this entropy current, concentrating particularly 
in $d=(3+1)$ space-time dimensions, where the Gauss-Bonnet term becomes topological.
%
%
%
\subsection{The Einstein-Gauss-Bonnet theory in $d=3+1$ }\label{ssec:gb3p1}
%
%
The Gauss-Bonnet theory in $(3+1)$ space-time dimensions needs a separate discussion. In this case, the Gauss-Bonnet term becomes a total derivative term and therefore it does not contribute to the equations of motion, i.e. $E_{vv}^{\text{HD}} = 0$ identically. However, if one uses the Wald entropy as the equilibrium definition of black hole entropy \eqref{entsch2}, there is a finite non-vanishing contribution to it even from the topological Gauss-Bonnet part of the Lagrangian. The Wald entropy density $s_{w}^{\text{HD}}$ (see \eqref{eq:waldentdensityHD}) for this case, is given by the Ricci scalar of the co-dimension-$2$ spatial slice of the horizon $\mathcal{H}_v$,
\begin{equation}
s_{w}^{\text{HD}} = 2 \, a_{gb} \, \mathcal{R} \, ,
\end{equation}
where $a_{gb}$ is the Gauss-Bonnet parameter appearing in \eqref{eq:gauss1}. 
Since $\mathcal{H}_v$, in this case, is a $2$-dimensional manifold, the integrated total entropy $S_{W}$ becomes the topological Euler number of $\mathcal{H}_v$. 

Once we consider dynamical black hole solutions in this theory and restrict ourselves to consider perturbations characterized by small amplitudes around a stationary configuration, 
the total integrated Wald entropy $S_W$ doesn't change with time as long as the perturbation is small and therefore, does not affect the topology of $\mathcal{H}_v$. 
However, if we consider the local Wald entropy density $s_{w}$ (without being integrated on the spatial slice $\mathcal{H}_v$), it does indeed 
change with time 
and therefore has a non-zero contribution to  $\partial_v (\sqrt{h}\, \widetilde B)$. 
With these in mind let us look at \eqref{eq:hvvzero}, 
which is the main result of this note and rewrite it here again for convenience
\begin{equation}\label{eq:hvvzero_again}
E_{vv}^{\text{HD}} \big\vert_\text{zero boost} \sim -\, \partial_v \left({1 \over\sqrt{h}}\partial_v \left(\sqrt{h}~ \widetilde B\right)\right)-\partial_v \left(\nabla_i J^i\right).
\end{equation}
From the above discussion it is clear that for $(3+1)$ space-time dimensions, the LHS of \eqref{eq:hvvzero_again} vanishes identically. However, the first term on the RHS is non-zero, making us wonder how to make sense of this equation if we had not included the second term on RHS involving the spatial entropy current. As we will see, both the terms on the RHS of the above equation are non-zero but they will precisely cancel each other, and that is how this equation will be satisfied. In other words, we are left with verifying that the RHS. of \eqref{eq:hvvzero_again} vanishes identically without using any on-shell gravity equations of motion, up to $\mathcal{O}(\epsilon^2)$ corrections.

We start by noting that the values of the coefficients $a_i$ presented in Table-(\ref{table:const2}) are achieved by explicitly computing the $E_{vv}^{\text{HD}}$ for different four derivative gravity theories and for our metric choice \eqref{Wallmet}, but most importantly, the results are not limited to the space-time dimensions we are working in. Therefore the same results (presented in Table-(\ref{table:const2})) holds for $(3+1)$-dimensional space-time as well. The specific values of these coefficients for Gauss-Bonnet theory turns out to be the following:
\begin{equation} \label{eq:aiforGB}
\begin{split}
&a_{16} = \, 2, \,  a_{17} = \,  -1, ~ a_i = 0 ~ (\text{for all} ~ i=1, \cdots , 15) \,, \\ \text{such that,} & \quad
E_{vv}^{\text{HD}}=~a_{16}\, T_{16} + a_{17}\, T_{17} = 2 \left( \mathcal{R}^{ij} - {1 \over 2} h^{ij} \mathcal{R} \right) \partial_v^2 h_{ij} \, . 
\end{split}
\end{equation}
However, for $2$-dimensional space-time one can show that the following relation is identically true, 
\begin{equation} \label{identity1}
 \mathcal{R}^{ij} - {1 \over 2} h^{ij} \mathcal{R} = 0 \, .
\end{equation}
This is true because, in $2$-dimensional space we could always choose a coordinate system where the metric is conformally flat and Einstein tensor vanishes on any $2$-dimensional conformally flat space-time.

Let us now consider the expression $(1/\sqrt{h})\, \partial_v(\sqrt{h}\, {\cal R})$. It is well-known that the linear variation of  Ricci scalar around any metric generates a term proportional to the Einstein tensor plus a total derivative term. Because of the fact mentioned above, without doing any further calculation, we could say 
\begin{equation} \label{eq:gb4did}
{1\over\sqrt{h}}\partial_v\left(\sqrt{h}~ {\cal R}\right) =  \left(\mathcal{R}^{ij} - {1 \over 2} h^{ij} \mathcal{R}\right)\left(\partial_v h_{ij} \right)+ \nabla_i Z^i\, , 
\end{equation}
where  $Z^i$ is some spatial current characterizing the total derivative term, which could be easily fixed as follows.
Using table Table-(\ref{table:const3}) we could find the list of values for $A_i, \, B_i$ for Gauss-Bonnet theory 
\begin{equation} \label{eq:AiBiGB}
A_5 = -2, \, B_2 =2, \, B_3 = -2, 
\end{equation} 
leading to the following expression for $Z^i$
\begin{equation}
Z^i =  2\nabla_j \left( K^{ij} -h^{ij \,}K\right)
\end{equation}
Again in $(3+1)$-dimensions, where $\{i,j\}$ indices run over $\{1,2\}$, 
the first term in the RHS of \eqref{eq:gb4did} identically vanishes. Therefore, we can immediately rewrite \eqref{eq:gb4did} as 
$${1\over\sqrt{h}}\partial_v\left(\sqrt{h}\, {\cal R}\right) - \nabla_i Z^i = 0, \, ~~(\text{only in} ~(3+1)\text{-dimensions}) $$ 
This looks exactly like a divergence of a four-current and identically vanishes in $(3+1)$-dimensions. In this particular case of $(3+1)$-dimensional space-time, the above expression  has exactly the same status as that of \eqref{eq:ppes}, or the structure multiplying the coefficient $A_2$ in the expression of $E_{vv}^{\text{HD}}$ \eqref{eq:finfirst}, see also \eqref{eq:hvvzero3} and \eqref{eq:tildeBrel}. In other words, in $(3+1)$-dimension we are free to add $\{{\cal R}, Z^i\}$ to the expression of entropy density and spatial entropy current respectively, with any arbitrary overall coefficient. Such an addition will not affect the ultra-local version of the second law or the physical process version of the first law and this is true for all theories as long as we are restricting ourselves to $(3+1)$ dimensions.
However, just like in case of $A_2$, the Wald entropy formalism fixes that arbitrary coefficient to a very specific value. 

To summarize, the main physical interpretation that one can draw from the arguments presented above is the following. For Gauss-Bonnet theory in $(3+1)$ dimensions $E_{vv}^{\text{HD}}$ vanishes identically and that is related to the fact that the total integrated Wald entropy $S_W$ is not changing due to time-dependent perturbations. 
This is because the Wald entropy $S_W$ in this case is given by topological Euler number of the 2-dimensional $\mathcal{H}_v$ 
and we are considering small amplitude approximation for the perturbations, 
which are too weak to change the topology of $\mathcal{H}_v$. 
However, even in that approximation, the local change of entropy density is not vanishing. 
This necessitates the introduction of the idea of a spatial entropy current, that quantifies the inflow or outflow of local entropy density 
and cancels the change in local entropy density, within any infinitesimal region in $\mathcal H_v$. This analysis, at least for the situation considered in this subsection, 
therefore plays an important role in motivating the need for a spatial entropy current.

\subsection{Comments on entropy current for higher boost terms in $E_{vv}^\text{HD}$} \label{ssec:entcurgenpro1}
Once we have analyzed the zero-boost terms, the next immediate question is 
to analyze the contribution of higher boost terms of $E_{vv}^{\text{HD}}$, to the entropy current \eqref{entcurkzero}.
As we have observed in \S\ref{sec:review}, the arguments in \cite{Wall:2015raa} for second law, works smoothly, for all the higher-boost terms in $E_{vv}^{\text{HD}}$. 
The contribution from these higher boost terms, to the total entropy falls within the class of JKM ambiguities, and they do not contribute to the physical process version of the first law.  
Unlike the zero-boost terms, nothing necessitates the existence of a spatial component of the current for these higher boost terms. Both the 
first and second law would remain valid, if we simply declare that these terms would just modify the entropy density as in \eqref{eq:scor}, and they do not affect the spatial 
components of the current. However, the spatial components of the current could still exist, even for the higher boost terms as we now demonstrate.

Before we proceed it is worth clarifying that we will not be doing an exhaustive classification of all such possible higher boost terms in $E_{vv}^{\text{HD}}$. Our aim here is just to present an argument based on analyzing a candidate term as an example that justifies the above-mentioned statement. We postpone a more detailed study of this aspect to future work.

Schematically, the higher boost terms have the following structure (see \eqref{eq:schemeHvv} or \eqref{eq:schemeHvvrep})
\begin{equation}\label{hibtyp}
E_{vv}^{\text{HD}} \big\vert_\text{higher boost} \sim \partial^2_v\left[\partial_r^k A^{(k)}~\partial^k_vB^{(k)}\right] + \mathcal{O} \left(\epsilon^2\right)\, ,
\end{equation}
where $A^{(k)}$ and $B^{(k)}$ are boost-invariant.
Now it turns out that the same higher boost term could be recast in different ways, up to corrections that are quadratic or higher-order, in the amplitude of the dynamics. 
This allows us to absorb certain higher boost terms (the ones that have at least one $\nabla_i$) either entirely  within the correction to entropy density, or partially in entropy density and partially in the spatial components of the current. Let us explain this ambiguity more specifically.

Consider a typical higher boost term in $E_{vv}^{\text{HD}}$, as in \eqref{hibtyp}, where the term $\partial_v^k B^{(k)}$ 
could be expressed as divergence of a spatial current with boost-weight $k>1$, 
$$\partial_v^k B^{(k)}\sim \vec\nabla\cdot \vec J^{(k)}, $$
and substituting it in the expression of $E_{vv}^{\text{HD}}$ we find
\begin{align}
 \label{eq:cuurhigh1}
&E_{vv}^{\text{HD}} \big\vert_\text{higher boost} \sim ~ \partial^2_v\left[\partial_r^k A^{(k)}~\vec\nabla\cdot \vec J^{(k)}\right] \\
\label{eq:cuurhigh2}
  &~~=\, \partial_v\bigg(\vec\nabla\cdot\left[\partial_v\left(\vec J^{(k)}~\partial_r^k A^{(k)}\right)\right]\bigg)+\partial_v^2\left[-\vec J^{(k)}\cdot\vec\nabla\left(\partial_r^k A^{(k)}\right)\right]+O\left(\epsilon^2\right) \, .
 \end{align} 
On one hand, from the first line of \eqref{eq:cuurhigh1}, we can conclude that the contribution of this term to entropy current is simply \footnote{The quantity $S^A$ has been introduced in \eqref{entcurkzero}. The subscript $k \geq 1$ in $S^A$ is to denote that this is the contribution from the higher boost terms in $E_{vv}^{\text{HD}}$ .}
\begin{equation} 
{\text{From \eqref{eq:cuurhigh1}: }} ~~S^{v}_{k \geq1} =-\, \partial_r^k A^{(k)}~~\vec\nabla\cdot \vec J^{(k)}, \quad S^i_{k \geq 1} = 0 \, . \label{hibencur1}
\end{equation}
with no spatial current.
On the other hand, from the second line \eqref{eq:cuurhigh2}, we may infer that, this term contributes to the spatial components of the entropy current, apart from the contribution to the entropy density, which is different from the previous case \eqref{hibencur1}. That is, we can write the contribution to entropy current also in the following way
\begin{equation}
{\text{From \eqref{eq:cuurhigh2}: }} ~~S^{v}_{k \geq1} =\, \vec J^{(k)}\cdot\vec\nabla\left(\partial_r^k A^{(k)}\right), \quad \vec S_{k \geq1}= -\,\partial_v\left(\vec J^{(k)}~\partial_r^k A^{(k)}\right) \, .
\label{hibencur2}
\end{equation}   
We would like to emphasize that the above manipulation, which is essentially an interchange of $\partial_v$ and $\vec\nabla$, crucially uses the fact that any term, generated due to the non zero commutator of these two types of derivatives, would be of higher-order in amplitude. This is because the commutator itself is of boost weight one, for higher 
boost terms,
\begin{equation}
\partial_v (\vec{\nabla}.\vec{J}^{k}) \sim \vec{\nabla}.(\partial_v \vec{J}^{k}) + (\partial_v \, \Gamma^i_{ij})\, {J^{(k)}}^j \sim \vec{\nabla}.(\partial_v \vec{J}^{k}) +\mathcal{O}(\epsilon^2) \, .
\end{equation}
Also note that, this is true only for the higher-boost terms, and in particular, not true 
for the boost invariant terms, for which the presence non-zero spatial entropy current was unambiguous. 
  
The two different choices of entropy density in \eqref{hibencur1} and \eqref{hibencur2}, are related by a total spatial derivative, as expected. This ensures that, in the integrated 
(weak) version of the second law, this difference would have no impact. However, in the ultra-local version, where we demand the entropy to be produced at every point in space and time, this difference is significant. This leads to an ambiguity in the definition of our entropy current, which cannot be fixed, merely from the transformation
property of $E_{vv}^{\text{HD}}$ under boost \eqref{rescale}. 

It is possible that, if we keep track of the higher order terms in amplitude expansion, this ambiguity may be removed. 
Alternatively, it is also possible that some suitable extension of the boost symmetry, like \eqref{genrepara}, which preserves our global choice of coordinates, 
might constraint the structure of our entropy current further, and consequently fix this ambiguity. We would like to explore this point further in our future work.

\section{Discussions and Future directions} \label{sec:disco}
In this note, we have demonstrated that the intricacies in the arguments involved in the proof of the physical process version of the first law, and the second law, naturally lead us to the notion of a spatial entropy current on the horizon. This spatial entropy current captures the inflow or outflow of entropy from any sub-region of $\mathcal H_v$ - the horizon $v$-slice. 
For most of our analysis in this note, we consider dynamical black holes which can be treated within the linearized approximation, where the amplitude of the `time' dependent metric fluctuations, about a given stationary black hole solution, is small. Under this approximation, we are able to establish that the entropy density and the spatial components of the entropy current, constructed through our algorithm, satisfy an ultra-local stronger version of the second law of black hole thermodynamics. 
The validity of this local form of the second law is ensured by the equations of motion for the higher derivative theories of gravity,
 and therefore, true for any metric that solves these classical equations, at the linearized level. 

The construction of our entropy current is not unique. 
All the ambiguities in defining the current can be traced back to the fact 
that there exist certain terms $\mathcal T_{\text{amb}}$ which can be simultaneous written in two ways. We can write 
$\mathcal T_{\text{amb}} = \frac{1}{\sqrt{h}} \partial_v \left( \sqrt{h} \mathcal J^v \right)$, but also we can write the same term 
as $\mathcal T_{\text{amb}} = -\nabla_i \mathcal J^i$, for some choice of $\mathcal J^v$ and some choice of $\mathcal J^i$. 
Obviously, it follows that $\mathcal J^v$ must have at least one spatial derivative, while $J^i$ must have at least one $v$-derivative. 
If we have such terms, appearing in the equation of motion as $\partial_v \mathcal T_{\text{amb}}$, then it becomes unclear 
whether to write it as $\mathcal T_{\text{amb}} = \frac{1}{\sqrt{h}} \partial_v \left( \sqrt{h} \mathcal J^v \right)$ and consider it to be 
a part of the entropy density. Or to write it as $\mathcal T_{\text{amb}} = -\nabla_i \mathcal J^i$ and interpret it to be 
being a part of the spatial components of the entropy current. A third possibility is to split this term up, into the entropy density 
and the entropy current. In \S\ref{ssec:entcurgenpro1}, we have discussed these kind of ambiguities in detail. 

Again, if we indeed have terms like $\mathcal T_{\text{amb}}$ which can be written in both these ways, 
we can always add a $0=\partial_v \left( \frac{1}{\sqrt{h}} \partial_v \left( \sqrt{h} \mathcal J^v \right) + \nabla_i \mathcal J^i \right) $, to the equation 
of motion, and subsequently include $\mathcal J^v$ and $\mathcal J^i$ into the definition of the entropy density and the spatial entropy current. 
Neither the equation of motion nor any of the laws of thermodynamics would be affected by this operation. This kind of ambiguity arises, for example, 
in the Einstein-Gauss-Bonnet theory in $3+1$ dimensions, discussed in detail in \S \ref{ssec:gb3p1} 
\footnote{Also see the ambiguity related to the parameter $A_2$, discussed in detail in \S\ref{ssec:entcurgenpro}}. 
If a term like $\mathcal T_{\text{amb}}$ is such that, $\mathcal J^v$ is non-zero on stationary solutions, then it would contribute to Wald entropy as well. In such a case, the ambiguity corresponding to this term may be removed by demanding that our entropy reduces to Wald entropy on stationary solutions. But if $\mathcal J^v$ vanishes in equilibrium then this additional criterion would remain ineffective in fixing it. 

It should also be noted that it may be possible to write down a particular term simultaneously in both the forms, only at the linearized order in perturbations. Such an equivalence may cease to be true once we proceed to consider corrections which are higher-order 
in amplitudes. In that case, these ambiguities would only be a linear order artifact and would disappear once we are able to construct the full non-linear 
current. However, some of these ambiguities of the entropy current may remain, even in the full non-linear construction. 

Having highlighted the ambiguities of the entropy density and the corresponding current, we must point out that, 
every member of this ambiguous class, have the property that the total entropy reduces to Wald entropy for stationary black holes. For the non-stationary dynamical black holes, all these entropy density and currents also satisfy a local second law. Hence, every such entropy density and entropy current are perfectly well defined macroscopic entities that can provide excellent effective thermodynamics description of the system. Some additional microscopic information is likely to make one of them special, 
and it can stand out as the correct definition of entropy density and entropy current away from equilibrium. 
Therefore, despite these ambiguities, it appears to us, that the notion of the spatial components of the entropy current 
and a local second law on a dynamical horizon is a concept of significant importance in the thermodynamic description of black holes.

This note is essentially a series of observations, on the evolution of black hole entropy in dynamical scenarios, in a specific set of examples of higher derivative theories of gravity. 
Through explicit calculations, we have been able to test our hypothesis about the spatial components of the entropy current, only in four derivative theories of gravity. This is a small step towards formulating an ultra-local version of the second law in gravitational theories 
(if it exists in a full non-perturbative sense) and deciphering all its physical ramifications. Clearly, there are several directions in which this work needs to be extended, so that a more complete picture of this whole mechanism may emerge. 
Here is a brief list of related questions, which we would like to investigate in the near future. 
\begin{enumerate}
\item The reader may have noticed that, throughout this note, we have used the word `time' under a quotation mark. 
This is because our `time' here is not really a parameter along a time-like vector field; 
rather, it is the affine parameter along a `distinguished' null direction that generates the event horizon.  
Therefore, the expression appearing in the local version of the second law is not the $d+1$ dimensional 
 \footnote{Remember if we are working in $D+1$ dimensional space-time, then the current 
is expected to have $d+1$ components, with $d=D-1$. This is because this current would be defined on the event horizon, which is a co-dimension one surface.}
covariant divergence of a covariant current. 
This is quite unlike the standard way in which the local version of the second law is expressed, for near-equilibrium dynamics of non-gravitational theories, 
where the $d+1$ dimensional Lorentz covariance is manifestly maintained.

On the event horizon, we do not have a time-like direction, so the question of Lorenz invariance does not arise here. However, our construction has used a specific choice of coordinates and physically we expect some form of invariance should exist once we choose a different coordinate system 
- for example, a different spatial slicing of the null generators. 

It would be extremely important to explore whether any such invariance exists and if it exists, then how does it control our construction. 
\item Another question related to the above is as follows. 
We have seen that our construction of the spatial entropy current mainly involves the `zero boost terms' in the equation of motion. For this construction to work, these `zero boost terms' were required to have a specific form. This requirement may be viewed as a set of constraints on the most general structure of
the relevant component of the equation of motion (see \S\ref{ssec:evvgenstr1} and \S \ref{ssec:entcurgenpro}). The physical origin of these constraints is, at the moment, unclear. 
We suspect that the reason behind these constraints could be the set of residual gauge invariance, expressed in \eqref{genrepara}, which is a generalization of the boost symmetry \eqref{rescale}. Whether this suspicion is true, or there is a completely different reason for these constraints must be investigated through explicit computations.

\item The `zero boost terms' in the equation of motion, which are central to our analysis in this note, are also 
relevant for the physical process version of the first law, and hence, control the definition of entropy in stationary situations.

Now, it is well known that Wald's formalism \cite{PhysRevD.48.R3427,Iyer:1994ys} also determines this same equilibrium entropy in a covariant fashion using the conserved Noether current corresponding to the diffeomorphism symmetry. 
It would be extremely interesting to clearly establish a connection between these two methods. 
In particular, if it is possible to identify our spatial current within Wald's construction, it would probably lead to a more satisfying covariant construction of the entropy current. This may help us arrive at an
abstract proof for the existence of this entropy current and the local second law, for any higher derivative theory of gravity.

In absence of any such concrete proof, it
would be quite useful to gather more data, simply by repeating the exercise presented in this note, for theories of gravity with 6 or more derivatives.
\item Another obvious generalization would be to extend our construction to non-linear order in amplitude $\epsilon$. This can potentially fix the ambiguity related to the construction of the spatial components of the entropy current, which arises for higher boost ($k \geq 0$) terms (see the discussion in \S\ref{ssec:entcurgenpro1}). 

But more importantly, it might provide us with further insights, which can help us formulate a non-perturbative proof of second law for higher derivative theories.
 For Einstein's theory, entropy production is ensured by the famous `horizon area increase theorem', which is proved for any dynamical situation, in full non-perturbative way. It would be nice to have a similar proof (or a clean counter-example) for higher derivative theories of gravity.
\item Naively, it might seem that, at non-linear order, we do not have to worry about the second law, since for Einstein's theory itself,
 the entropy production takes place at quadratic order in amplitude. Now because higher derivative corrections are always suppressed compared to the leading order piece corresponding to Einstein's theory, they cannot reverse the sign which guarantees entropy production.
  
However, if we are interested in an ultra-local form, then during a non-trivial `time' evolution, the contribution to entropy due to Einstein's theory could vanish locally at a given point. Then, for the question of entropy production and the second law, we must take the higher-derivative corrections seriously. 
See \cite{Bhattacharyya:2016xfs} for the construction of entropy in dynamical black holes for the Einstein-Gauss-Bonnet theory, 
where these subtle issues have been addressed. 
The construction of entropy in \cite{Bhattacharyya:2016xfs} did not yield a second law, for the most generic dynamical situation. 
But, in \cite{Bhattacharyya:2016xfs} this idea of a spatial entropy current was not used. 
It would be very interesting to revisit \cite{Bhattacharyya:2016xfs}, and check if the obstruction is resolved when the spatial entropy current is incorporated into the statement of the second law. 
\item Within the framework of gauge gravity duality, a precise correspondence exists between slowly varying fluctuations of a black hole and the hydrodynamic fluctuations of the boundary fluid. Since the boundary fluid dynamics comes equipped with a local entropy current, there exists a
dual of this current, for the black hole in the bulk \cite{Bhattacharyya:2008xc, Chapman:2012my, Eling:2012xa}. This dual also constitutes a gravitational entropy current, in this particular context. 
In \cite{Bhattacharyya:2008xc}, the construction has been done for two derivative Einstein's theories. While in \cite{Chapman:2012my, Eling:2012xa}, it has been extended to higher derivative theories of gravity, following Wald's formalism of Noether current. 
All these constructions use the derivative expansion extensively and their validity is restricted to this particular case of fluid-gravity correspondence.  
Therefore, although these constructions of the entropy current relate to horizon dynamics, it subtly uses the asymptotic AdS conditions, which ensures black-brane solutions exist and the fluid-gravity correspondence could be formulated in a clean fashion. For example, the entropy current constructed in these papers is clearly a (3+1) dimensional current (for 5 bulk dimension) with one component (the entropy density) clearly along with a time-like direction. This is achieved by lifting the null coordinate along the horizon, to the time-like direction of the boundary through the fluid-gravity map. This time-like direction also serves to formulate an unambiguous statement of the second law, in terms of the divergence of this entropy current. 

On the other hand, our construction is completely confined to the horizon, it does not have any time-like direction, to begin with. As we have explained before, in absence of any Lorentz symmetry it is not straight-forward to interpret our result as a covariant `four'-current. Also, we do not need any assumption about the asymptotic structure of spatial infinity.

Our construction looks quite different from what has been done in \cite{Bhattacharyya:2008xc, Chapman:2012my, Eling:2012xa}. But it is also clear that there must be some relation between these two constructions. This question is a topic of our ongoing investigation. 

\item Very recently, one candidate entropy current has been constructed in \cite{Dandekar:2019hyc}, for Gauss-Bonnet theory within the framework of membrane-gravity duality, 
in an expansion in inverse powers of space-time dimension $D$. This is a duality that gives a precise correspondence between the dynamics of a membrane (a time-like hyper-surface, embedded in flat space-time) and that of the horizon, in the large $D$ expansion. Unlike our construction, which does not rely on any duality, in \cite{Dandekar:2019hyc} the entropy current has been constructed in the dual picture of the membrane. 
Their entropy current is entirely confined within the membrane and has the usual property of Lorentz invariance. In their case, the 
non-negative divergence of the entropy current follows from  
the membrane dynamics governed by those membrane equations, which have been derived from the dual gravity picture.

Moreover, within their approximation, the authors of \cite{Dandekar:2019hyc} have also shown that the existence of a Killing vector is a consequence of no entropy production.
They have also demonstrated that the charge corresponding to this conserved entropy current reduces to the well-known expression of Wald entropy, in a stationary situation.

It would be extremely interesting to see how the entropy current of \cite{Dandekar:2019hyc} compares with ours. 
In particular, we would like to explore, if this membrane-gravity duality can be used to formulate a principle which can fix the ambiguities of the entropy density and spatial entropy current.

\end{enumerate}

\acknowledgments 
We would like to thank Parthajit Biswas, Milan Patra, and Shuvayu Roy for initial collaboration, several 
useful discussions and many important inputs. 
We would also like to thank Arjun Bagchi, Shamik Banerjee, Pallab Basu, Diptarka Das, Sayan Kar, Loganayagam R., Mangesh Mandlik, 
Shiraz Minwalla, Sudipta Mukherji, Arunabha Saha, Sudipta Sarkar, Tarun Sharma,
Nilakash Sorokhaibam and Yogesh Kumar Srivastava for several useful discussions. 
NK and JB would like to acknowledge hospitality during an academic visit at NISER Bhubaneswar, where much of this work had been carried out. 
%
\appendix
%
\section{A general stationary metric can have $v$ dependent components}\label{app:vdep}

In a black hole usually, the Killing generators of the horizon cannot be affinely parametrized maintaining the Killing conditions. In other words, the components of the stationary metric are independent of the Killing coordinate - $\tau$, but they are not independent of the affine parameter $v$ along the generators of the Killing horizons. Though  in a stationary metric with a Killing horizon, the Killing vector field - $\partial_\tau$  and the affinely parametrized null generators $\partial_v$ are proportional to each other and there exists a precise relation between them. 
In this appendix, we shall use this relation to fix the $v$ dependence of the stationary metric. 

 More precisely, we would like to determine how the components of a generic stationary metric, written in the gauge of \eqref{Wallmet}, could depend on the $v$-coordinate.

Consider a generic stationary black hole with a Killing horizon, i.e, there exists a coordinate $\tau$ such that
\begin{enumerate}
\item All metric components are independent of $\tau$
\item $\partial_\tau$ is time-like everywhere outside the horizon.
\item $\partial_\tau$ becomes null on the event horizon.
\end{enumerate}
Now we could do exactly the same construction as in case of the metric \eqref{Wallmet}, the only difference being that now the coordinates on the horizon would be $\partial_\tau$ and $\partial_i$, instead of the affinely parametrized $\partial_v$. Let $\rho$ be the coordinate that denotes distances away from the horizon. Now also we could choose $\rho$ to be the affine parameter along the set of null geodesics, intersecting the horizon at fixed angles with $\partial_\tau$ and $\partial_i$ and labelled by the coordinates of the  intersection point.  Following the same logic as before, the metric in $\tau$, $x^i$ and $\rho$ coordinate will have almost the same structure as that of \eqref{Wallmet}. The $\tau\tau$ and $\tau i$ components of the metric will again vanish on the horizon ($\rho =0$) owing to the fact that it is a null hyper-surface. But since $\partial_\tau$ is not affinely parametrized, unlike \eqref{Wallmet}, the first $\rho$ derivative of the $(\tau\tau)$ component of the metric (let us denote it by $g_{\tau\tau} (\rho,x^i)$) will not vanish on the horizon. However, for stationary black holes, $\partial_\rho g_{\tau\tau}$ is related to the temperature of the black hole and the zeroth law of Black hole mechanics 
%
%
ensures $\bigg[\partial_\rho g_{\tau\tau}\vert_{\rho=0}\equiv C\bigg]$ is a constant, i.e., independent of the spatial coordinates $x^i$s. Putting all these facts together we finally write the most general stationary metric in our gauge.
\begin{equation}\label{stamet}
\begin{split}
ds^2 = 2~ d\tau~ d\rho - \left(\rho ~C +\rho^2 X(\rho)\right)~ d\tau^2 + 2 \rho ~\omega_i(\rho)~ d\tau~ dx^i + h_{ij}(\rho) ~dx^i dx^j
\end{split}
\end{equation}
Now we have to transform this metric to the gauge of \eqref{Wallmet} where the null coordinate along the horizon is an affine parameter of the null generators $v$. Our final goal is to find out how the metric components of an arbitrary stationary metric will depend on $v$.

The coordinate transformation which fulfills this objective is given by 
\begin{equation}\label{eq:coord}
\begin{split}
\rho = {C\over 2}r~v ,~~\tau = {2\over C}\log\left(C~v\over 2\right)
\end{split}
\end{equation}
The metric in the new coordinate takes the following form 
\begin{equation}\label{stametwg}
\begin{split}
ds^2 = 2 ~dv~dr - r^2 X\left(C r v/2\right) dv^2 + r\omega_i\left(C r v/2\right) ~dv~dx^i + h_{ij}\left(C r v/2\right)~dx^i~dx^j
\end{split}
\end{equation}
To get the above metric, we have crucially used the fact that $C$ is independent of $v$ and $x^i$ ($C$ is independent of $\rho$ by construction).

The most important noteworthy feature of this metric \eqref{stametwg} is that the metric components 
are explicitly dependent on the $v$-coordinate, although it describes a stationary black hole because 
it is a mere coordinate transformation of the most general stationary metric \eqref{stamet}. However, though imposing the condition of stationarity on the general form of the metric \eqref{Wallmet}, \emph{does not} imply that the metric functions $X$, $\omega_i$ and $h_{ij}$ should be independent of the $v$ coordinate, there are some constraints on the $v$ dependence of the stationary metrics. Here the metric components never depend on $r$ and $v$ independently, but always on the product $r v$. In other words, on any metric of the form \eqref{stametwg}, with components depending only on the product $r v$, we could always apply the inverse of the coordinate transformation \eqref{eq:coord} to take it to a form where redefined coordinate - $\tau$ is manifestly the Killing coordinate.

Now for the proof of second law, it is crucial that the $\partial_v$ of entropy vanishes on stationary black holes attained at $v\rightarrow\infty$. Naively the form of the stationary metric \eqref{stametwg}  contradicts this step of the argument.
But note that any $\partial_v$ derivative on the metric components in \eqref{stametwg}, will also bring down a factor of $r$ and therefore will vanish on $\mathcal H$ (the hyper-surface at $r=0$),
unless there is also one $\partial_r$ derivative present along with every $\partial_v$ derivative. 
Thus we may conclude that, the terms of the form $\left((\partial_r \partial_v)^m P\right)$, where $P$ is a function of the metric components in \eqref{Wallmet} and their $\nabla_i$ derivatives (without any $\partial_r$ or $\partial_v$ derivatives), can be non-zero on a generic Killing horizon. Note that, all such terms are invariant under the $\lambda$ scaling \eqref{rescale}. It also implies that the terms of the 
form $\left(  \partial^n_r \partial^m_v P \right)$, with $m>n$ must vanish on a Killing horizon. This is because, as is apparent from \eqref{stametwg}, the higher number of $v$-derivatives would give rise to factors of $r$, which will force the entire term to zero on the $r=0$ hyper-surface.

\section{Arguments leading to 
vanishing of $T_{vv}$ on any Killing horizon}\label{app:boostinv}
In this section we would like to argue that the $vv$ component of the matter stress tensor  vanishes on Killing horizons.

We shall use the boost transformation property of $T_{vv}$ to reach this conclusion. Note that just like the gravity part of the equation of motion (i.e., $E_{vv}$ or $E_{vv}^\text{HD}$) , matter stress tensor itself is a covariant with nice transformation properties under any coordinate transformation, in particular  the $\lambda$ scaling described in equation \eqref{rescale}. $T_{vv}$ should transform exactly the way $E_{vv}$ or $E_{vv}^\text{HD}$ transforms, namely
\begin{equation}\label{eq:Tvvtransform}
T_{vv}\rightarrow T_{\tilde v\tilde v} = {1\over \lambda^2} T_{vv}
\end{equation}

Now we shall consider only those stress tensors that are regular on the event horizons at least in those coordinate systems where the full dynamical metric is regular, everywhere apart from the black hole singularity. This is certainly the case in the coordinate system we have chosen in our metric \eqref{Wallmet}. It follows that $T_{vv}$ must admit a Taylor series expansion around the horizon at $r=0$.  Equation \eqref{rescale} and equation \eqref{eq:Tvvtransform}  together suggest the following expansion for $T_{vv}$
\begin{equation}\label{eq:Tvvexpand}
T_{vv}={1\over v^2}\sum_{k=0}^\infty (r v)^k~ w^{(k)}(\vec x)
\end{equation}
where $w^{(k)}$ s are scalar functions of only the spatial coordinates $\{x^i\}$, (i,e., they are both boost invariant and also invariant under any coordinate transformation that mixes only the $\{x^i\}$ coordinates among themselves). Exactly on the horizon only the leading terms of the above expansion will contribute.
\begin{equation}\label{eq:Tvvhorizon}
T_{vv}\big\vert_\text{horizon}={w^{(0)}(\vec x)\over v^2} 
\end{equation}
Note that both  \eqref{eq:Tvvexpand} and \eqref{eq:Tvvhorizon} do not need any stationarity for their validity.

Now let us specialize to stationary cases. Here we have a Killing vector ($\partial_\tau$). All relevant fields including the matter fields are independent of this $\tau$ coordinate and the same is true for $T_{vv}$, as well. In terms of equation it implies
\begin{equation}\label{eq:Tvvstation}
\begin{split}
\partial_\tau T_{vv} \big\vert_\text{stationary}=0\\
\end{split}
\end{equation}
From equation \eqref{eq:coord} it follows
\begin{equation}\label{eq:ff}
\partial_\tau = {C\over 2} \left(v~\partial_v -r~\partial_r\right)
\end{equation}
Equation \eqref{eq:ff} clearly contradicts equation \eqref{eq:Tvvstation} unless $w^{(k)} =0$ for every $k$. It follows that $T_{vv}$ vanishes identically on any configuration with a Killing vector.

We would like to emphasize that in the above arguments the key elements are 
\begin{enumerate}
\item The existence of the event horizon (or more precisely  a null hyper-surface at $r=0$) so that the horizon-adapted coordinate choice in the metric \eqref{Wallmet} and  consequently the boost symmetry is meaningful.
\item Stationarity or the existence of a Killing vector, which is proportional to the null generators of the horizon.
\end{enumerate}
 We have not used the fact that $T_{vv}$ is stress tensor, neither the fact that the field configuration (including the metric) satisfies any particular equation. What we have argued is that whenever there is one Killing vector field, the $vv$ component of any covariant tensor identically vanishes in the vicinity of the horizon (where the Taylor expansion in equation \eqref{eq:Tvvexpand} makes sense) and it is a completely off-shell statement.

\section{Conventions, notations and useful formulae}\label{app:notcon}
In this appendix we summarize our conventions, write down various notations and collectively represent several important formulae that we have used in the  note.
\begin{itemize}
\item The coordinate choice:
\begin{align*} 
x^\mu =& \, \text{The full space-time coordinates in} ~(d+1)\text{-dimensions}: \, \{v,\, r, \, x^i\}, \\
v=& \, \text{The Eddington–Finkelstein type time coordinate}, \\
r=& \, \text{The radial coordinate}, \\
x^i=& \, \text{The}~(d-1)~\text{spatial coordinates},
\end{align*}
\item The choice for the space-time metric:
\begin{equation}
\begin{split}
ds^2 =& \, 2\, dv \, dr - r^{2} \, X(r,v,x^{i})\, dv^{2}+2 \, r \, \omega_{i}(r, v,x^{i})\, dv\, dx^{i} \\
& +\, h_{ij}(r,v,x^{i})\, dx^{i}dx^{j}
\end{split}
\end{equation}
\item Useful notations and conventions:
\begin{enumerate}
\item $\mathcal{H} =$ The co-dimension one horizon, which we choose to be at the radial coordinate $r=0$,
\item $\mathcal{H}_v =$ The co-dimension two, constant $v$-slice of the  horizon,
\item $h =  $ Determinant of the induced metric, $h_{ij}$, on $\mathcal{H}$,
\item The total integrated Wald entropy at equilibrium is defined as $$S_{W}= -\,2\, \pi \int_{\mathcal{H}_v}d^{d-2}x \,\sqrt{h}\, \frac{\partial \mathcal{L}}{\partial R_{\mu \nu \rho \sigma}} \, \epsilon_{\mu \nu}\epsilon_{\rho \sigma} =-\,2\, \pi \int_{\mathcal{H}_v}d^{d-2}x \,\sqrt{h}\, s_w \, ,$$
where $\epsilon_{\mu \nu}= $ Bi-normal to $\mathcal{H}_v$,
\item $s_w=\frac{\partial \mathcal{L}}{\partial R_{\mu \nu \rho \sigma}} \, \epsilon_{\mu \nu}\epsilon_{\rho \sigma}=$ The Wald entropy density,
\item $S_{W}^{\text{HD}},\, s_w^{\text{HD}} =$ Contributions to integrated Wald entropy ($S_{W}$) and Wald entropy density ($s_w$) from the higher derivative part of the gravity Lagrangian $\mathcal{L}^{\text{HD}}$. \newline It can be shown that: $s_w = 1 +s_w^{\text{HD}}$, such that for Einstein gravity one obtains $s_w = 1$.
\item The density of time variation of Wald entropy is denoted by $\vartheta$, defined as 
$$\partial_v S_W = \int_{\mathcal{H}_v} d^{d-2}x \,\sqrt{h}\,\vartheta \,;   \quad \vartheta = \vartheta_{E} + \vartheta^{\text{HD}},$$
such that $\vartheta_{E} = {1 \over \sqrt{h}} \partial_v (\sqrt{h}) =$ contribution from Einstein gravity, and $\vartheta^{\text{HD}} ={1 \over \sqrt{h}} \partial_v (\sqrt{h}\, s_w^{\text{HD}}) = $ contribution from higher derivative part of the Lagrangian $\mathcal{L}^{\text{HD}}$, 
\item $s_{c} =$ correction to the entropy density which vanish on stationary solutions,
\item $E_{vv}^{\text{HD}} = $ `vv' component of the equation of motion, getting contribution only from the higher derivative part of the Lagrangian $\mathcal{L}^{\text{HD}}$,
\end{enumerate} 
\item Useful definitions:
\begin{enumerate}
\item The extrinsic curvatures of the horizon $\mathcal{H}$: \\ 
(a). ~$\mathcal{K}_{ij}= {1 \over 2}\partial_{v}h_{ij}$; \qquad  $\mathcal{K}^{ij}= -{1 \over 2}\partial_{v}h^{ij}$ ,\\ 
(b). ~$\overline{\mathcal{K}}_{ij}= {1 \over 2}\partial_{r}h_{ij}$; \qquad  $\overline{\mathcal{K}}^{ij}= -{1 \over 2}\partial_{r}h^{ij}$.
\item The trace of the extrinsic curvatures:\\
(a). ~$\mathcal{K} = {1 \over 2}h^{ij}\partial_{v}h_{ij}=\frac{1}{\sqrt{h}}\partial_{v}\sqrt{h}$ , \\ 
(b). ~$\overline{\mathcal{K}} = {1 \over 2} h^{ij}\partial_{r}h^{ij}=\frac{1}{\sqrt{h}}\partial_{r}\sqrt{h}$ .
\end{enumerate}
\item Expressions for the components of Riemann tensors, Ricci tensors and Ricci scalar on the horizon:
\begin{equation}
\begin{split}
R_{rvrv} &= X +\frac{1}{4} \omega^2\\  
R_{rvri} &= -\partial_r \omega_i+\frac{1}{2} \omega^j \Kn_{ij}\\  
R_{rvvi} &= -\frac{1}{2}\left(\partial_v \omega_i+\omega^j \Kt_{ij}\right)\\  
R_{rirj} &= -\partial_r \Kn_{ij}+ \Kn_{ik}\Kn^k_j\\  
R_{rivj} &= -\partial_r \Kt_{ij} + \frac{1}{2}\nabla_j \omega_i - \frac{1}{4}\omega_i\omega_j + \Kn_{jk} \Kt^{k}_i\\  
R_{vivj} &= -\partial_v \Kt_{ij}+ \Kt_{ik}\Kt^k_j\\  
R_{ijvk} &=\nabla_j \Kt_{ik} -  \nabla_i \Kt_{jk}-\frac{1}{2}\omega_i {\cal K}_{jk}+\frac{1}{2}\omega_j {\cal K}_{ik}\\  
R_{ijrk} &= \left(\nabla_j - \frac{1}{2}\omega_j\right) \Kn_{ik} - \left(\nabla_i - \frac{1}{2}\omega_i\right) \Kn_{jk}\\  
R_{ijkl} &=\mathcal{R}_{ijkl} - \Kn_{ik}\Kt_{jl} - \Kt_{ik}\Kn_{jl}+ \Kn_{il}\Kt_{jk} + \Kt_{il}\Kn_{jk}
\end{split}
\end{equation}
where $\nabla_i$ is the covariant derivative with respect to the induced metric $h_{ij}$.
\item Expressions for the components of Ricci tensors on the horizon:
\begin{equation}
\begin{split}
& R_{rr} = -\partial_{r} \Kn - \Kn_{ij} \, \Kn^{ij}\\  
& R_{rv} = - X-\frac{1}{2} \, \omega^2- \partial_r \Kt -  \Kn_{ij} \, \Kt^{ij} + \frac{1}{2} \, \nabla^i \,  \omega_i\\  
& R_{ri} = \partial_r \omega_i - \frac{1}{2} \, \Kn_i^j \,  \omega_j + \left(\nabla_j - \frac{1}{2} \, \omega_j\right)\Kn^j_i - \left(\nabla_i -\frac{1}{2} \,  \omega_i\right) \Kn \\ 
& R_{vv} = -\partial_v \Kt - \Kt_{ij} \, \Kt^{ij} \\  
& R_{vi} = -\frac{1}{2} \, \partial_v \omega_i - \frac{1}{2} \, \Kt^j_i  \, \omega_j +\left(\nabla_j+\frac{1}{2} \, \omega_j\right) \Kt^j_i -\left(\nabla_i+\frac{1}{2} \, \omega_i\right) \Kt\\  
& R_{ij} =\mathcal{R}_{ij} - 2  \, \partial_r \Kt_{ij} +\frac{1}{2}\left(\nabla_j \omega_i +\nabla_i \omega_j -  \omega_i \, \omega_j\right)- \Kn \,  \Kt_{ij} -\Kt \,  \Kn_{ij}\\    & ~~~~~~~~ +2  \left(\Kn_{ik} \, \Kt^k_j + \Kn_{jk} \, \Kt^k_i\right)  
\end{split}
\end{equation}
\item Expressions for the Ricci scalar on the horizon:
\begin{equation}
\begin{split}
R={\cal R}-2X-\frac{3}{2}~\omega^2-4~\partial_r {\cal K}+2(\nabla\cdot\omega)-2~\bar{\cal K}_ {ij}{\cal K}^ {ij}-2~{\cal K} \, \overline{\cal K}
\end{split}
\end{equation}
\end{itemize}

\section{Detailed expressions}

\subsection{Expressions of Riemann tensors and Ricci tensors off the horizon}
As we will compute the `vv'-component of the equations of motion $E_{vv}^{\text{HD}}$, we will need the following expressions for the components of  Riemann tensors and Ricci tensor calculated off the horizon, i.e. without imposing $r=0$,
{{
\begin{equation}
\begin{split}
R_{rvrv} &=X +\frac{1}{4} \omega^2\\
R_{rviv}&=\frac{1}{2}\nabla_i(2rX+r^{2}\partial_r X)+\frac{1}{2}\partial_v (\omega_i+r\partial_{r}\omega_{i})+\frac{r^{2}}{2}(2X+r\partial_r X)\omega^j\bar{\cal K}_{ij}\\
 &~-r(\partial_v \omega_j)\bar{\cal K}^j_i-\frac{r^{2}}{2}(\nabla_jX )\bar{\cal K}^j_i  +\frac{r}{4}(\omega^{j}(\omega_{j}+r\partial_{r}\omega_{j}))(\omega_i+r\partial_{r}\omega_{i})\\
 &~-\frac{r}{4}(\omega^j+r\partial_{r}\omega^j)(\nabla_j \omega_i-\nabla_i \omega_j-\cal {K}_{ij}) \\
4\, R_{vivj}&=r^{2}\left(\nabla^m \omega_i\right)\left(\nabla_m\omega_j\right)+r^{2}\left(\nabla_i \omega^m\right)\left(\nabla_j \omega_m\right)-r^{2}(\nabla_m \omega_i)(\nabla_j \omega^m)\\  
&~-r^{2}(\nabla_m \omega_j)(\nabla_i \omega^m) -2r(\nabla_m \omega_i){\cal K}^m_j-2r(\nabla_m \omega_j){\cal K}_i^m+4~{\cal K}_{im}{\cal K}^m_j\\  
&~+r^2(\nabla_i \omega^m){\cal K}_{jm}+2r(\nabla_j \omega^m){\cal K}_{im}+2~\bar{\cal K}_{ij}\big[r^2\omega^{2}(2rX+r^{2}\partial_r X)\\  
&~+r^{2}X(2rX+r^{2}\partial_r X)-r^{3}(\omega\cdot\nabla)X-2~r^{2}\omega^m(\partial_v \omega_m)-(\partial_v r^{2}X)\big]\\  
&~+~(\omega_j+r\partial_{r}\omega_{i})\left[2~r\omega^m{\cal K}_{im}-r^{2}(\omega\cdot\nabla)\omega_i+r^{2}\omega^m(\nabla_i \omega_m)+(\nabla_i r^{2}X)\right]\\  
&~+4r^{2}~(r^{2}X+r^{r}\omega^2)(\omega_i+r\partial_{r}\omega_{i})(\omega_j+r\partial_{r}\omega_{j})\\  
&~+~(\omega_i+r\partial_{r}\omega_{i})\left[2~r\omega^m{\cal K}_{jm}-r^{2}(\omega\cdot\nabla)\omega_j+r^{2}\omega^m(\nabla_j \omega_m)+(\nabla_j r^{2}X)\right]\\  
&~-r^{2}(2X+r\partial_r X)\left[\nabla_i \omega_j+\nabla_j \omega_i\right]\\  
&~+2r^{2}\nabla_i\nabla_j X+2(2rX+r^{2}\partial_r X){\cal K}_{ij}-4~\partial_v{\cal K}_{ij}\\  
&~+2r~\nabla_i(\partial_v \omega_j)+2~r\nabla_j(\partial_v \omega_i)\\
\end{split}
\end{equation}
\begin{equation}
\begin{split}
R_{vv}&=r^{2}(X+\omega^2)\bigg[\frac{1}{2}(2X+4rX+r^{2}\partial_{r}^{2}X)+\frac{1}{2}(\omega^{i}+r\partial_{r}\omega^{i})(\omega_{i}+r\partial_{r}\omega_{i})\\ 
&~+\frac{1}{2}\bar{\cal K}(2rX+r^{2}\partial_r X)\bigg]-r^{2}(\omega\cdot\nabla)(2X+r\partial_rX) -r\omega^i(\partial_v(\omega_i+r\partial_{r}\omega_{i}))\\ 
&~-r^{3}(2X+r\partial_r X)(\omega^i\bar{\cal K}_{ij}\omega^j)+2~r^{2}(\partial_v \omega_i)\omega_j\bar{\cal K}^{ij}+r^{3}(\nabla_j X)\omega_i\bar{\cal K}^{ij}\\ 
&~-\frac{r^{2}}{2}(\omega^i(\omega_{i}+r\partial_{r}\omega_{i}))^2+r^{2}\omega^i((\omega+r\partial_{r}\omega)\cdot\nabla)\omega_i\\ 
&~-r^{2}(\omega^i+r\partial_{r}\omega^{i})(\omega\cdot\nabla)\omega_i+\frac{r^{2}}{2}(\nabla^i \omega^j)(\nabla_i \omega_j)-\frac{r^{2}}{2}(\nabla_j \omega^i)(\nabla_i \omega^j)\\ 
&~-\partial_v {\cal K}-{\cal K}_{ij}{\cal K}^{ij}+\frac{r^{2}}{2}(\nabla^2 X)+\frac{1}{2}(2rX+r^{2}\partial_r X){\cal K}\\ 
&~+r\nabla^i(\partial_v \omega_i)-\frac{r^{2}}{2}\bar{\cal K}(\omega\cdot\nabla)X-\frac{r^{2}}{2}\bar{\cal K}(\partial_v X)-r^{2}\bar{\cal K}\omega^i (\partial_v \omega_i)\\ 
&~+\frac{r^{2}}{2}((\omega+r\partial_{r}\omega)\cdot\nabla)X-\frac{r^{2}}{2}(2X+r\partial_r X)(\nabla\cdot {\omega})
\end{split}
\end{equation}
}}

\subsection{Relevant terms on the horizon $\mathcal{H}$, to compute $E_{vv}^{\text{HD}}$ for different theories}

\begin{equation}
\begin{split}
R^{\alpha}_{v} \,  R_{\alpha v}=&~(\partial_{v}\mathcal{K})\left(2 \, X + \omega^{2} - \nabla \cdot \omega +2 \, \partial_{v}\bar{\mathcal{K}} \right) \\  
{R_{v}}^{\alpha \beta \gamma} \, R_{v \alpha \beta \gamma} =&~ \omega^{i} \omega^{j} \partial_{v}\mathcal{K}_{ij} -2 \, \nabla^{i}\omega^{j} \partial_{v}\mathcal{K}_{ij} +\, 4 \, \partial_{v}\overline{\mathcal{K}}_{ij} \, \partial_{v}\mathcal{K}^{ij}\\  
R^{\alpha \beta} \, R_{v\alpha v \beta}=& - \left(X+ \frac{\omega^{2}}{4}\right)(\partial_{v}\mathcal{K}) - \mathcal{R}^{ij} \, \partial_{v}\mathcal{K}_{ij} -  (\nabla^{i} \omega^{j}) \partial_{v}\mathcal{K}_{ij} \\   &   + \frac{1}{2}\omega^{i}\omega^{j} \partial_{v}\mathcal{K}_{ij}   +2 \, \partial_{r}\mathcal{K}_{ij} \, \partial_{v}\mathcal{K}^{ij} \\ 
D_{v}D_{v} R =&~\partial_{v}^{2} \mathcal{R} - 2 \, \partial_{v}^{2}X - 3 \,  \omega^{i} \partial_{v}^{2} \omega_{i}  + 3 \,  \omega^{i}\omega^{j} \partial_{v}\mathcal{K}_{ij}  \\   &
+ 2 \,  \partial_{v}^{2}(\nabla \cdot \omega)-4 \, \partial_{r}\partial_{v}^{2}\mathcal{K}-2 \, \overline{\mathcal{K}}_{ij}\partial_{v}^{2}\mathcal{K}^{ij} \\ 
& -2 \, \overline{\mathcal{K}} \, \partial_{v}^{2}\mathcal{K} 
 -4 \, \partial_{v}\overline{\mathcal{K}}_{ij} \, \partial_{v}\mathcal{K}^{ij}-4 \, \partial_{v}\mathcal{K} \, \partial_{v}\mathcal{\bar{K}}\\ 
D_{\alpha}D^{\alpha} R_{vv}=&\left(\frac{3}{2} \, \omega^{2}-  \nabla\cdot \omega+4 \, X\right)(\partial_{v}\mathcal{K}) + (\omega \cdot \nabla)\partial_{v}\mathcal{K} \\
& - 2 \omega^{i} \, \nabla^{j}(\partial_{v} \mathcal{K}_{ij})-2 \, \partial_r\partial_{v}^{2}\mathcal{K}   - \omega^{i}  \, \partial_{v}^{2} \omega_{i} - \nabla^{2}(\partial_{v}\mathcal{K})\\    
& +2 \,  \nabla^{i} (\partial_{v}^{2}\omega_{i})-\overline{\mathcal{K}}\partial_{v}^{2}\mathcal{K} -4 \, \partial_{r}\mathcal{K}_{ij} \, \partial_{v}\mathcal{K}^{ij} \, .
\end{split}
\end{equation}
Note that, to obtain the last two expressions above we need to first evaluate $D_{\mu}D_{\nu}R_{\alpha\beta}$ where the indices $\mu, \nu$ runs over the full space-time coordinates: $v,\, r, \, x^i$ and $D_{\mu}$ is covariant derivative with respect to the full space-time metric $g_{\mu\nu}$. 

\subsection{Ricci scalar square Theory}
Following the discussions in sections \S\ref{ssec:4dercor} and \S\ref{sssec:ricciscal} here we write down the detailed expressions for various quantities for the Ricci Scalar squared theory. 

The `vv'-component of $E^{\text{HD}}_{\mu\nu}$ from \eqref{evvhdth1}
\begin{equation}
\begin{split}
E^{\text{HD}}_{vv}&=  a_{1} \bigg[\partial_{v}\left(\frac{1}{\sqrt{h}}\partial_{v}(\sqrt{h}\, (\, 3  \, \omega^{2} - 4  \, (\nabla \cdot \omega) + 4  \, X\, ) \right) - 2 \, \mathcal{R} \, \partial_{v} \mathcal{K} - 2 \, \partial_{v}^{2}\mathcal{R} \\
& +8 \, \partial_{r}\partial_{v}^{2}\mathcal{K}+4 \, \bar{\mathcal{K}}_ {ij} \, \partial_{v}^{2}\mathcal{K}^ {ij}+4 \, \bar{\mathcal{K}} \, \partial_{v}^{2}\mathcal{K}+8\, \partial_{v}\bar{\mathcal{K}}_ {ij} \, \partial_{v}\mathcal{K}^ {ij}+16 \, \partial_{v}\mathcal{K} \, \partial_{v}\mathcal{\bar{K}}\bigg] + \mathcal{O}[\epsilon^{2}]\, .
\end{split}
\end{equation}
From \eqref{eq:WaldRicci} we know the Wald entropy density as
\begin{equation}
s^{\text{HD}}_w = 2\, a_{1}\, R\, ,
\end{equation}
and therefore, we immediately obtain 
\begin{equation}
\begin{split}
\partial_{v}\left(\frac{1}{\sqrt{h}}\partial_{v} \left(\sqrt{h}\, s^{\text{HD}}_w \right)\right)=& 2 \, a_{1} \, \partial_{v}\bigg(\frac{1}{\sqrt{h}}\,  \partial_{v}\bigg(\sqrt{h}\, \big({\cal R}-2\, X-\frac{3}{2}\, \omega^2-4\, \partial_r {\cal K} \\
& +2(\nabla\cdot\omega)-2\, \bar{\cal K}_ {ij}{\cal K}^ {ij}-2\, {\cal K}\bar{\cal K} \big)\bigg)\bigg).
\end{split}
\end{equation}
We can now use \eqref{eq:esdiff} and after some algebraic manipulation we obtain 
\begin{equation}
{\mathbb{E}^{\text{HD}}_{vv}}^* = \mathcal{O}[\epsilon^{2}].
\end{equation}
Finally, comparing with \eqref{eq:Estar} we see that for Ricci scalar squared theory there is no spatial entropy current
\begin{equation}
J^i = 0.
\end{equation}

\subsection{Ricci tensor squared Theory}
The `vv'-component of $E^{\text{HD}}_{\mu\nu}$ for Ricci tensor squared theory, following \eqref{evvhdth1} as discussed in \S\ref{sssec:ricciten}, comes out to be
\begin{equation}
\begin{split}
E^{\text{HD}}_{vv}=  a_{2} \bigg[&\partial_{v}\bigg(\frac{1}{\sqrt{h}} \, \partial_{v}\bigg(\sqrt{h} \, \big(\omega^{2}+2 \, X-2 \, \nabla \cdot \omega+2 \, \partial_{v}\bar{\mathcal{K}}+\bar{\mathcal{K}} \, \mathcal{K}+2 \, \bar{\mathcal{K}}_{ij} \, \mathcal{K}^{ij}\big)\bigg) \bigg)  \\
&+\partial_{v}\big(\nabla_{i} \, \big(2 \,h^{ij} \,\partial_v \omega_j+ \omega^i \, \mathcal{K}-h^{ij} \, \nabla_{j} \mathcal{K}-2\, \omega_j \,\mathcal{K}^{ij}\big)\big)\\
& -2 \, \partial_{v}\big(\nabla_{i}\nabla_{j}\big(\mathcal{K}^{ij}-\mathcal{K} \, h^{ij}\big)\big)\bigg]
\end{split}
\end{equation}
From \eqref{eq:WaldRicciT} we recognize that the Wald entropy density coming from the higher derivative part of the Lagrangian is 
\begin{equation}
s^{\text{HD}}_w = 2 \,a_2 \, R_{rv}
\end{equation}
and therefore we compute
\begin{equation}
\begin{split}
\partial_{v}\left(\frac{1}{\sqrt{h}}\partial_{v} \left(\sqrt{h}\, s^{\text{HD}}_w \right)\right)= 2 \, a_{2} \, \partial_{v}\bigg(\frac{1}{\sqrt{h}}\,  \partial_{v}\bigg(\sqrt{h}\,\bigg(&   -X-\frac{1}{2}\,  \omega^2- \partial_r \mathcal{K} \\
& -\bar{\mathcal{K}}_{ij}\,\mathcal{K}^{ij}  +\frac{1}{2} \, \nabla^i \omega_i\bigg)\bigg)\bigg).
\end{split}
\end{equation}
Using the definition as given in \eqref{eq:esdiff} we calculate the following 
\begin{equation}
{\mathbb{E}^{\text{HD}}_{vv}}^* = a_2 \, \partial_{v}\bigg[\frac{1}{\sqrt{h}}\,  \partial_{v}\big(\sqrt{h}\, \bar{\mathcal{K}} \, \mathcal{K}\big)+ \nabla_{i}\big(h^{ij} \, \nabla_{j} \mathcal{K} + h^{ij} \,\partial_v \omega_j- 2\,\nabla_{j}\mathcal{K}^{ij} \big)\bigg], 
\end{equation}
where to derive this we have used the identity
\begin{equation}
\partial_{v}\bigg[\frac{1}{\sqrt{h}}\,  \partial_{v}\big(\sqrt{h}\, \nabla \cdot \omega \big) \bigg] = \nabla_{i}\big( h^{ij} \,\partial_v \omega_j -2 \, \mathcal{K}^{ij} \omega_j +\omega^i \, \mathcal{K} \big)
\end{equation}

Therefore the spatial entropy current turns out to be
\begin{equation}
\begin{split}
J^{v} =& - s^{\text{HD}}_w - \, a_2\,  \bar{\mathcal{K}} \, \mathcal{K}\, ,\\
J^{i} =& \, a_{2} \,\big(2\,\nabla_{j}\mathcal{K}^{ij} - h^{ij} \, \nabla_{j} \mathcal{K} -h^{ij} \,\partial_v \omega_j \big).
\end{split}
\end{equation}

\subsection{Riemann tensor squared Theory}

Following the same steps as followed in the previous subsections for the cases of Ricci scalar squared and Ricci tensor squared theory, we compute the `vv'-component of equations of motion for Riemann tensor squared theory, previously discussed in \S\ref{sssec:riesq}, as follows
\begin{equation}
\begin{split}
E^{\text{HD}}_{vv}=  a_{3} \bigg[&\partial_{v}\bigg(\frac{1}{\sqrt{h}} \, \partial_{v}\bigg(\sqrt{h} \, \big(\omega^{2}+4 \, X-4 \, \nabla \cdot \omega+4 \, \bar{\mathcal{K}}_{ij} \, \mathcal{K}^{ij}\big)\bigg) \bigg)  \\
&+4\, \partial_{v}\big(\nabla_{i} \, \big(2 \,h^{ij} \,\partial_v \omega_j+ \omega^i \, \mathcal{K}-h^{ij} \, \nabla_{j} \mathcal{K}-2\, \omega_j \,\mathcal{K}^{ij}\big)\big)\\
& -4 \, \partial_{v}\big(\nabla_{i}\nabla_{j}\big(\mathcal{K}^{ij}-\mathcal{K} \, h^{ij}\big)\big)\bigg]
\end{split}
\end{equation}
We take note of the fact that in this case the Wald entropy density for the higher derivative part of the Lagrangian is 
\begin{equation}
s^{\text{HD}}_w = -4 \, a_{3} \, R_{rvrv}
\end{equation}
and using this we compute 
\begin{equation}
\begin{split}
\partial_{v}\left(\frac{1}{\sqrt{h}}\partial_{v} \left(\sqrt{h}\, s^{\text{HD}}_w \right)\right)= -4 \, a_{3} \, \partial_{v}\bigg(\frac{1}{\sqrt{h}}\,  \partial_{v}\bigg(\sqrt{h}\,\bigg(X+ \frac{1}{4}\,  \omega^2\bigg)\bigg)\bigg).
\end{split}
\end{equation}
Next we compute ${\mathbb{E}^{\text{HD}}_{vv}}^*$ defined in \eqref{eq:esdiff} as given below,
\begin{equation}
{\mathbb{E}^{\text{HD}}_{vv}}^* = 4 \, a_{3} \, \partial_{v}\bigg[\frac{1}{\sqrt{h}}\,  \partial_{v}\big(\sqrt{h}\, \bar{\mathcal{K}}^{ij} \, \mathcal{K}_{ij}\big)+ \nabla_{i}\big(h^{ij} \,\partial_v \omega_j- \nabla_{j}\mathcal{K}^{ij} \big)\bigg].
\end{equation}

Finally we are now at a stage to write down the expressions for the components of the entropy current 
\begin{equation}
\begin{split}
J^{v} =& - s^{\text{HD}}_w - 4\, a_3\,\bar{\mathcal{K}}^{ij} \, \mathcal{K}_{ij}\, ,\\
J^{i} =& \, 4 \, a_{3} \,\big(\nabla_{j}\mathcal{K}^{ij} -h^{ij} \,\partial_v \omega_j\big)
\end{split}
\end{equation}

%
%
%
%
%
%
\bibliographystyle{JHEP}
\bibliography{WaldEntropyCurrent}
\end{document}